\documentclass[a4paper,11pt]{amsart}

\usepackage{amssymb, amsmath, amsthm, xcolor, amstext, graphicx, float, appendix, mathrsfs, mathabx, dsfont}

\usepackage{tikz}
\usepackage[all,cmtip]{xy}
\usepackage{caption}
\usepackage[colorlinks=true, linkcolor=black, breaklinks=true, citecolor=blue, urlcolor=red]{hyperref} 
\usepackage[english]{babel}

\numberwithin{equation}{section}

\usetikzlibrary{decorations.markings}

\newtheorem{theo}{Theorem}[section]

\newtheorem{corollary}[theo]{Corollary}
\newtheorem{prop}[theo]{Proposition}

\theoremstyle{definition} 
\newtheorem{definition}[theo]{Definition}

\newtheorem{remark}[theo]{Remark}

\newcommand{\textb}{\textcolor{blue}}
\newcommand{\textr}{\textcolor{red}}

\newcommand{\textg}{\textcolor{green}}

\makeatletter
\DeclareRobustCommand{\cev}[1]{%
  \mathpalette\do@cev{#1}%
}
\newcommand{\do@cev}[2]{%
  \fix@cev{#1}{+}%
  \reflectbox{$\m@th#1\vec{\reflectbox{$\fix@cev{#1}{-}\m@th#1#2\fix@cev{#1}{+}$}}$}%
  \fix@cev{#1}{-}%
}
\newcommand{\fix@cev}[2]{%
  \ifx#1\displaystyle
    \mkern#23mu
  \else
    \ifx#1\textstyle
      \mkern#23mu
    \else
      \ifx#1\scriptstyle
        \mkern#22mu
      \else
        \mkern#22mu
      \fi
    \fi
  \fi
}

\makeatother

\DeclareMathOperator{\Ad}{Ad}

\DeclareMathOperator{\Aut}{Aut}

\DeclareMathOperator{\flip}{flip}

\DeclareMathOperator{\Homeo}{Homeo}

\DeclareMathOperator{\id}{id}

\DeclareMathOperator{\supp}{supp}

\DeclareMathOperator{\tr}{tr}
\DeclareMathOperator{\Tr}{Tr}

\DeclareMathOperator{\vol}{vol}

\newcommand{\act}{\curvearrowright}
\newcommand{\tca}{\curvearrowleft}
\newcommand{\scrA}{\mathscr A}\newcommand{\fA}{\mathfrak A}
\newcommand{\cB}{\mathcal B}\newcommand{\scrB}{\mathscr B}\newcommand{\fB}{\mathfrak B}
\newcommand{\ub}{\textup{b}}

\newcommand{\C}{\mathds C}
\newcommand{\Cyl}{\textup{Cyl}}
\newcommand{\uc}{\textup{c}}
\newcommand{\D}{\mathbf D}
\newcommand{\ud}{\textup{d}}
\newcommand{\cE}{\mathcal E}
\newcommand{\cF}{\mathcal F}\newcommand{\scrF}{\mathscr F}\newcommand{\fr}{\textup{fr}}

\newcommand{\fg}{\mathfrak g}
\newcommand{\h}{\textup{hol}}
\newcommand{\cH}{\mathcal H}
\newcommand{\fI}{\mathfrak I}

\newcommand{\scrL}{\mathscr L}

\newcommand{\uL}{\textup{L}}

\newcommand{\ltwo}{\ell^{2}}\newcommand{\li}{\ell^{\infty}}
\newcommand{\cM}{\mathcal{M}}\newcommand{\scrM}{\mathscr{M}}
\newcommand{\N}{\mathds{N}}
\newcommand{\ot}{\otimes}

\newcommand{\ovt}{\overline\otimes}
\newcommand{\cO}{\mathcal O}
\newcommand{\cP}{\mathcal P}

\newcommand{\R}{\mathds{R}}\newcommand{\cR}{\mathcal R}\newcommand{\uR}{\textup{R}}\newcommand{\Rot}{\textup{Rot}}
\newcommand{\fS}{\mathbf{S}}

\newcommand{\cS}{\mathcal{S}}
\newcommand{\fT}{\mathfrak T}\newcommand{\T}{\mathds T}

\newcommand{\utriv}{\textup{triv}}
\newcommand{\cU}{\mathcal U}

\newcommand{\uY}{\textup{Y}}
\newcommand{\Z}{\mathds{Z}}

\begin{document}

\tikzset{->-/.style={decoration={
  markings,
  mark=at position #1 with {\arrow{latex}}},postaction={decorate}}}

\title[Canonical quantization of Yang-Mills theory]{Canonical quantization of 1+1-dimensional Yang-Mills theory: an operator-algebraic approach}
\thanks{Both authors were partially supported by European Research Council Advanced Grant 669240 QUEST. AS is supported by Alexander-von-Humboldt Foundation through a Feodor Lynen Return Fellowship.
AB is supported by a UNSW starting grant.}
\author{Arnaud Brothier and Alexander Stottmeister}
\address{Arnaud Brothier\\ School of Mathematics and Statistics, University of New South Wales, Sydney NSW 2052, Australia}
\email{arnaud.brothier@gmail.com\endgraf
\url{https://sites.google.com/site/arnaudbrothier/}}
\address{Alexander Stottmeister\\Mathematical Institute, University of M{\"u}nster, Einsteinstra{\ss}e 62, 48149 M{\"u}nster, Germany}
\email{alexander.stottmeister@gmail.com}

\begin{abstract}
We present a mathematically rigorous canonical quantization of Yang-Mills theory in 1+1 dimensions (YM$_{1+1}$) by operator-algebraic methods. The latter are based on Hamiltonian lattice gauge theory and multi-scale analysis via inductive limits of $C^{*}$-algebras which are applicable in arbitrary dimensions. The major step, restricted to one spatial dimension, is the explicitly construction of the spatially-localized von Neumann algebras of time-zero fields in the time gauge in representations associated with scaling limits of Gibbs states of the Kogut-Susskind Hamiltonian. We relate our work to existing results about YM$_{1+1}$ and its counterpart in Euclidean quantum field theory (YM$_{2}$). In particular, we show that the operator-algebraic approach offers a unifying perspective on results about YM$_{1+1}$ obtained by Dimock as well as Driver and Hall, especially regarding the existence of dynamics. Although our constructions work for non-abelian gauge theory, we obtain the most explicit results in the abelian case by applying the results of our recent companion article. In view of the latter, we also discuss relations with the construction of unitary representations of Thompson's groups by Jones. To understand the scaling limits arising from our construction, we explain our findings via a rigorous adaptation of the Wilson-Kadanoff renormalization group, which connects our construction with the multi-scale entanglement renormalization ansatz (MERA). Finally, we discuss potential generalizations and extensions to higher dimensions ($d+1\geq 3$).
\end{abstract}

\maketitle

\tableofcontents

\section{Introduction}
\label{sec:intro}
Quantum gauge theory and more specifically quantum Yang-Mills theory is an essential building block of the standard model of particle physics encompassing, apart from the Higgs field, all fundamental interactions. But, despite an enormous amount of work leading to an ever growing deep understanding of its physical and mathematical structure, a completely rigorous definition beyond the perturbative regime remains elusive. While the success of the perturbative approach is based on renormalizability \cite{tHooftRegularizationAndRenormalization} and asymptotic freedom \cite{GrossUltravioletBehaviorOf, PolitzerReliablePerturbativeResults}, progress \cite{DuerrAbInitionDetermination} in the non-perturbative regime relies on the lattice formulation \cite{WilsonConfinement, KogutHamiltonianFormulation} and its amenability to numerical methods. Also the rigorous construction of the theory is easier to access in the lattice approach compared to a quantization of the classical continuum theory because (finite) lattices provide ultraviolet (UV) and infrared (IR) cutoffs that render the path integral (in the Euclidean framework) \cite{OsterwalderGaugeFieldTheories, SeilerGaugeTheoriesAs} or the Hamiltonian (in the time-zero formulation) \cite{BorgsLatticeYangMills,TomboulisFiniteTemperatureSU(2), BorgsConfinementDeconfinementAnd} well-defined. In both cases, the construction of the aimed for continuum quantum theory is relegated to the difficult problem of taking a non-trivial scaling limit of a lattice model.\\[0.1cm]
To date the Euclidean framework has seen a multitude of remarkable results and constitutes the most advanced attack on establishing the existence of the continuum limit in dimensions 2, 3 and 4. In two dimension, the results are most complete \cite{BrydgesConstructionOfQuantized1, BrydgesConstructionOfQuantized2, BrydgesConstructionOfQuantized3} (for the $U(1)$-Higgs model) and \cite{FineQuantumYangMillsSphere, FineQuantumYangMillsRiemann, WittenOnQuantumGauge, DriverYM2, GrossTwoDimensionalYang, SenguptaQuantumGaugeTheory, SenguptaTheYangMills, AshtekarSU(N)QuantumYang, FleischhackANewType} (for pure Yang-Mills fields), and there is still ongoing progress \cite{LevyYangMillsMeasure, LevyTwoDimensionalMarkovian, LevyTheMasterField, NguyenQuantumYangMills, DriverThreeProofsOf, DriverTheMakeenkoMigdal}. In three or more dimensions, the situation much more involved and less clear, a famous success being the proof ultraviolet stability (in finite volume) for pure (non-abelian) quantum Yang-Mills theory by the renormalization group \cite{WilsonTheRenormalizationGroupKondo} inspired approaches of Ba{\l}aban \cite{BalabanHiggsQuantumFields1, BalabanHiggsQuantumFields2, BalabanHiggsQuantumFields3, BalabanPropagatorsAndRenormalization1, BalabanPropagatorsAndRenormalization2, BalabanAveragingOperationsFor, BalabanRenormalizationGroupApproach1, BalabanRenormalizationGroupApproach2} and Federbush \cite{BattleDivergenceFreeVector, FederbushAPhaseCell1, FederbushAPhaseCell2, FederbushAPhaseCell3, FederbushAPhaseCell4, FederbushAPhaseCell5, FederbushAPhaseCell6} as well as the continuum approach of Magnen, Rivasseau and S{\'e}n{\'e}or \cite{MagnenConstructionOfYM4}. A related important achievement, is the construction of the 3-dimensional $U(1)$-Higgs model by King \cite{KingTheU(1)Higgs1, KingTheU(1)Higgs2} based on Ba{\l}aban's methods. The existence of the scaling limit of pure $U(1)$ Yang-Mills theory (electromagnetic fields) was established by Gross \cite{GrossConvergenceOfU(1)} and Driver \cite{DriverConvergenceOfU(1)}. Recently, Dimock \cite{DimockTheRenormalizationGroup1, DimockTheRenormalizationGroup2, DimockTheRenormalizationGroup3} has reviewed Ba{\l}aban's approach and demonstrated its validity beyond gauge theories, namely for $\Phi^{4}_{3}$. Other recent work deals with large $N$-limits \cite{tHooftAPlanarDiagram} as well as weak-coupling asymptotics \cite{ChatterjeeRigorousSolutionOf, ChatterjeeThe1NExpansion, BasuSO(N)LatticeGauge, ChatterjeeTheLeadingTerm, ChatterjeeWilsonLoopsIn}.\\[0.1cm]
In contrast, the Hamiltonian (or $\tau$-continuum) formulation of lattice gauge theory is mathematically less well-developed. Notable recent work on an operator-algebraic formulation is due to Kijowski and Rudolph \cite{KijowskiOnTheGauss, JarvisOnTheStructure} and the existence of the thermodynamic limit has been proved by Rudolph and Grundling \cite{GrundlingQCDOnAn, GrundlingDynamicsForQCD}. Moreover, there are works on the continuum theory of pure Yang-Mills fields by Dimock \cite{DimockCanonicalQuantizationOf} as well as Driver and Hall \cite{DriverYangMillsTheory} in one spatial dimension.\\[0.1cm]
This said, we initiate in the present article an analysis of Hamiltonian lattice gauge theory with the aim of understanding the problem of scaling limits from an operator-algebraic perspective. To this end, we intend to provide a rather complete analysis of the problem in the case of 1+1 dimensional (pure) Yang-Mills theory to complement the theory's exhaustive treatment in the Euclidean framework. Since we construct the model in terms of algebras of localized field operators in the spirit of \cite{GlimmBosonQuantumField}, we hope to advance the understanding of gauge theories in the setting of algebraic quantum field theory \cite{HaagAnAlgebraicApproach, HaagLocalQuantumPhysics} beyond the abelian case, see, for example, \cite{BuchholzLocalityAndTheGauge, BuchholzLocalityAndThe, BuchholzThePhysicalState,FredenhagenChargedStatesIn, MorchioChiralSymmetryBreaking, LoeffelholzMathematicalStructureOf, HollandsRenormalizedQuantumYangMills, StrocchiNonPerturbativeFoundations, FredenhagenBatalinVilkoviskyFormalism2, BeniniACAlgebra, BuchholzTheUniversalC} for results about gauge theories (perturbative and non-perturbative). Our method for dealing with scaling limits is a multi-scale analysis based on well-known inductive-limit techniques from the theory of operator algebras \cite{KadisonFundamentalsOfThe2, TakesakiTheoryOfOperator3, BlackadarOperatorAlgebras} which have been previously considered in loop quantum gravity models \cite{StottmeisterCoherentStatesQuantumIII}, see also for a related approach \cite{AriciQuantumLatticeGauge}. In a wider context, it is interesting to note that our results are directly related to recent work on tensor networks, matrix product states and the multi-scale entanglement renormalization ansatz, see, for example, \cite{BuyensMatrixProductStates, BuyensTensorNetworksFor, DittrichDecoratedTensorNetwork, MilstedQuantumYangMills, OsborneContinuumLimitsOf}. In view of our companion article \cite{BrothierConstructionsOfConformal}, we point out the connection with Jones' construction of unitary representations of Thompson's groups \cite{JonesSomeUnitaryRepresentations, JonesANoGo, JonesScaleInvariantTransfer} special to 1+1 dimensions, see also \cite{BrothierPythagoreanRepresentationsOf}. Building on the functorial machinery of Jones, we consider groups as well as $C^{*}$-algebras instead of Hilbert spaces as a target category. This leads to (covariant) actions of Thompson's groups on the (localized) inductive limit of lattice field $C^{*}$-algebras, which can be related to previously considered unitary representations via the Gel'fand-Naimark-Segal (GNS) construction. The covariant actions of Thompson's groups on the localized inductive-limit algebras resembles (in a discrete way) the action of the orientation preserving diffeomorphisms on conformal nets, see, for example, \cite{BuchholzTheCurrentAlgebra, GabbianiOperatorAlgebrasAnd, BrunettiModularStructureAnd} and also cp.~\cite{OsborneQuantumFieldsFor, KlieschContinuumLimitsOf}. The actions in the context of groups have just been used by one of the authors in a very different context \cite{BrothierHaagerupPropertyFor}.\\[0.1cm]
The article is structured as follows:\\[0.1cm]
In section \ref{sec:hlgt}, we present the necessary notions for an operator-algebraic treatment of Hamiltonian lattice gauge theory in the time gauge. While section \ref{sec:basics} explains the basic constructions on a single (regular) finite lattice in $\R^{d}$, section \ref{sec:scales} introduces a multi-scale treatment in terms of dyadic refinements of a given finite lattice and associated projective and inductive limit constructions. Section \ref{sec:fstruct} states some general structural results necessary for the following sections.\\[0.1cm]
In section \ref{sec:ym11}, we almost exclusively focus on the case of one spatial dimension. We show how Jones' actions of Thompson's groups appear in this context in section \ref{sec:thompson} and depict useful dualities in section \ref{sec:duality}. Following this, we define specific Gibbs states related to the Kogut-Susskind Hamiltonian, the heat-kernel and dual heat-kernel states, and analyze their compatibility with the multi-scale analysis in section \ref{sec:can11}. The structure of the limit von-Neumann algebras of gauge fields arising from the multi-scale treatment in unison with the limit of said Gibbs states and the question of dynamics for the field algebras is elucidated in section \ref{sec:timezero}. The reduction from fields to observables is discussed in section \ref{sec:gaugeobs}. \\[0.1cm]
In section \ref{sec:rgp}, we review our constructions and results from the perspective of the Wilson-Kadanoff renormalization group. We argue that renormalization group theory adapted to the operator-algebraic context provides a clear picture of the multi-scale analysis of section \ref{sec:scales} and the compatibility of states discussed in section \ref{sec:can11}. Moreover, we exemplify the notion of scaling limit and symmetry enhancement by means of the dual heat-kernel states in section \ref{sec:wiener}. \\[0.1cm]
In section \ref{sec:con}, we discuss some open questions, potential directions for future research, specifically the connection between 1+1-dimensional Yang-Mills theory and the $O(2)$-quantum rotor model as well as extension of the heat-kernel and dual heat-kernel states to higher dimensions.

\section{Hamiltonian lattice gauge theory: Kinematics}
\label{sec:hlgt}
In the following, $G$ denotes a unimodular, locally compact group that plays the role of the structure group of classical gauge theory. If $G$ is a Lie group, we will denote its Lie algebra by $\fg$ and the dual by $\fg^{*}$. Direct sums, $\bigoplus_{i\in I}G_{i}$, and direct products, $\bigtimes_{j\in J}G_{j}$, always carry their natural topologies, and all maps are continuous, or smooth in the case of Lie groups, unless stated otherwise. As usual, $C(G)$ are the continuous functions, $C_{b}(G)$ are the bounded ones, $C_{0}(G)$ are those vanishing at infinity, and $C_{c}(G)$ are those of compact support with their common locally convex topologies.\\[0.1cm]
Although, we introduce some analogues of classical object, e.g.~the phase space of lattice gauge fields, we focus on the quantum theoretical setup. A notationally similar treatment can be found in \cite{StottmeisterCoherentStatesQuantumIII} where also the relation with the classical setup and quantization maps is explained, cp.~\cite{KijowskiOnTheGauss, GrundlingQCDOnAn, GrundlingDynamicsForQCD, AriciQuantumLatticeGauge}. We refrain from (re)stating proofs, but only refer to the appropriate places in the literature.

\subsection{Basic constructions}
\label{sec:basics}
Let us recall some basic constructions used in Hamiltonian lattice gauge theory (cf.~\cite{KogutHamiltonianFormulation, KogutTheLatticeGauge, CreutzQuarksGluonsLattices}) that are required throughout the remainder of the article. For sake of simplicity, we assume that the spatial manifold is the standard cube, $\Sigma = [0,L]^{d}\subset\R^{d}$ (free boundary conditions) in Euclidean space, or the torus, $\Sigma = \T_{L}^{d}=\R^{d}/(L\Z)^{d}$ (periodic boundary conditions), of length $L>0$. We work with finite, cubic, axis-aligned, oriented\footnote{The directions of next-neighbor links are assumed to be fixed.} lattices $\gamma\subset\Sigma\cap(a_{\gamma}\Z)^{d}$, with spacing $0<a_{\gamma}<L$. We use $|\gamma|$ to refer to the underlying unoriented lattice of $\gamma$. Although, more general formulations w.r.t.~a suitable class of (embedded) oriented graphs $\gamma\in\Gamma\subset\R^{d}$ are conceivable (cf.~\cite{StottmeisterCoherentStatesQuantumIII, LaneryProjectiveLimitsOf2}), we only comment on this possibility at the appropriate instances to keep the presentation more accessible. Clearly, $L$ is an IR cutoff while the $a_{\gamma}$'s serve as UV cutoffs. To fix notation let us make the following definition.

\begin{definition}
\label{def:cells}
Given a finite, cubic, axis-aligned, oriented lattice $\gamma\subset\Sigma\cap(a_{\gamma}\Z)^{d}$, we can treat it as a $d$-dimensional cell complex. We denote by $V(\gamma)$ the set of its vertices or $0$-cells, by $E(\gamma)$ the set of its edges or $1$-cells, and by $F(\gamma)$ the set of its faces or $2$-cells. In accordance with this identification, we use the standard notation for boundaries, e.g.~for an oriented edge $e\in E(\gamma)$, $\partial e_{\pm}\in V(\gamma)$ are the boundary vertices, where $+$ refers to the endpoint of $e$ and $-$ to its starting point. Moreover, we will use the notation $e^{-1}$ to indicate a reversal of orientation.  
\end{definition}

Next, we define the space of (discrete) gauge-field configurations and the corresponding gauge group associated with a lattice $\gamma\in\Sigma$. The former is modeled as the collection of $G$-valued parallel transports or holonomies along the edges, $e\in E(\gamma)$, while the latter induces changes of reference points in the thought-of affine $G$-fibres over the vertices, $v\in V(\gamma)$.

\begin{definition}
\label{def:gaugefields}
For $\gamma$ as above, put $G_{v}=G_{e}=G$ for $v\in V(\gamma), e\in E(\gamma)$. We call the direct products,
\begin{align*}
\scrA^{E(\gamma)} & = \bigtimes_{e\in E(\gamma)}G_{e}, & G^{V(\gamma)} & = \bigtimes_{v\in V(\gamma)}G_{v},
\end{align*}
its space of gauge-field configurations\footnote{We use the symbol $\scrA^{E(\gamma)}$ to denote the space of gauge-field configuration in accordance with common notation for connections in principal bundles. Moreover, we emphasize in this way that $\scrA^{E(\gamma)}$ has only an affine group structure.} and its group of gauge transformations respectively. Associated with these objects, we have left and right multiplications,
\begin{align*}
L ^{(\gamma)} & : G^{E(\gamma)}\times \scrA^{E(\gamma)} \longrightarrow \scrA^{E(\gamma)}, \\
R^{(\gamma)} & : G^{E(\gamma)}\times \scrA^{E(\gamma)} \longrightarrow \scrA^{E(\gamma)},
\end{align*}
and the action of the gauge group,
\begin{align*}
\tau^{(\gamma)} : G^{V(\gamma)}\times \scrA^{E(\gamma)} & \longrightarrow \scrA^{E(\gamma)}, \\
 (\{g_{v}\}_{v\in V(\gamma)},\{g_{e}\}_{e\in E(\gamma)}) & \longmapsto \{g_{\partial e_{+}}g_{e}g^{-1}_{\partial e_{-}}\}_{e\in E(\gamma)}.
\end{align*}
\end{definition}

Since we work in a Hamiltonian setting, we need to extend the space of gauge-field configurations by a suitable set of conjugate momenta. This is achieved in terms of either the left $L^{(\gamma)}$ or the right $R^{(\gamma)}$ multiplications, which arise naturally in this context (cf.~\cite{CreutzQuarksGluonsLattices}, also \cite{StottmeisterCoherentStatesQuantumIII} where similar notations are used and a classical-quantum correspondence is given). Strictly speaking, it is more common to invoke infinitesimal versions of $L^{(\gamma)}$ or $R^{(\gamma)}$, which is possible if $G$ is a locally compact Lie group making the cotangent bundle\footnote{The isomorphism works either by left or right translation depending on the choice for the momenta.} $T^{*}\scrA^{E(\gamma)}\cong (G\times\fg^{*})^{E(\gamma)}$ a suitable candidate for the extension in that case. But, it turns out that this is largely unnecessary for the quantum theory, and we work with the integrated form $G^{E(\gamma)}\times\scrA^{E(\gamma)}$. The left and right multiplications satisfy a simple equivariance relation w.r.t.~the gauge action,
\begin{align}
\label{eq:gaugeequiv}
\tau^{(\gamma)}_{\{g_{v}\}_{v\in V(\gamma)}}(L^{(\gamma)}_{\{h_{e}\}_{e\in E(\gamma)}}(\{g_{e}\}_{e\in E(\gamma)})) & = L^{(\gamma)}_{\{\alpha_{g_{\partial e_{+}}(h_{e})}\}_{e\in E(\gamma)}}(\tau_{\{g_{v}\}_{v\in V(\gamma)}}(\{g_{e}\}_{e\in E(\gamma)})), \\ \nonumber
\tau^{(\gamma)}_{\{g_{v}\}_{v\in V(\gamma)}}(R^{(\gamma)}_{\{h_{e}\}_{e\in E(\gamma)}}(\{g_{e}\}_{e\in E(\gamma)})) & = R^{(\gamma)}_{\{\alpha_{g_{\partial e_{-}}(h_{e})}\}_{e\in E(\gamma)}}(\tau_{\{g_{v}\}_{v\in V(\gamma)}}(\{g_{e}\}_{e\in E(\gamma)})),
\end{align}
where $\{g_{v}\}_{v\in V(\gamma)}\in G^{V(\gamma)}, \{g_{e}\}_{e\in E(\gamma)}\in\scrA^{E(\gamma)},\{h_{e}\}_{e\in E(\gamma)}\in G^{E(\gamma)}$, and $\alpha:G\times G\rightarrow G$ is the conjugation. Utilizing this observation, we define the (integrated) form of the phase space of gauge fields.

\begin{definition}
\label{def:phasespacegauge}
Given $\gamma$ as before with $G_{e}=G$ for $e\in E(\gamma)$, we call the direct product,
\begin{align*}
\Pi^{E(\gamma)} & = \bigtimes_{e\in E(\gamma)}\underbrace{G_{e}\times G_{e}}_{=\Pi_{e}} = G^{E(\gamma)}\times\scrA^{E(\gamma)},
\end{align*}
together with action of the first on the second factor by $L^{(\gamma)}$ or $R^{(\gamma)}$ its phase space of gauge fields. Depending on the choice of left or right multiplication, we define lifts of the action $\tau^{(\gamma)}$ of the gauge group $G^{V(\gamma)}$ to $\Pi^{E(\gamma)}$ according to \eqref{eq:gaugeequiv}:
\begin{align*}
\tilde{\tau}^{(\gamma)}_{L/R} : G^{V(\gamma)}\times \Pi^{E(\gamma)} & \longrightarrow \Pi^{E(\gamma)}, \\
 (\{g_{v}\}_{v\in V(\gamma)},\{(h_{e},g_{e})\}_{e\in E(\gamma)}) & \longmapsto (\{\alpha_{g_{\partial e_{\pm}}}(h_{e}),g_{\partial e_{+}}g_{e}g^{-1}_{\partial e_{-}}\}_{e\in E(\gamma)}).
\end{align*}
Additionally, we denote the restriction of $\tilde{\tau}^{(\gamma)}_{L/R}$ to the first factor of $\Pi^{E(\gamma)}$ by $\alpha^{(\gamma)}_{\partial_{\pm}}$.
\end{definition}

In the terminology of \cite{WilliamsCrossedProductsOf}, $(\Pi^{E(\gamma)},L^{(\gamma)})$ is a locally compact (left) transformation group. Since left, $L^{(\gamma)}$, and right, $R^{(\gamma)}$, multiplications are equivalent via conjugation, $\alpha$, and inversion, $(\ .\ )^{-1}$, we work with $(\Pi^{E(\gamma)},L^{(\gamma)})$ in the remainder of the article -- unless explicitly stated otherwise. Therefore, we drop the reference to $L^{(\gamma)}$ in the following. Given the transformation group $\Pi^{E(\gamma)}$, we may consider its associated $C^{*}$-dynamical system,
\begin{align}
\label{eq:gaugecrossprod}
(C_{0}(\scrA^{E(\gamma)}),G^{E(\gamma)},(L^{(\gamma)}_{(\ .\ )^{-1}})^{*}),
\end{align}
consisting of the $C^{*}$-algebra $C_{0}(\scrA^{E(\gamma)})$ (equipped with the $||\!\ .\!\ ||_{\infty}$-norm) and $(L^{(\gamma)}_{(\ .\ )^{-1}})^{*})$ is the strongly continuous representation of $G^{E(\gamma)}$ into the automorphism group of the former via pullback of the left multiplication. The $C^{*}$-dynamical system for the pullback of the right multiplication is denoted analogously. 
There is a natural covariant representation, $(M^{(\gamma)},\lambda^{(\gamma)})$, of \eqref{eq:gaugecrossprod} via the point-wise multiplication,
\begin{align}
\label{eq:multrep}
M^{(\gamma)} : C_{0}(\scrA^{E(\gamma)}) & \longrightarrow B(L^{2}(\scrA^{E(\gamma)})),
\end{align}
and the (unitary) left regular representation
\begin{align}
\label{eq:leftreg}
\lambda^{(\gamma)} : G^{E(\gamma)} & \longrightarrow \cU(L^{2}(\scrA^{E(\gamma)})).
\end{align}
Here, $L^{2}(\scrA^{E(\gamma)})$ is defined w.r.t.~a choice of invariant Haar measure, $m_{G}$, on $G$ and its product extension, $m^{(\gamma)}_{G}$, to $G^{\times|E(\gamma)|}$. Generically, it is not possible to replace $C_{0}(\scrA^{E(\gamma)})$ by $C_{b}(\scrA^{E(\gamma)})$ in \eqref{eq:gaugecrossprod} because the map $G^{E(\gamma)}\to \Aut(C_{b}(\scrA^{E(\gamma)}))$ might not be continuous \cite{VriesEquivariantEmbeddingsOf}.\\
Comparing the $C^{*}$-dynamical systems associated with two lattices, $\gamma, \gamma'$, that agree up to orientation, we make the following definition that reflects the natural behavior of holonomies in principal fibre bundles.

\begin{definition}
\label{def:orient}
Consider two edges, $e\in E(\gamma), e'\in E(\gamma')$, such that $e' = e^{-1}$. We identify $G_{e}, G_{e'}$ via the group inversion,
\begin{align*}
(\ .\ )^{-1} : G_{e}=G & \longrightarrow G=G_{e'}, \\
g & \longmapsto g^{-1},
\end{align*}
i.e.~$g_{e'} = g_{e^{-1}} = g^{-1}_{e}$. We extend this to a homeomorphism,
\begin{align*}
\iota^{\gamma}_{\gamma'} : \scrA^{E(\gamma)} & \longrightarrow \scrA^{E(\gamma')},
\end{align*}
for $|\gamma|=|\gamma'|$ in the obvious way. 
\end{definition}

This implies the elementary observation \cite{StottmeisterCoherentStatesQuantumIII}:
\begin{prop}
\label{prop:orient}
Given $e,e'$ as in the previous definition \ref{def:orient}. Then, we have an isomorphism of transformation groups,
\begin{align*}
(\Pi_{e'},L^{(\{e'\})}_{(\ .\ )}) & = (\Pi_{e},R^{(\{e\})}_{(\ .\ )^{-1}}),
\end{align*}
as well as of $C^{*}$-dynamical systems:
\begin{align*}
(C_{0}(G_{e'}),G_{e'},(L^{(e')}_{(\ .\ )^{-1}})^{*}) & \cong (C_{0}(G_{e}),G_{e},(R^{(e)}_{(\ .\ )})^{*}),
\end{align*}
where we consider the edges $e,e'$ as lattices by themselves.
\end{prop}

It is important to note that the actions $L_{(\ .\ )^{-1}}^{*}$ and $R_{(\ .\ )}^{*}$ are in general not identical even for abelian groups $G$ because of the inversion.\\
\\
Next, we introduce the kinematical algebras of time-zero gauge fields over finite lattices.

\begin{definition}
\label{def:gaugealg}
The $C^{*}$-algebra $B(L^{2}(\scrA^{E(\gamma)}))$ is called the algebra of time-zero gauge fields, or shortly the field algebra, over $\gamma\subset\Sigma$.
\end{definition}

Clearly, for a Lie group $G$, the covariance relations,
\begin{align}
\label{eq:Gccr}
M^{(\gamma)}((L^{(\gamma)}_{h^{-1}})^{*}f) & = \lambda^{(\gamma)}_{h}M^{(\gamma)}(f)\lambda^{(\gamma)}_{h^{-1}}, & f\in C_{0}(\scrA^{E(\gamma)}), h\in G^{E(\gamma)}
\end{align}
resembles the integrated form of the canonical commutation relations corresponding to the natural symplectic structure on $T^{*}\scrA^{E(\gamma)}$.\\
\\
Given a field algebra, $B(L^{2}(\scrA^{E(\gamma)}))$, the (strictly continuous) unitary representation,
\begin{align}
\label{eq:gaugeunitary}
U_{\tau^{(\gamma)}} : G^{V(\gamma)} & \longrightarrow \cU(L^{2}(\scrA^{E(\gamma)})),
\end{align}
induced by the action, $\tau^{(\gamma)}$, lifts to a $*$-automorphic action on the field algebra by
\begin{align}
\label{eq:gaugeaut}
\Ad_{U_{\tau^{(\gamma)}}} : G^{V(\gamma)} & \longrightarrow \Aut(B(L^{2}(\scrA^{E(\gamma)}))).
\end{align}
Thus, a natural definition of the action of the gauge group on the kinematical algebras is:

\begin{definition}
\label{def:gaugeactionalg}
The $*$-automorphic action \eqref{eq:gaugeaut} is called the action of the gauge group, $G^{V(\gamma)}$, on the field algebra, $B(L^{2}(\scrA^{E(\gamma)}))$.
\end{definition}

We note, that \eqref{eq:gaugeaut} implements the lifted action, $\tilde{\tau}^{(\gamma)}$, associated with $\Pi^{E(\gamma)}$ in the sense of the following covariance relations:
\begin{align*}
\Ad_{U_{\tau^{(\gamma)}}(g)}(M^{(\gamma)}(f)) & = M(\tau^{(\gamma)}(g^{-1})^{*}f), \\
\Ad_{U_{\tau^{(\gamma)}}(g)}(\lambda^{(\gamma)}_{h}) & = \lambda^{(\gamma)}_{\alpha^{(\gamma)}_{\partial_{+}(g)}(h)} = \Ad_{\lambda^{(\gamma)}_{\{g_{\partial e_{\pm}}\}_{e\in E(\gamma)}}}(\lambda^{(\gamma)}_{h}),
\end{align*}
where $f\in C_{0}(\scrA^{E(\gamma)})$, $g\in G^{V(\gamma)}$, and $h\in G^{E(\gamma)}$.\\
\\
An especially versatile parametrization of a weakly dense $*$-subalgebra of the field algebra is given in terms of $C_{c}(\Pi^{E(\gamma)})\subset C_{c}(G^{E(\gamma)},C_{0}(\scrA^{E(\gamma)}))$ viewed as a subalgebra of (left) convolution operators \cite{WilliamsCrossedProductsOf} of the transformation group $C^{*}$-algebra \eqref{eq:gaugecrossprod}:
\begin{align}
\label{eq:convkernel}
\lambda^{(\gamma)}(F) & = \int_{G^{E(\gamma)}}dh\ \!M^{(\gamma)}(F(h))\lambda^{(\gamma)}_{h}\, & F\in C_{c}(\Pi^{E(\gamma)}),
\end{align}
where $dh$ is a short-hand notation for $dm^{(\gamma)}_{G}(h)$. For obvious reasons, we call $F$ the convolution kernel of $\lambda^{(\gamma)}(F)$. The action of the gauge group becomes especially transparent on convolution kernels:
\begin{align}
\label{eq:gaugeconv}
\Ad_{U_{\tau^{(\gamma)}}(g)}(\lambda^{(\gamma)}(F)) & = \lambda^{(\gamma)}(\tilde{\tau}^{(\gamma)}_{L}(g^{-1})^{*}F).
\end{align}
Denoting by $\delta_{h}\in C'_{b}(G^{E(\gamma)})$ the delta distribution at $h\in G^{E(\gamma)}$, we infer that the basic field operators given by \eqref{eq:multrep} \& \eqref{eq:leftreg} are obtained, in the sense of distributions, from convolution kernels in $C'_{b}(G^{E(\gamma)})\otimes C_{b}(\scrA^{E(\gamma)})$, where we refrain from specifying a completion of the algebraic tensor product for simplicity.
\begin{align}
\label{eq:basickernels}
\lambda^{(\gamma)}(\delta_{1}\otimes f) & = M^{(\gamma)}(f), \\ \nonumber
\lambda^{(\gamma)}(\delta_{h}\otimes 1) & = \lambda^{(\gamma)}_{h},
\end{align}
where $f\in C_{0}(\scrA^{E(\gamma)})$. These equalities have a rigorous interpretation in the sense of zeroth-order pseudo-differential operators on $L^{2}(\scrA^{E(\gamma)})$ (cf.~\cite{RuzhanskyPseudoDifferentialOperators, StottmeisterCoherentStatesQuantumII}).\\
\\
We conclude this subsection with an extended comment relating our choice of field algebra to existing ones in the literature.

\begin{remark}
\label{rem:gaugealg}
Starting from the transformation group $\Pi^{E(\gamma)}$, there is another natural choice of $C^{*}$-algebra to embody the covariance condition \eqref{eq:Gccr}: the $C^{*}$-crossed product or transformation group $C^{*}$-algebra of $G^{\times|E(\gamma)|}$,
\begin{align}
\label{eq:Gcross}
 C_{0}(\scrA^{E(\gamma)})\rtimes_{(L^{(\gamma)}_{(\ .\ )^{-1}})^{*}}G^{E(\gamma)}.
 \end{align} 
The latter is naturally isomorphic to the algebra of compact operators, $K(L^{2}(\scrA^{E(\gamma)}))$, via the integrated form of the covariant representation $(M^{(\gamma)},\lambda^{(\gamma)})$ (cf.~\cite{WilliamsCrossedProductsOf}). Thus, the algebra $B(L^{2}(\scrA^{E(\gamma)}))$ is the multiplier algebra of the former (in the $C^{*}$-context, cf.~\cite{WilliamsCrossedProductsOf}), but it also corresponds to the double dual or, similarly, to the von Neumann closure \cite{BratteliOperatorAlgebrasAnd1}. Now, $K(L^{2}(\scrA^{E(\gamma)}))$ has been chosen as the field algebra in other recent operator-algebraic treatments of lattice gauge gauge theory \cite{JarvisOnTheStructure, GrundlingQCDOnAn, AriciQuantumLatticeGauge}  because of its regular representation theory. But in another article by one of the authors \cite{StottmeisterCoherentStatesQuantumIII}, it has already been argued for the need to unitaly extend this algebra to obtain a coherent Weyl-quantization w.r.t.~a certain class of lattices. Another related point of view in favor of extending $K(L^{2}(\scrA^{E(\gamma)}))$ due to renormalization group considerations is discussed in section \ref{sec:rgp}. Concerning the representation theory of the field algebra, $B(L^{2}(\scrA^{E(\gamma)}))$, we note that, as long as we are dealing with strictly continuous (or normal) states, no essential difference arises because of duality. In a certain sense \cite{BlackadarOperatorAlgebras}, $B(L^{2}(\scrA^{E(\gamma)}))$ resembles a maximal choice for the field algebra of $\gamma\subset\Sigma$, and it has the major advantage of being unital over $K(L^{2}(\scrA^{E(\gamma)}))$ which makes various constructions, e.g.~infinite tensor products, simpler to handle (cf.~\cite{GrundlingQCDOnAn}). Moreover, since, in the course of the article, we provide the field algebra for each $\gamma\subset\Sigma$ with a normal state $\omega^{(\gamma)}$ and consider its weak closure in the induced Gel'fand-Naimark-Segal (GNS) representation, the choice of field algebra will not be essential, if it at least contains a weakly dense subset for $B(L^{2}(\scrA^{E(\gamma)}))$. Notably, the same maximal choice of field algebra has be advocated for in the more recent article \cite{GrundlingDynamicsForQCD} as compared to \cite{GrundlingQCDOnAn} to allow for the definition of dynamics of the Kogut-Susskind Hamiltonian in the thermodynamic limit w.r.t.~the strong-coupling vacuum representation, cf.~remark \ref{rem:compactstate} and the discussion following \eqref{eq:KShamiltonian} below.
\end{remark}

\subsection{Multi-scale analysis and inductive limits}
\label{sec:scales}
We now discuss how to relate the field algebras, $B(L^{2}(\scrA^{E(\gamma)}))$, in the sense of multi-scale analysis w.r.t.~a directed set of lattices, $\Gamma_{\D,L}$, constructed from dyadic refinements of the basic unoriented lattice $|\gamma_{0}|=\Sigma\cap(L\Z)^{d}$. Here, $\D$ denotes the set of dyadic rationals in $[0,1)$ (if necessary we include $1$ in $\D$ for convenience).

\subsubsection{The dyadic scale of lattices}
\label{sec:dyadiclat}

\begin{definition}
\label{def:lattices}
The dyadic sequence of unoriented lattices $|\Gamma_{\D,L}|$ is given by the collection of unoriented lattices:
\begin{align}
\label{eq:dyadiclat}
|\gamma_{N}| & = \Sigma\cap(2^{-N}L\Z)^{d}, & N\in\N_{0}.
\end{align}
The set $\Gamma_{\D,L}$ consist of those lattices that are oriented versions of those in $|\Gamma_{\D,L}|$, i.e.~$\gamma\in\Gamma_{\D,L}$ iff $|\gamma|\in|\Gamma_{\D,L}|$. We call $\Gamma_{\D,L}$ the dyadic scale of lattices, and $N$ the level of $\gamma$ if $|\gamma|=|\gamma_{N}|$.
\end{definition}

$|\Gamma_{\D,L}|$ is ordered according to decreasing spacing $a_{\gamma_{N}} = 2^{-N}L>2^{-N'}L=a_{\gamma_{N'}}$ for $N<N'$. This ordering agrees with the subset relation, $|\gamma_{N}|\subset|\gamma_{N'}|$, that results from the operations of deleting, composing, and inverting edges of a lattice, $\gamma_{N'}$, to arrive at a coarser lattice, $\gamma_{N}$. The elementary operations required for our considerations are made precise in the next definition. Although, a more general treatment is possible, we only refer to the literature \cite{BaezSpinNetworksIn, LaneryProjectiveLoopQuantum1, StottmeisterCoherentStatesQuantumIII} in this respect, as it is unnecessary below.

\begin{definition}
\label{def:edgeop}
Given a lattice $\gamma'\subset\Sigma$, the composition, $e = e'_{2}\circ e'_{1}$, of two edges $e'_{1},e'_{2}\in E(\gamma)$ is defined if $\partial_{+}e'_{1} = \partial_{-}e'_{2}$ (or $\partial_{+}e'_{2} = \partial_{-}e'_{1}$) in $V(\gamma)$. The resulting edge, $e$, is given by the concatenation of $e'_{1}$ and $e'_{2}$ along their common vertex. The orientation of $e$ is that induced by $e'_{1}$ and $e'_{2}$. Thus, we obtain a new lattice $\gamma$ from $\gamma'$ by replacing the subset $\{e'_{1},e'_{2}\}\subset E(\gamma')$ by $\{e\}$.\\
Obviously, the deletion of an edge, $e'\in E(\gamma')$, is given by removing it from $\gamma'$. The new lattice $\gamma$ has the set of edges $E(\gamma)=E(\gamma')\setminus\{e'\}$.\\
Furthermore, we agree that any vertex, $v\in V(\gamma')$, that is not a boundary of a (non-trivial) edge is removed to arrive at $V(\gamma)$.
\end{definition}

\begin{figure}[t]
	\begin{tikzpicture}
	
	
	\foreach \x in {0,1} 
		\foreach \y in {0,...,2}	
		{
			\draw[->-=0.6, thick] (1+\x,4+\y) to (2+\x,4+\y);
			\draw[->-=0.6, thick] (1+\y,4+\x) to (1+\y,5+\x);
		}
	\draw[->-=0.6, thick, red] (1,6) to (2,6);
	\draw[->-=0.6, thick, red] (2,6) to (3,6);
	\foreach \v in {0,...,2} 
		\foreach \w in {0,...,2}
		{
			\filldraw (1+\v,4+\w) circle (2pt);
		}
	\draw[red, fill=red] (2,6) circle (2pt);
		
	\draw[->, thick] (3.75,5) to (4.25,5);
	
	\foreach \x in {0,1} 
		\foreach \y in {0,1}	
		{
			\draw[->-=0.6, thick] (5+\x,4+\y) to (6+\x,4+\y);
		}
	\draw[->-=0.6, thick, red] (5,6) to (7,6);
	\foreach \x in {0,1} 
		\foreach \y in {0,...,2}	
		{
			\draw[->-=0.6, thick] (5+\y,4+\x) to (5+\y,5+\x);
		}
	\foreach \v in {0,...,2} 
		\foreach \w in {0,...,2}
		{
			\filldraw (5+\v,4+\w) circle (2pt);
		}
	\draw[red, fill=red] (6,6) circle (2pt);
		
	\draw (9,5) node[align=left]{(\textb{composition})};
	
	
	\foreach \x in {0,1} 
		\foreach \y in {0,...,2}	
		{
			\draw[->-=0.6, thick] (1+\x,1+\y) to (2+\x,1+\y);
			\draw[->-=0.6, thick] (1+\y,1+\x) to (1+\y,2+\x);
		}
		
		\draw[->-=0.6, thick, red] (2,2) to (2,3);
	
	\foreach \v in {0,...,2} 
		\foreach \w in {0,...,2}
		{
			\filldraw (1+\v,1+\w) circle (2pt);
		}
		
	\draw[->, thick] (3.75,2) to (4.25,2);
	
	\foreach \x in {0,1} 
		\foreach \y in {0,...,2}	
		{
			\draw[->-=0.6, thick] (5+\x,1+\y) to (6+\x,1+\y);
		}
	\foreach \x in {0,1} 
		\foreach \y in {0,2}	
		{
			\draw[->-=0.6, thick] (5+\y,1+\x) to (5+\y,2+\x);
		}
	\draw[->-=0.6, thick] (6,1) to (6,2);
	\foreach \v in {0,...,2} 
		\foreach \w in {0,...,2}
		{
			\filldraw (5+\v,1+\w) circle (2pt);
		}
	
	\draw (9,2) node[align=left]{(\textb{deletion})};

\end{tikzpicture}
\caption{Illustration of the two basic coarsening operations on lattices in dimension $d=2$.}
\label{fig:edgeop}
\end{figure}
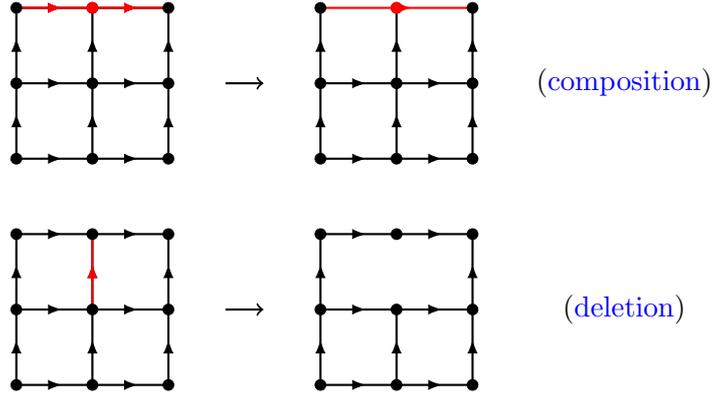

We introduce two partial orders on $\Gamma_{\D,L}$ that reflect the ambiguity due to the notion of left and right multiplications on $G$ as well as their equivalence via inversion and conjugation (cf.~\cite{LaneryProjectiveLoopQuantum1, StottmeisterCoherentStatesQuantumIII}).

\begin{definition}
\label{def:po}
Given the set $\Gamma_{\D,L}$, we define two relations $\subset_{\uL}$ and $\subset_{\uR}$. For $\gamma, \gamma'\in\Gamma_{\D,L}$, we say
\begin{itemize}
	\item[(a)] $\gamma\subset_{\uL}\gamma'$ if $|\gamma|\subset|\gamma'|$ and
	\begin{align*}
	\forall e\in E(\gamma) : \exists n\in\N_{0}, e'\cup\{e'_{k}\}^{n}_{k=1}\subset E(\gamma'), \{s_{k}\}^{n}_{k=1}\subset\{\pm1\}^{\times n} : e = e'\circ (e'_{1})^{s_{1}}\circ ... \circ (e'_{n})^{s_{n}}.
	\end{align*}
	\item[(b)] $\gamma\subset_{\uR}\gamma'$ if $|\gamma|\subset|\gamma'|$ and
	\begin{align*}
	\forall e\in E(\gamma) : \exists n\in\N_{0}, e'\cup\{e'_{k}\}^{n}_{k=1}\subset E(\gamma'), \{s_{k}\}^{n}_{k=1}\subset\{\pm1\}^{\times n} : e = (e'_{1})^{s_{1}}\circ ... \circ (e'_{n})^{s_{n}}\circ e'.
	\end{align*}
\end{itemize}
\end{definition}

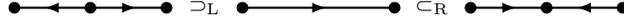
\begin{figure}[t]
	\begin{tikzpicture}
	
	\draw[->-=0.6, thick] (2,1) to (1,1);
	\draw[->-=0.6, thick] (2,1) to (3,1);
	\filldraw (1,1) circle (2pt) (2,1) circle (2pt) (3,1) circle (2pt);
	
	\draw (3.5,1) node{\scriptsize $\supset_{\uL}$};
	
	\draw[->-=0.55, thick] (4,1) to (6,1);
	\filldraw (4,1) circle (2pt) (6,1) circle (2pt);
	
	\draw (6.5,1) node{\scriptsize $\subset_{\uR}$};
	
	\draw[->-=0.6, thick] (7,1) to (8,1);
	\draw[->-=0.6, thick] (9,1) to (8,1);
	\filldraw (7,1) circle (2pt) (8,1) circle (2pt) (9,1) circle (2pt);
		
\end{tikzpicture}
\caption{Illustration of the partial orders $\subset_{\uL}$ and $\subset_{\uR}$ for the elementary edge composition.}
\label{fig:basicpo}
\end{figure}

It follows immediately from the definition of $\subset_{\uL/\uR}$ that we have:

\begin{prop}
\label{prop:po}
The relations $\subset_{\uL}$ and $\subset_{\uR}$ turn $\Gamma_{\D,L}$ into a directed set.
\end{prop}

Definition \ref{def:po} is motivated by the geometrical intuition that $\scrA^{E(\gamma)}$ represents the collection of possible $G$-valued holonomies along the edges $e\in E(\gamma)$ of a lattice $\gamma$: for $\gamma_{N}\subset_{\uL/\uR}\gamma_{N'}$, the holonomies of $\gamma_{N}$ are obtained from those of $\gamma_{N'}$ by inducing the orientation from the edge at final vertex or at the initial vertex, e.g.~for $e = e_{2}\circ e^{-1}_{1}$, we have either $e\subset_{\uL}\{e_{1},e_{2}\}$ or $e\subset_{\uR}\{e^{-1}_{1},e^{-1}_{2}\}$ (cp.~figure \ref{fig:basicpo}).

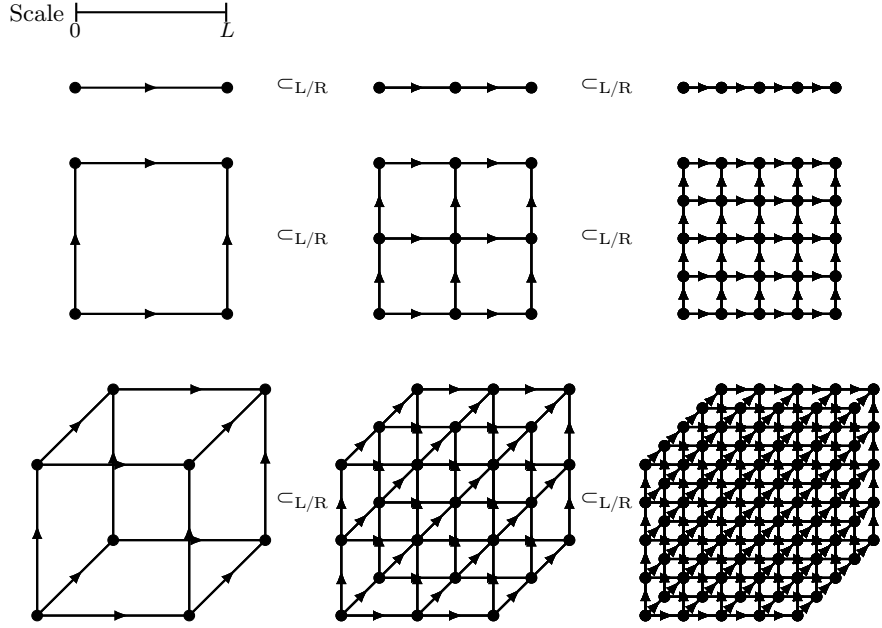
\begin{figure}[t]
	\begin{tikzpicture}
	
	\draw[|-|, thick] (1,8) node[left]{{\footnotesize Scale}} node[below]{{\scriptsize $0$}} to (3,8) node[below]{{\scriptsize $L$}};
	
	\draw[->-=0.55, thick] (1,7) to (3,7);
	\filldraw (1,7) circle (2pt) (3,7) circle (2pt);
	
	\draw (4,7) node{\scriptsize $\subset_{\uL/\uR}$};
	
	\foreach \x in {0,1} 
		\foreach \v in {0,...,2}	
		{
			\draw[->-=0.6, thick] (5+\x,7) to (6+\x,7);
			\filldraw (5+\v,7) circle (2pt);
		}
	
	\draw (8,7) node{\scriptsize $\subset_{\uL/\uR}$};
	
	\foreach \x in {0,...,3} 
		\foreach \v in {0,...,4}	
		{
			\draw[->-=0.75, thick] (9+0.5*\x,7) to (9.5+0.5*\x,7);
			\filldraw (9+0.5*\v,7) circle (2pt);
		}
	
	\foreach \x in {0} 
		\foreach \y in {0,1}
		\foreach \v in {0,1} 	
		{
			\draw[->-=0.55, thick] (1+2*\x,4+2*\y) to (3+2*\x,4+2*\y);
			\draw[->-=0.55, thick] (1+2*\y,4+2*\x) to (1+2*\y,6+2*\x);
			\filldraw (1+2*\y,4+2*\v) circle (2pt);
		}
	
	\draw (4,5) node{\scriptsize $\subset_{\uL/\uR}$};
	
	\foreach \x in {0,1} 
		\foreach \y in {0,...,2}
		\foreach \v in {0,...,2} 	
		{
			\draw[->-=0.6, thick] (5+\x,4+\y) to (6+\x,4+\y);
			\draw[->-=0.6, thick] (5+\y,4+\x) to (5+\y,5+\x);
			\filldraw (5+\y,4+\v) circle (2pt);
		}
	
	\draw (8,5) node{\scriptsize $\subset_{\uL/\uR}$};
	
	\foreach \x in {0,...,3} 
		\foreach \y in {0,...,4}
		\foreach \v in {0,...,4} 	
		{
			\draw[->-=0.75, thick] (9+0.5*\x,4+0.5*\y) to (9.5+0.5*\x,4+0.5*\y);
			\draw[->-=0.75, thick] (9+0.5*\y,4+0.5*\x) to (9+0.5*\y,4.5+0.5*\x);
			\filldraw (9+0.5*\y,4+0.5*\v) circle (2pt);
		}

	\foreach \x in {0} 
		\foreach \y in {0,1}
		\foreach \z in {0,1}
		\foreach \v in {0,1}
		{
			\draw[->-=0.6, thick] (0.5+2*\x+\z,0+2*\y+\z) to (2.5+2*\x+\z,0+2*\y+\z);
			\draw[->-=0.6, thick] (0.5+2*\y+\z,0+2*\x+\z) to (0.5+2*\y+\z,2+2*\x+\z);
			\draw[->-=0.6, thick] (0.5+2*\y,0+2*\z) to (1.5+\x+2*\y,1+\x+2*\z);
			\filldraw (0.5+2*\y+\z,0+2*\v+\z) circle (2pt);
		}
		
	\draw (4,1.5) node{\scriptsize $\subset_{\uL/\uR}$};
		
	\foreach \x in {0,1} 
		\foreach \y in {0,...,2}
		\foreach \z in {0,...,2}
		\foreach \v in {0,...,2}
		{
			\draw[->-=0.6, thick] (4.5+\x+0.5*\z,0+\y+0.5*\z) to (5.5+\x+0.5*\z,0+\y+0.5*\z);
			\draw[->-=0.6, thick] (4.5+\y+0.5*\z,0+\x+0.5*\z) to (4.5+\y+0.5*\z,1+\x+0.5*\z);
			\draw[->-=0.7, thick] (4.5+0.5*\x+\y,0+0.5*\x+\z) to (5+0.5*\x+\y,0.5+0.5*\x+\z);
			\filldraw (4.5+\y+0.5*\z,0+\v+0.5*\z) circle (2pt);
		}
		
	\draw (8,1.5) node{\scriptsize $\subset_{\uL/\uR}$};
	
	\foreach \x in {0,...,3} 
		\foreach \y in {0,...,4}
		\foreach \z in {0,...,4}
		\foreach \v in {0,...,4}
		{
			\draw[->-=0.75, thick] (8.5+0.5*\x+0.25*\z,0+0.5*\y+0.25*\z) to (9+0.5*\x+0.25*\z,0+0.5*\y+0.25*\z);
			\draw[->-=0.75, thick] (8.5+0.5*\y+0.25*\z,0+0.5*\x+0.25*\z) to (8.5+0.5*\y+0.25*\z,0.5+0.5*\x+0.25*\z);
			\draw[->-=0.8, thick] (8.5+0.25*\x+0.5*\y,0+0.25*\x+0.5*\z) to (8.75+0.25*\x+0.5*\y,0.25+0.25*\x+0.5*\z);
			\filldraw (8.5+0.5*\y+0.25*\z,0+0.5*\v+0.25*\z) circle (2pt);
		}
		
\end{tikzpicture}
\caption{Three elements, $\gamma_{0}\subset_{\uL/\uR}\gamma_{1}\subset_{\uL/\uR}\gamma_{2}$, of the partially ordered set of lattices, $(\Gamma_{\D,L},\subset_{\uL/\uR})$, in dimensions $d=1,2,3$.}
\label{fig:basicref}
\end{figure}

\begin{remark}
\label{rem:cofinal}
It is helpful to note that $(\Gamma_{\D,L},\subset_{\uL/\uR})$ contains the cofinal sequence $\{\gamma_{N}\}_{N\in\N_{0}}$, that is inductively constructed in the following way:
\begin{itemize}
	\item[(1)] For $N=0$, choose any $\gamma$ such that $|\gamma| = |\gamma_{0}|$, and put $\gamma_{0}=\gamma$. For example, choose the orientation of $\gamma$ according to the canonical positive orientation of $\R^{d}$.
	\item[(2)] Let us assume, we have constructed the sequence up to $N$. Firstly, we decompose all $e\in\gamma_{N}$ as $e = e'_{2}\circ (e'_{1})^{-1}$ (in case of $\subset_{\uL}$) or $e=(e'_{2})^{-1}\circ e'_{1}$ (in case of $\subset_{\uR}$) such that $|e'_{2}|, |e'_{1}|$ are unoriented edges of $|\gamma_{N+1}|$. Secondly, we add edges $e'$ (with arbitrary orientation) corresponding to unoriented edges $|e'|$ of $|\gamma_{N+1}|$ that were not obtained in the first step. The resulting oriented lattice $\gamma_{N+1}$ is of the required form.
\end{itemize}
\end{remark}

In the following, we primarily work with the directed set $(\Gamma_{\D,L},\subset_{\uL})$ as it fits naturally with our choice of transformation group $(\Pi^{E(\gamma)},L^{(\gamma)})$. Therefore, we drop the reference to $\subset_{\uL}$ unless necessary to avoid confusion.

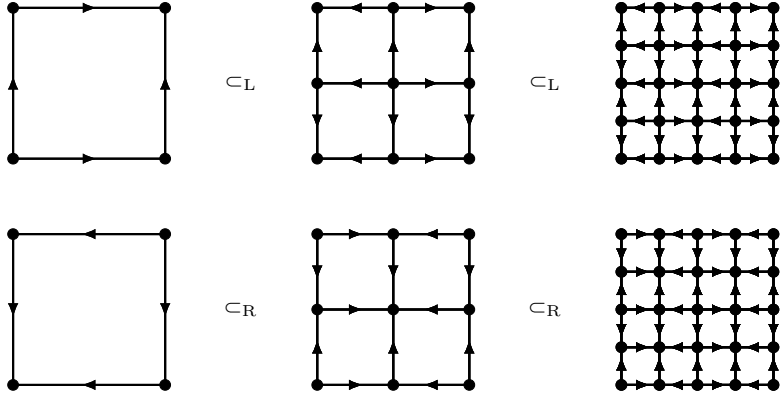
\begin{figure}[t]
	\begin{tikzpicture}
	
	
	\foreach \x in {0} 
		\foreach \y in {0,1}
		\foreach \v in {0,1} 	
		{
			\draw[->-=0.55, thick] (1+2*\x,4+2*\y) to (3+2*\x,4+2*\y);
			\draw[->-=0.55, thick] (1+2*\y,4+2*\x) to (1+2*\y,6+2*\x);
			\filldraw (1+2*\y,4+2*\v) circle (2pt);
		}
	
	\draw (4,5) node{\scriptsize $\subset_{\uL}$};
	
	\foreach \x in {1} 
		\foreach \y in {0,...,2}
		\foreach \v in {0,...,2} 	
		{
			\draw[->-=0.6, thick] (5+\x,4+\y) to (6+\x,4+\y);
			\draw[->-=0.6, thick] (7-\x,4+\y) to (6-\x,4+\y);
			\draw[->-=0.6, thick] (5+\y,4+\x) to (5+\y,5+\x);
			\draw[->-=0.6, thick] (5+\y,6-\x) to (5+\y,5-\x);
			\filldraw (5+\y,4+\v) circle (2pt);
		}
	
	\draw (8,5) node{\scriptsize $\subset_{\uL}$};
	
	\foreach \x in {1,3} 
		\foreach \y in {0,...,4}
		\foreach \v in {0,...,4} 	
		{
			\draw[->-=0.75, thick] (9+0.5*\x,4+0.5*\y) to (9.5+0.5*\x,4+0.5*\y);
			\draw[->-=0.75, thick] (11-0.5*\x,4+0.5*\y) to (10.5-0.5*\x,4+0.5*\y);
			\draw[->-=0.75, thick] (9+0.5*\y,4+0.5*\x) to (9+0.5*\y,4.5+0.5*\x);
			\draw[->-=0.75, thick] (9+0.5*\y,6-0.5*\x) to (9+0.5*\y,5.5-0.5*\x);
			\filldraw (9+0.5*\y,4+0.5*\v) circle (2pt);
		}
		
	\foreach \x in {0} 
		\foreach \y in {0,1}
		\foreach \v in {0,1} 	
		{
			\draw[->-=0.55, thick] (3-2*\x,1+2*\y) to (1-2*\x,1+2*\y);
			\draw[->-=0.55, thick] (1+2*\y,3-2*\x) to (1+2*\y,1-2*\x);
			\filldraw (1+2*\y,1+2*\v) circle (2pt);
		}
	
	\draw (4,2) node{\scriptsize $\subset_{\uR}$};
	
	\foreach \x in {1} 
		\foreach \y in {0,...,2}
		\foreach \v in {0,...,2} 	
		{
			\draw[->-=0.6, thick] (6+\x,1+\y) to (5+\x,1+\y);
			\draw[->-=0.6, thick] (6-\x,1+\y) to (7-\x,1+\y);
			\draw[->-=0.6, thick] (5+\y,2+\x) to (5+\y,1+\x);
			\draw[->-=0.6, thick] (5+\y,2-\x) to (5+\y,3-\x);
			\filldraw (5+\y,1+\v) circle (2pt);
		}
	
	\draw (8,2) node{\scriptsize $\subset_{\uR}$};
	
	\foreach \x in {1,3} 
		\foreach \y in {0,...,4}
		\foreach \v in {0,...,4} 	
		{
			\draw[->-=0.75, thick] (9.5+0.5*\x,1+0.5*\y) to (9+0.5*\x,1+0.5*\y);
			\draw[->-=0.75, thick] (10.5-0.5*\x,1+0.5*\y) to (11-0.5*\x,1+0.5*\y);
			\draw[->-=0.75, thick] (9+0.5*\y,1.5+0.5*\x) to (9+0.5*\y,1+0.5*\x);
			\draw[->-=0.75, thick] (9+0.5*\y,2.5-0.5*\x) to (9+0.5*\y,3-0.5*\x);
			\filldraw (9+0.5*\y,1+0.5*\v) circle (2pt);
		}
		
\end{tikzpicture}
\caption{Construction of a cofinal sequence, $\{\gamma_{N}\}_{N\in\N_{0}}$, for $\subset_{\uL}$ and $\subset_{\uR}$ in dimension $d=2$.}
\label{fig:cofinal}
\end{figure}

In line with our interpretation of the family of transformation groups $\{\Pi^{E(\gamma)}\}_{\gamma\in\Gamma_{\D,L}}$ as phase spaces of gauge fields, we introduce inductive and projective systems on $\{\scrA^{E(\gamma)}\}_{\gamma\in\Gamma_{\D,L}}$ respectively $\{G^{E(\gamma)}\}_{\gamma\in\Gamma_{\D,L}}$ that yield an inductive system of unital, injective $*$-morphisms (see proposition \ref{prop:convsys}),
\begin{align}
\label{eq:convsys}
\varsigma^{\gamma}_{\gamma'} : C'_{b}(G^{E(\gamma)})\otimes C_{b}(\scrA^{E(\gamma)}) & \longrightarrow C'_{b}(G^{E(\gamma')})\otimes C_{b}(\scrA^{E(\gamma')}), & \gamma\subset_{\uL}\gamma',
\end{align}
on the family of spaces of convolution kernels. To this end, we define projections and injections corresponding to the elementary operations on lattices (see definition \ref{def:edgeop}).

\begin{definition}
\label{def:groupop}
Consider a lattice $\gamma$ and two edges, $e_{1}, e_{2}\in E(\gamma)$, with $\partial_{+}e_{1} = \partial_{-}e_{2}$. Given the composition $e = e_{2}\circ e_{1}$, we define a projection via group multiplication,
\begin{align*}
p^{\{e_{1},e_{2}\}}_{\{e\}} : G_{e_{2}}\times G_{e_{1}} & \longrightarrow G_{e}, \\
(g_{e_{2}},g_{e_{1}}) & \longmapsto g_{e} = g_{e_{2}}g_{e_{1}},
\end{align*}
and an injection via trivial extension,
\begin{align*}
j^{\{e\}}_{\{e_{1},e_{2}\}} : G_{e} & \longrightarrow G_{e_{2}}\times G_{e_{1}}, \\
h_{e} & \longmapsto (h_{e_{2}},h_{e_{1}}) = (h_{e},1_{G}).
\end{align*}
Corresponding to the deletion of an edge, $e\in E(\gamma)$, we define a projection to the trivial group,
\begin{align*}
p^{\{e\}}_{\emptyset} : G_{e} & \longrightarrow \{1\}, \\
g_{e} & \longmapsto 1,
\end{align*}
and an injection of the trivial group,
\begin{align*}
j^{\emptyset}_{\{e\}} : \{1\} & \longrightarrow G_{e}, \\
1 & \longmapsto h_{e}=1_{G}.
\end{align*}
\end{definition}

As noted above, we can generate any $\gamma\in\Gamma_{\D,L}$ from some $\gamma'\supset_{\uL}\gamma$ via a sequence consisting of the elementary operations of edge composition, edge deletion, and edge inversion (cp.~definition \ref{def:orient} for the latter). Thus, because of associativity of group multiplication and projection in direct products, we have (cp.~\cite{StottmeisterCoherentStatesQuantumIII}):

\begin{prop}
\label{prop:groupsys}
Given two lattices $\gamma,\gamma'\in\Gamma_{\D,L}$ such that $\gamma\subset_{\uL}\gamma'$, there is a unique projection,
\begin{align*}
p^{\gamma'}_{\gamma} : \scrA^{E(\gamma')} & \longrightarrow \scrA^{E(\gamma)},
\end{align*}
composed out of a sequence of the elementary projections of definitions \ref{def:groupop} and \ref{def:orient}. Three projections $p^{\gamma'}_{\gamma}, p^{\gamma''}_{\gamma'}, p^{\gamma''}_{\gamma}$ satisfy the transitivity condition,
\begin{align*}
p^{\gamma''}_{\gamma} & = p^{\gamma''}_{\gamma'}\circ p^{\gamma'}_{\gamma},
\end{align*}
if $\gamma\subset_{\uL}\gamma'\subset_{\uL}\gamma''$.
Furthermore, there is a unique injection,
\begin{align*}
j^{\gamma}_{\gamma'} : G^{E(\gamma)} & \longrightarrow G^{E(\gamma')},
\end{align*}
composed out of a sequence of the elementary injections of definitions \ref{def:groupop} and \ref{def:orient}. Three injections $j^{\gamma}_{\gamma'}, j^{\gamma'}_{\gamma''}, j^{\gamma}_{\gamma''}$ satisfy the transitivity condition,
\begin{align*}
j^{\gamma}_{\gamma''} & = j^{\gamma'}_{\gamma''}\circ j^{\gamma}_{\gamma'},
\end{align*}
if $\gamma\subset_{\uL}\gamma'\subset_{\uL}\gamma''$.
\end{prop}

As an immediate consequence, we deduce (cp.~\cite{ThiemannModernCanonicalQuantum, StottmeisterCoherentStatesQuantumIII, BrothierConstructionsOfConformal}, also \cite{TatsuumaOnGroupTopologies, HiraiInductiveLimitsOf}):

\begin{corollary}
\label{cor:groupsys}
$\{\{\scrA^{E(\gamma)}\}_{\gamma\in\Gamma_{\D,L}},\{p^{\gamma'}_{\gamma}\}_{\gamma\subset_{\uL}\gamma'}\}$ is a projective system of topological spaces, and we have projective limit:
\begin{align*}
\varprojlim_{\gamma\in\Gamma_{\D,L}}\scrA^{E(\gamma)} & = \overline{\scrA}.
\end{align*}
Moreover, $\{\{G^{E(\gamma)}\}_{\gamma\in\Gamma_{\D,L}},\{j^{\gamma}_{\gamma'}\}_{\gamma\subset_{\uL}\gamma'}\}$ is an inductive system of topological groups, and we have an inductive limit:
\begin{align*}
\varinjlim_{\gamma\in\Gamma_{\D,L}}G^{E(\gamma)} & = \underline{G},
\end{align*}
together with an
action,
\begin{align*}
\underline{L} : \underline{G} & \longrightarrow \Homeo(\overline{\scrA}).
\end{align*}
\end{corollary}

The action, $\underline{L}$, of the previous corollary is characterized by actions at finite level because $\Gamma_{\D,L}$ is directed \cite{BrothierConstructionsOfConformal}:
\begin{align}
\label{eq:projgroupaction}
(\underline{L}_{\underline{h}}(\overline{g}))_{\gamma} & = p^{\gamma''}_{\gamma}(L_{j^{\gamma'}_{\gamma''}(h_{\gamma'})}(g_{\gamma''})), & \gamma\subset_{\uL}\gamma''\supset_{\uL}\gamma'
\end{align}
for $\overline{g}\in\overline{\scrA}$, $\underline{h}\in\underline{G}$. Here, $g_{\gamma} = p_{\gamma}(\overline{g})$ is an abbreviation involving the canonical projection $p_{\gamma}:\overline{\scrA}\rightarrow\scrA^{E(\gamma)}$, and $\underline{h} = j^{\gamma'}(h_{\gamma'})$ for the canonical injection $j^{\gamma}:G^{E(\gamma)}\rightarrow\underline{G}$.\\
See \cite[Section 2]{BrothierConstructionsOfConformal} for explicit descriptions of the limits $\underline G, \overline{\scrA}$ and $\underline L$.

The definition of the inductive and projective systems is motivated by their compatibility w.r.t.~the distinction of configuration and momentum variables captured by the transformation groups, $\{\Pi^{E(\gamma)}\}_{\gamma\in\Gamma_{\D,L}}$(cf.~\cite{StottmeisterCoherentStatesQuantumIII} for further details):

\begin{corollary}
\label{cor:groupactionsys}
Given two lattices as in definition \ref{def:groupop}, their transformation groups, $\Pi^{E(\gamma)}, \Pi^{E(\gamma')}$, satisfy the identities,
\begin{align*}
p^{\gamma'}_{\gamma}(L_{j^{\gamma}_{\gamma'}(h)}(g)) & = L_{h}(p^{\gamma'}_{\gamma}(g)),
\end{align*}
where $(h,p^{\gamma'}_{\gamma}(g))\in\Pi^{E(\gamma)}$ and $(j^{\gamma}_{\gamma'}(h),g)\in \Pi^{E(\gamma')}$.
\end{corollary}

We also have a projective system on the family of gauge groups, $\{G^{V(\gamma)}\}_{\gamma\in\Gamma_{\D,L}}$, that is compatible with the inductive and projective systems on $\{\scrA^{E(\gamma)}\}_{\gamma\in\Gamma_{\D,L}}$ respectively $\{G^{E(\gamma)}\}_{\gamma\in\Gamma_{\D,L}}$ (cf.~\cite{ThiemannModernCanonicalQuantum, StottmeisterCoherentStatesQuantumIII}).

\begin{prop}
\label{prop:gaugesys}
Given two lattices, $\gamma\subset_{\uL}\gamma'$, we consider the projection,
\begin{align*}
q^{\gamma'}_{\gamma} : G^{V(\gamma')} & \longrightarrow G^{V(\gamma)}, \\
\{g_{v'}\}_{v'\in V(\gamma')} & \longmapsto \{g_{v'}\}_{v'\in V(\gamma)},
\end{align*}
by restriction of the set of vertices $V(\gamma)\subset V(\gamma')$. Then, we have a projective system,\\ $\{\{G^{V(\gamma)}\}_{\gamma\in\Gamma_{\D,L}},\{q^{\gamma}_{\gamma'}\}_{\gamma\subset_{\uL}\gamma'}\}$, of topological groups and a projective limit:
\begin{align*}
\varprojlim_{\gamma\in\Gamma_{\D,L}}G^{V(\gamma)} & = \overline{G},
\end{align*}
together with a (continuous) 
action,
\begin{align*}
\overline{\tau} : \overline{G} & \longrightarrow \Homeo(\overline{\scrA}).
\end{align*}
\end{prop}

The action, $\overline{\tau}$, of the previous proposition is characterized by:
\begin{align}
\label{eq:projgaugeaction}
\overline{\tau}(\overline{g}')(\overline{g})_{\gamma} & = \tau^{(\gamma)}(g'_{\gamma})(g_{\gamma}),
\end{align}
for $\overline{g}\in\overline{\scrA}$, $\overline{g}'\in\overline{G}$. Here, $g_{\gamma} = p_{\gamma}(\overline{g})$ as before, and $g'_{\gamma} = q_{\gamma}(\overline{g}')$ is an abbreviation involving the canonical projection $q_{\gamma}:\overline{G}\rightarrow G^{V(\gamma)}$. Moreover, we have the following compatibility with the family of transformation groups, $\{\Pi^{E(\gamma)}\}_{\gamma\in\Gamma_{\D,L}}$, in view of corollary \ref{cor:groupactionsys}:

\begin{corollary}
\label{cor:gaugeactionsys}
Given two lattices as in definition \ref{def:groupop}, the action of the gauge groups, $G^{V(\gamma)}, G^{V(\gamma')}$, on the respective transformation groups, $\Pi^{E(\gamma)}, \Pi^{E(\gamma')}$, satisfy the identities,
\begin{align*}
\tau^{(\gamma)}(q^{\gamma'}_{\gamma}(g'))(p^{\gamma'}_{\gamma}(L_{j^{\gamma}_{\gamma'}(h)}(g))) & = L_{\alpha^{(\gamma)}_{\partial_{+}(q^{\gamma'}_{\gamma}(g'))}(h)}(p^{\gamma'}_{\gamma}(\tau^{(\gamma')}(g')(g))),
\end{align*}
where $(h,p^{\gamma'}_{\gamma}(g'))\in\Pi^{E(\gamma)}$, $(j^{\gamma}_{\gamma'}(h),g')\in \Pi^{E(\gamma')}$, and $g'\in G^{V(\gamma')}$.
\end{corollary}

\subsubsection{The dyadic scale of field algebras}
\label{sec:dyadicalg}
Combining the inductive and projective systems from proposition \ref{prop:groupsys}, corollary \ref{prop:groupsys}, and using corollary \ref{cor:groupactionsys}, we obtain the inductive system \eqref{eq:convsys} of unital, injective $*$-morphisms:

\begin{prop}
\label{prop:convsys}
Given two lattices $\gamma,\gamma'\in\Gamma_{\D,L}$ such that $\gamma\subset_{\uL}\gamma'$, there is a unique unital, injective $*$-morphism,
\begin{align*}
\varsigma^{\gamma}_{\gamma'} : C'_{b}(G^{E(\gamma)})\otimes C_{b}(\scrA^{E(\gamma)}) & \longrightarrow C'_{b}(G^{E(\gamma')})\otimes C_{b}(\scrA^{E(\gamma')}), & \gamma\subset_{\uL}\gamma',
\end{align*}
defined by:
\begin{align*}
\lambda^{(\gamma')}(\varsigma^{\gamma}_{\gamma'}(F)) & = \int_{G^{E(\gamma)}}dh\ \!M^{(\gamma')}((p^{\gamma'}_{\gamma})^{*}(F(h)))\lambda^{(\gamma')}_{j^{\gamma}_{\gamma'}(h)}\, & F\in C_{c}(\Pi^{E(\gamma)}).
\end{align*}
As a consequence of the corresponding properties of the injections and projections of proposition \ref{prop:groupsys}, three morphisms $\varsigma^{\gamma}_{\gamma'}, \varsigma^{\gamma'}_{\gamma''}, \varsigma^{\gamma}_{\gamma''}$ satisfy the transitivity condition,
\begin{align*}
\varsigma^{\gamma}_{\gamma''} & = \varsigma^{\gamma'}_{\gamma''}\circ \varsigma^{\gamma}_{\gamma'},
\end{align*}
if $\gamma\subset_{\uL}\gamma'\subset_{\uL}\gamma''$.
\end{prop}

From corollary \ref{cor:gaugeactionsys} we infer the compatibility with the action of the projective system of gauge groups.

\begin{corollary}
\label{cor:gaugeactionconvsys}
Given two lattices as in previous proposition \ref{prop:convsys}, we have the identities,
\begin{align*}
\tilde{\tau}^{(\gamma')}_{L}(g^{-1})^{*}\varsigma^{\gamma}_{\gamma'}(F) & = \varsigma^{\gamma}_{\gamma'}(\tilde{\tau}^{(\gamma')}_{L}(q^{\gamma'}_{\gamma}(g)^{-1})^{*}F),
\end{align*}
where $F\in C'_{b}(G)\otimes C_{b}(G)$ and $g\in G^{V(\gamma')}$.
\end{corollary}

Now, we generalize the maps $\{\varsigma^{\gamma}_{\gamma'}\}_{\gamma,\gamma'\in\Gamma_{\D,L}}$ to a family of unital, injective $*$-morphism among the field algebras,
\begin{align}
\label{eq:algsys}
\alpha^{\gamma}_{\gamma'} : B(L^{2}(\scrA^{E(\gamma)})) & \longrightarrow B(L^{2}(\scrA^{E(\gamma')})), & \gamma\subset_{\uL}\gamma'.
\end{align}
For this purpose, we notice that the elementary injections and projections of definition \ref{def:groupop} lead to the following $*$-morphisms on the level of convolution kernels (cf.~\cite{StottmeisterCoherentStatesQuantumIII}):
\begin{align}
\label{eq:algop}
\alpha^{\{e\}}_{\{e_{1},e_{2}\}}(\lambda^{(\{e\})}(F)) & = \lambda^{(\{e_{1},e_{2}\})}(\varsigma^{\{e\}}_{\{e_{1},e_{2}\}}(F)) = U_{\uL}(\lambda^{(\{e_{2}\})}(F)\otimes\mathds{1}_{e_{1}})U^{*}_{\uL}, \\ \nonumber
\alpha^{\{e_{2}\}}_{\{e_{1},e_{2}\}}(\lambda^{(\{e_{2}\})}(F)) & = \lambda^{(\{e_{1},e_{2}\})}(\varsigma^{\{e_{2}\}}_{\{e_{1},e_{2}\}}(F)) = \lambda^{(\{e_{2}\})}(F)\otimes\mathds{1}_{e_{1}},
\end{align}
where $F\in C'_{b}(G)\otimes C_{b}(G)$. The unitary $U_{\uL}\in\cU(L^{2}(G_{e_{2}}\times G_{e_{1}}))$ is defined by\footnote{There is a corresponding unitary, $U_{\uR}$, for the relation $\subset_{\uR}$ defined by \begin{align*}
(U_{\uR}\psi)(g_{e_{2}},g_{e_{1}}) & = \psi(g_{e_{1}},g_{e_{2}}g_{e_{1}}), & \psi\in C_{c}(G_{e_{2}}\times G_{e_{1}}).
\end{align*}}:
\begin{align}
\label{eq:funitary}
(U_{\uL}\psi)(g_{e_{2}},g_{e_{1}}) & = \psi(g_{e_{2}}g_{e_{1}},g_{e_{1}}), & \psi\in C_{c}(G_{e_{2}}\times G_{e_{1}}).
\end{align}
By duality, the $*$-morphisms \eqref{eq:algop} represent refining operations w.r.t.~the directed set of lattices, $\Gamma_{\D,L}$. The reversal of the orientation of an edge as implemented in definition \ref{def:orient} leads to the elementary $*$-morphism (cp.~also proposition \ref{prop:orient}):
\begin{align}
\label{eq:algorient}
\alpha^{\{e\}}_{\{e^{-1}\}}(\lambda^{(\{e\})}(F)) & = \lambda^{(\{e^{-1}\})}(\varsigma^{\{e\}}_{\{e^{-1}\}}(F)) = U_{\iota}\lambda^{\{e\}}(F)U^{*}_{\iota},
\end{align}
where $F\in C'_{b}(G)\otimes C_{b}(G)$. The unitary $U_{\iota}\in\cU(L^{2}(G_{e}))$ is defined by:
\begin{align}
\label{eq:orientunitary}
(U_{\iota}\phi)(g_{e}) & = \phi(g^{-1}_{e}), & \phi\in C_{c}(G_{e}).
\end{align}

\begin{remark}
\label{rem:funitary}
It is interesting to note that the unitary $U_{\uL}$ is directly related to the multiplicative unitary, $W$, typically invoked in the context of quantum groups \cite{KustermansLocallyCompactQuantum}:
\begin{align*}
\hat{\Delta}(\lambda(f)) & = W^{*}(\lambda(f)\otimes\mathds{1})W = \lambda^{\otimes 2}(U_{\uL}^{*}(\delta_{1}\otimes f)),
\end{align*}
where $\hat{\Delta}$ is the coproduct of the reduced group $C^{*}$-algebra, $C^{*}_{\lambda}(G)$.
\end{remark}

In analogy with proposition \ref{prop:groupsys}, we find:

\begin{prop}
\label{prop:algsys}
Given two lattices $\gamma,\gamma'\in\Gamma_{\D,L}$ such that $\gamma\subset_{\uL}\gamma'$, there is a unique unital, injective $*$-morphism,
\begin{align*}
\alpha^{\gamma}_{\gamma'} : B(L^{2}(\scrA^{E(\gamma)})) & \longrightarrow B(L^{2}(\scrA^{E(\gamma')})), & \gamma\subset_{\uL}\gamma',
\end{align*}
composed out of a sequence of elementary morphisms given in equations \eqref{eq:algop} \& \eqref{eq:algorient}. As a consequence of the corresponding properties of the morphisms of proposition \ref{prop:convsys}, three morphisms $\alpha^{\gamma}_{\gamma'}, \alpha^{\gamma'}_{\gamma''}, \alpha^{\gamma}_{\gamma''}$ satisfy the transitivity condition,
\begin{align*}
\alpha^{\gamma}_{\gamma''} & = \alpha^{\gamma'}_{\gamma''}\circ \alpha^{\gamma}_{\gamma'},
\end{align*}
if $\gamma\subset_{\uL}\gamma'\subset_{\uL}\gamma''$.
\end{prop}

Using corollary \ref{cor:gaugeactionconvsys} and equation \eqref{eq:gaugeconv}, we deduce the compatibility with the action of the projective system of gauge groups.

\begin{corollary}
\label{cor:gaugeactionalgsys}
Given two lattices as in previous proposition \ref{prop:algsys}, we have the identities,
\begin{align*}
\Ad_{U_{\tau^{(\gamma')}}(g)}(\alpha^{\gamma}_{\gamma'}(a)) & = \alpha^{\gamma}_{\gamma'}(\Ad_{U_{\tau^{(\gamma)}}(q^{\gamma'}_{\gamma}(g))}(a)),
\end{align*}
where $a\in B(L^{2}(\scrA^{E(\gamma)}))$ and $g\in G^{V(\gamma')}$.
\end{corollary}

Now, standard results \cite{KadisonFundamentalsOfThe2, EvansQuantumSymmetriesOn} concerning inductive limits of $C^{*}$-algebras provide us with a limit of the multi-scale family of field algebras, $\{B(L^{2}(\scrA^{E(\gamma)}))\}_{\gamma\in\Gamma_{\D,L}}$.

\begin{corollary}
\label{cor:algsys}
$\{\{B(L^{2}(\scrA^{E(\gamma)}))\}_{\gamma\in\Gamma_{\D,L}},\{\alpha^{\gamma}_{\gamma'}\}_{\gamma\subset_{\uL}\gamma'}\}$ is a directed system of $C^{*}$-algebras, and we have an inductive limit:
\begin{align*}
\varinjlim_{\gamma\in\Gamma_{\D,L}}B(L^{2}(\scrA^{E(\gamma)})) & = \fA_{\D,L}.
\end{align*}
Moreover, the compatibility with the action by the projective system of gauge groups, $\{G^{V(\gamma)}\}_{\gamma\in\Gamma_{\D,L}}$, given in corollary \ref{cor:gaugeactionalgsys} leads to an automorphic action,
\begin{align*}
\alpha_{\overline{\tau}} : \overline{G} & \longrightarrow \Aut(\fA_{\D,L}).
\end{align*}
\end{corollary}

In analogy with \cite{JonesANoGo}, we call $\fA_{\D,L}$ the semi-continuum field algebra. This terminology is justified by the fact that the latter is associated with dyadic standard cube (or torus), $(L\D)^{d}$, of scale $L>0$. Thus, as $(L\D)^{d}$ is a dense subset of $\Sigma$, continuous gauge-field configurations, $\Phi\in C(\Sigma,G)$, are uniquely defined by their restrictions $\Phi_{\D,L}=\Phi_{|(L\D)^{d}}$.

\subsection{Further structure}
\label{sec:fstruct}
We conclude this section on general operator-algebraic constructions for lattice gauge theory by providing some structural results of the semi-continuum field algebra, $\fA_{\D,L}$, that are of particular relevance for the other sections below.

\subsubsection{Locality structure of the semi-continuum field algebra}
\label{sec:localalgsys}

A interesting observation concerning the elementary refining operations \eqref{eq:algop} and the edge inversion \eqref{eq:algorient} is that these provide the semi-continuum field algebra, $\fA_{\D,L}$, with a locality structure.

\begin{definition}
\label{def:locallat}
For an open star domain $\cS\subset\Sigma$, we consider its intersections,
\begin{align*}
\gamma_{\cS} & = \gamma\cap\cS, & \gamma\in\Gamma_{\D,L},
\end{align*}
where we agree on the convention that any edge $e\in E(\gamma)$ with $e\cap\partial\cS\neq\emptyset$ is not considered part of $\gamma_{\cS}$. Thus, any $e\in\gamma_{\cS}$ is completely inside $\cS$. We call $\Gamma_{\D,L}(\cS) = \{\gamma_{\cS}\ |\ \exists\gamma\in\Gamma_{\D,L}:\gamma_{\cS}=\gamma\cap\cS\}$ the dyadic scale of lattices subordinate to $\cS$. 
\end{definition}

\begin{figure}[t]
	\begin{tikzpicture}
	
	
	\foreach \x in {0,...,7} 
		\foreach \y in {0,...,8}	
		{
			\draw[->-=0.6, dashed, thin, gray] (1+\x,1+\y) to (2+\x,1+\y);
			\draw[->-=0.6, dashed, thin, gray] (1+\y,1+\x) to (1+\y,2+\x);
		}
	\foreach \v in {0,...,8} 
		\foreach \w in {0,...,8}
		{
			\filldraw[gray] (1+\v,1+\w) circle (1pt);
		}
	
	\draw[thick] (3,3) to[out=150, in=250] (2,5);
	\draw[thick] (2,5) to[out=70,in=190] (4,7);
	\draw[thick] (4,7) to[out=10,in=80] (8,6);
	\draw[thick] (8,6) to[out=260,in=120] (7,3);
	\draw[thick] (7,3) to[out=300,in=330] (3,3);
	
	\foreach \x in {0,1} 
		\foreach \y in {0,...,2}	
		{
			\draw[->-=0.6, thick] (4+\x,4+\y) to (5+\x,4+\y);
			\draw[->-=0.6, thick] (4+\y,4+\x) to (4+\y,5+\x);
		}
	\foreach \v in {0,...,2} 
		\foreach \w in {0,...,2}
		{
			\filldraw (4+\v,4+\w) circle (2pt);
		}
		
	
	\draw[->-=0.6, thick] (4,3) to (4,4);
	\draw[->-=0.6, thick] (5,3) to (5,4);
	\draw[->-=0.6, thick] (6,3) to (6,4);
	\draw[->-=0.6, thick] (3,4) to (3,5);
	\draw[->-=0.6, thick] (3,5) to (3,6);
	\draw[->-=0.6, thick] (5,6) to (5,7);
	\draw[->-=0.6, thick] (6,6) to (6,7);
	\draw[->-=0.6, thick] (7,6) to (7,7);
	\draw[->-=0.6, thick] (7,5) to (7,6);
	
	\filldraw (4,3) circle (2pt) (5,3) circle (2pt) (6,3) circle (2pt) (3,4) circle (2pt) (3,5) circle (2pt) (3,6) circle (2pt) (5,7) circle (2pt) (6,7) circle (2pt) (7,7) circle (2pt) (7,6) circle (2pt) (7,5) circle (2pt);
	
	
	\draw[->-=0.6, thick] (4,3) to (5,3);	
	\draw[->-=0.6, thick] (5,3) to (6,3);
	\draw[->-=0.6, thick] (3,4) to (4,4);
	\draw[->-=0.6, thick] (3,5) to (4,5);
	\draw[->-=0.6, thick] (3,6) to (4,6);
	\draw[->-=0.6, thick] (5,7) to (6,7);
	\draw[->-=0.6, thick] (6,7) to (7,7);
	\draw[->-=0.6, thick] (6,6) to (7,6);
	\draw[->-=0.6, thick] (6,5) to (7,5);
		
	\draw[gray] (1.5,1.5) node{$\gamma$};
	\draw (7.5,2.5) node{$\cS$};
	\draw (5.5,3.5) node{$\gamma_{\cS}$};

\end{tikzpicture}
\caption{Illustration of the localization of lattices in dimension $d=2$.}
\label{fig:locallat}
\end{figure}
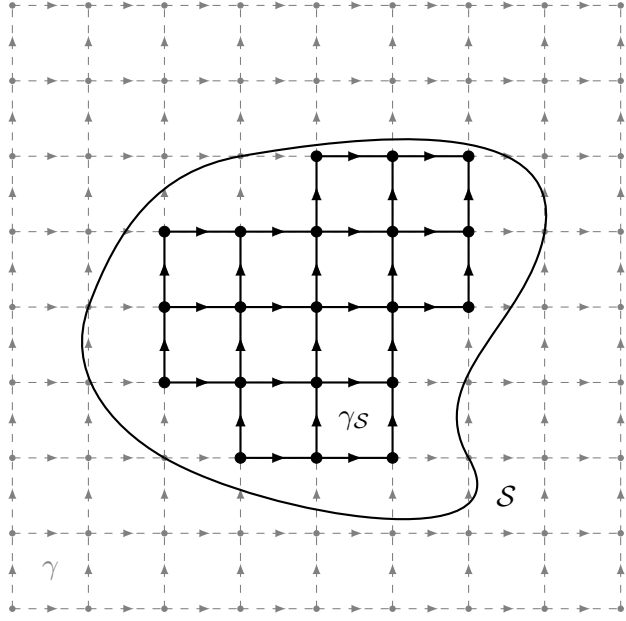

It is clear from the constructions in the previous subsection that $\Gamma_{\D,L}(\cS)$ inherits being directed w.r.t.~$\subset_{\uL}$ (or $\subset_{\uR}$) from $\Gamma_{\D,L}$, and we have the analogues of projective and inductive systems given in corollaries \ref{cor:groupsys} \& \ref{cor:algsys} as well as \ref{prop:gaugesys} \& \ref{prop:convsys}, together with the limits:
\begin{align}
\label{eq:localsys}
\varprojlim_{\gamma_{\cS}\in\Gamma_{\D,L}(\cS)}\scrA^{E(\gamma_{\cS})} & = \overline{\scrA}(\cS), & \varinjlim_{\gamma_{\cS}\in\Gamma_{\D,L}(\cS)}G^{E(\gamma_{\cS})} & = \underline{G}(\cS), &
\varprojlim_{\gamma_{\cS}\in\Gamma_{\D,L}(\cS)}G^{V(\gamma_{\cS})} & = \overline{G}(\cS), \\ \nonumber
&  & \varinjlim_{\gamma_{\cS}\in\Gamma_{\D,L}(\cS)}B(L^{2}(\scrA^{E(\gamma_{\cS})})) & = \fA_{\D,L}(\cS). &  &
\end{align}

\begin{definition}
\label{def:localalg}
For an open star domain\footnote{For $\Sigma=\T^{d}_{L}$ we assume $\cS$ to be contractible.} $\cS\subset\Sigma$, we call $\fA_{\D,L}(\cS)$ the local semi-continuum field algebra (in $\cS$).
\end{definition}

Since any $\gamma_{\cS}$ results from some $\gamma\in\Gamma_{\D,L}$ via the elementary operations of edge composition, removal, or inversion, such that for $\gamma_{\cS}\subset_{\uL}\gamma$, our constructions provide us with unital, injective $*$-morphisms,
\begin{align}
\label{eq:lattalginclusion}
\alpha^{\gamma_{\cS}}_{\gamma} : B(L^{2}(\scrA^{E(\gamma_{\cS})})) & \longrightarrow B(L^{2}(\scrA^{E(\gamma)})).
\end{align}
Loosely speaking, $\alpha^{\gamma_{\cS}}_{\gamma}(a)$ corresponds to the extension of $a\in B(L^{2}(\scrA^{E(\gamma_{\cS})}))$ by the identity operator on $\gamma\setminus\gamma_{\cS}$. Because of the transitivity condition for subsequent refinements of lattices (see proposition \ref{prop:algsys}), there are unital, injective $*$-morphisms \cite{EvansQuantumSymmetriesOn},
\begin{align}
\label{eq:localalginclusion}
\alpha^{\cS} : \fA_{\D,L}(\cS) & \longrightarrow \fA_{\D,L}.
\end{align}
Thus, with a slight abuse of notation we may regard the local semi-continuum field algebras as $C^{*}$-subalgebras of semi-continuum field algebra, $\fA_{\D,L}(\cS)\subset\fA_{\D,L}$. Moreover, we can repeat the argument for any set inclusion $\cS\subset\cS'\subset\Sigma$, such that we end up with corresponding unital, injective $*$-morphisms:
\begin{align}
\label{eq:localalgsys}
\alpha^{\gamma_{\cS}}_{\gamma'_{\cS'}} : B(L^{2}(\scrA^{E(\gamma_{\cS})})) & \longrightarrow B(L^{2}(\scrA^{E(\gamma'_{\cS'})})), & \gamma_{\cS}\subset_{\uL}\gamma'_{\cS'}, \\ \nonumber
\alpha^{\cS}_{\cS'} : \fA_{\D,L}(\cS) & \longrightarrow \fA_{\D,L}(\cS'), &  &
\end{align}
that satisfy transitivity conditions analogous to those stated in proposition \ref{prop:algsys}. As an immediate consequence of unitality and the form of the elementary refining operations \eqref{eq:algop}, we have:
\begin{corollary}
\label{cor:localalgsys}
The family of algebras,
\begin{align*}
\{\fA_{\D,L}(\cS)\ |\ \cS\subset\Sigma\textup{ is an open star-domain}\},
\end{align*}
is an isotonuous, local net of $C^{*}$-subalgebras of $\fA_{\D,L}$, i.e.
\begin{align*}
\fA_{\D,L}(\cS) & \subset \fA_{\D,L}(S'), & \cS & \subset \cS', \\
[\fA_{\D,L}(\cS),\fA_{\D,L}(\cS')] & = \{0\}, & \cS \cap \cS' & \neq \emptyset.
\end{align*}
\end{corollary}

The corollary shows that $\fA_{\D,L}$ will be a quasi-local algebra \cite{BratteliOperatorAlgebrasAnd1}, if the the uniform closure of the local semi-continuum field algebras is equivalent to the semi-continuum field algebra,
\begin{align}
\label{eq:quasilocalalg}
\fA_{\D,L} & = \bigcup_{\cS}\fA_{\D,L}(\cS).
\end{align}

\begin{remark}
\label{rem:localalg}
Clearly, it is essential for the definition of the locality structure of $\fA_{\D,L}$ that the refining $*$-morphisms \eqref{eq:algsys} are unital. For example, if we represented our basic operations by non-unital $*$-morphisms,
\begin{align*}
\alpha^{\{e\}}_{\{e_{1},e_{2}\}}(\lambda^{(\{e\})}(F)) & = \Ad_{u}(\lambda^{(\{e\})}(F)), \\
\alpha^{\{e_{2}\}}_{\{e_{1},e_{2}\}}(\lambda^{(\{e_{2}\})}(F)) & = \Ad_{v}(\lambda^{\{e_{2}\}}(F)),
\end{align*}
where $u:L^{2}(G_{e})\rightarrow L^{2}(G_{e_{2}}\times G_{e_{1}})$, $v:L^{2}(G_{e_{2}})\rightarrow L^{2}(G_{e_{2}}\times G_{e_{1}})$ are partial isometries, we would loose the locality structure because,
\begin{align*}
\alpha^{\{e\}}_{\{e_{1},e_{2}\}}(\mathds{1}_{e}) & = p_{u}, \\
\alpha^{\{e_{2}\}}_{\{e_{1},e_{2}\}}(\mathds{1}_{e_{2}}) & = p_{v},
\end{align*}
are in general not central. Here, $p_{u}=uu^{*}$ and $p_{v} = vv^{*}$ are the final projections of $u$ and $v$. For example, in \cite{AriciQuantumLatticeGauge, GrundlingQCDOnAn} the $*$-morphism $\alpha^{\{e_{2}\}}_{\{e_{1},e_{2}\}}$ is represented by $v(\phi_{e_{2}}) = \phi_{e_{2}}\otimes1_{e_{1}}$ ($\phi_{e_{2}}\in L^{2}(G_{e_{2}})$, $G$ is assumed to be compact), which leads to:
\begin{align*}
\alpha^{\{e_{2}\}}_{\{e_{1},e_{2}\}}(\lambda^{(\{e_{2}\})}(F)) & = \lambda^{(\{e_{2}\})}(F)\otimes(p_{1})_{e_{1}},
\end{align*}
with the projection onto the constant unit function $p_{1}:L^{2}(G)\rightarrow L^{2}(G)$.
\end{remark}

\subsubsection{Representations and dyadic scales of Hilbert spaces}
\label{sec:dyadichilbert}

To connect our abstract operator-algebraic constructions to the realm of Hilbert spaces, and thus to the common framework of quantum physics \cite{KogutTheLatticeGauge, KogutAnIntroductionTo, CreutzQuarksGluonsLattices} (cp.~also \cite{JonesANoGo}), we invoke the GNS construction w.r.t.~to coherent families of states, $\{\omega^{(\gamma)}\}_{\gamma\in\Gamma_{\D,L}}$. More precisely, we consider families of positive, linear functional satisfying the coherence conditions:
\begin{align}
\label{eq:statecoherence}
\omega^{(\gamma')}\circ\alpha^{\gamma}_{\gamma'} & = \omega^{(\gamma)}, & \gamma & \subset_{\uL}\gamma',
\end{align}
w.r.t.~the directed system of  $C^{*}$-algebras \eqref{eq:algsys}. By standard reasoning \cite{KadisonFundamentalsOfThe2}, a family of states satisfying said coherence condition induces an inductive system of Hilbert spaces,
\begin{align}
\label{eq:hilbertsys}
R^{\gamma}_{\gamma'} : \cH_{\gamma} & \longrightarrow \cH_{\gamma'}, & (R^{\gamma}_{\gamma'})^{*}R^{\gamma}_{\gamma'} & = \mathds{1}_{\cH_{\gamma}}, & \gamma & \subset_{\uL}\gamma', \\ \nonumber
 &  & R^{\gamma'}_{\gamma''}R^{\gamma}_{\gamma'} & = R^{\gamma}_{\gamma''}, & \gamma & \subset_{\uL}\gamma'\subset_{\uL}\gamma'',
\end{align}
together with coherent family of $*$-representations,
\begin{align}
\label{eq:hilbertrepsys}
\pi_{\gamma} : B(L^{2}(\scrA^{E(\gamma)})) & \longrightarrow B(\cH_{\gamma}), & \gamma & \subset_{\uL}\gamma',
\end{align}
and cyclic vectors,
\begin{align}
\label{eq:vacuumsys}
\{\Omega_{\gamma}\}_{\gamma\in\Gamma_{\D,L}},
\end{align}
Moreover, we have the following relations among the various objects and maps:
\begin{align}
\label{eq:coherentsys}
\pi_{\gamma'}(\alpha^{\gamma}_{\gamma'}(a))R^{\gamma}_{\gamma'} & = R^{\gamma}_{\gamma'}\pi_{\gamma}(a), \\ \nonumber
R^{\gamma}_{\gamma'}\Omega_{\gamma} & = \Omega_{\gamma'}, \\ \nonumber
\omega_{\gamma}(a^{*}b) & = (\pi_{\gamma}(a)\Omega_{\gamma},\pi_{\gamma}(b)\Omega_{\gamma})_{\cH_{\gamma}},
\end{align}
where $a,b\in B(L^{2}(\scrA^{E(\gamma)}))$, $\gamma\subset_{\uL}\gamma'$.\\
\\
At this point, it is important to point out that the first line of \eqref{eq:coherentsys} is not equivalent to the condition,
\begin{align}
\label{eq:adjointsys}
\pi_{\gamma'}(\alpha^{\gamma}_{\gamma'}(a)) & = R^{\gamma}_{\gamma'}\pi_{\gamma}(a)(R^{\gamma}_{\gamma'})^{*},
\end{align}
for $a\in B(L^{2}(\scrA^{E(\gamma)}))$ because generically $p_{\gamma'} = R^{\gamma}_{\gamma'}(R^{\gamma}_{\gamma'})^{*}\neq\mathds{1}_{\cH_{\gamma'}}$. Condition \eqref{eq:adjointsys} has been used in \cite{AriciQuantumLatticeGauge} (cp.~remark \ref{rem:localalg}) as definition of $\alpha^{\gamma}_{\gamma'}$, and in a dual sense (on density matrices) in \cite{JonesScaleInvariantTransfer, LangHamiltonianRenormalizationI} (see section \ref{sec:rgp}).

\begin{remark}
In case, we would like to avoid the use of a distinguished family of states $\{\omega^{(\gamma)}\}_{\gamma\in\Gamma_{\D,L}}$, we may ask for the existence of an inductive family of Hilbert spaces \eqref{eq:hilbertsys} and an associated family of representations \eqref{eq:hilbertrepsys} supplemented by the first condition of \eqref{eq:coherentsys} \cite{KadisonFundamentalsOfThe2}. Moreover, we may localize everything w.r.t.~the locality structure discussed in the previous subsection \ref{sec:localalgsys}.
\end{remark}

To conclude our general outline of inductive systems of Hilbert space accompanying the dyadic scale of field algebras, $\{B(L^{2}(\scrA^{E(\gamma)}))\}_{\gamma\in\Gamma_{\D,L}}$, we note that the former lead to suitable representations of the semi-continuum field algebra $\fA_{\D,L}$ as well (cf.~\cite{KadisonFundamentalsOfThe2}).

\begin{prop}
\label{prop:statelimit}
Given a coherent family of states, $\{\omega^{(\gamma)}\}_{\gamma\in\Gamma_{\D,L}}$, as above, there is unique (projective) limit state,
\begin{align*}
\varprojlim_{\gamma\in\Gamma_{\D,L}}\omega^{(\gamma)} & = \omega_{\D,L},
\end{align*}
together with a representation,
\begin{align*}
\pi_{\D,L} : \fA_{\D,L} & \longrightarrow B(\cH_{\D,L}),
\end{align*}
on the inductive limit Hilbert space,
\begin{align*}
\varinjlim_{\gamma\in\Gamma_{\D,L}}\cH_{\gamma} & = \cH_{\D,L},
\end{align*}
with cyclic vector $\Omega_{\D,L}\in\cH_{\D,L}$. Moreover, we have:
\begin{align*}
\pi_{\D,L}(\alpha^{\gamma}(a))R^{\gamma} & = R^{\gamma}\pi_{\gamma}(a), & R^{\gamma}\Omega_{\gamma} & = \Omega_{\D,L}, & \omega_{\D,L}(b^{*}c) & = (\pi_{\D,L}(b)\Omega_{\D,L},\pi_{\D,L}(c)\Omega_{\D,L})_{\cH_{\D,L}},
\end{align*}
for $a\in B(L^{2}(\scrA^{E(\gamma)}))$ and $c,b\in\fA_{\D,L}$. Here, $\alpha^{\gamma}:B(L^{2}(\scrA^{E(\gamma)}))\rightarrow\fA_{\D,L}$ and $R^{\gamma}:\cH_{\gamma}\rightarrow\cH_{\D,L}$ are the natural maps associated with the inductive limit construction.
\end{prop}

\begin{definition}
\label{def:fieldvna}
In view of the previous proposition \ref{prop:statelimit}, we call the weak closure,
\begin{align*}
\scrM_{\D,L} & = \pi_{\D,L}(\fA_{\D,L})'',
\end{align*}
the semi-continuum von-Neumann field algebra, and similarly,
\begin{align*}
\scrM_{\D,L}(\cS) & = \pi_{\D,L}(\alpha^{\cS}(\fA_{\D,L}(\cS)))'',
\end{align*}
the local semi-continuum von-Neumann field algebra (in $\cS$).
\end{definition}

We return to the discussion of specific representations and weak-closures of the semi-continuum field algebra in sections \ref{sec:ym11} to \ref{sec:ymd1}. Nevertheless, it is instructive to take a closer look at the coherence conditions \eqref{eq:statecoherence} for the elementary refining operations \eqref{eq:algop}. The nature of these conditions becomes especially evident for convolution operators $\lambda^{(\gamma)}(F)$, $F\in C_{c}(\Pi^{E(\gamma)})$, together with the assumption that each finite-level state is normal, $\omega^{(\gamma)}(\ .\ ) = \Tr(T_{\gamma}\ .\ )$, for $T_{\gamma} = \lambda^{(\gamma)}(F_{T_{\gamma}})\in B(L^{2}(\scrA^{E(\gamma)}))_{*}$:
\begin{align}
\label{eq:elementarystatecoherence1}
 & \omega^{(\{e_{1},e_{2}\})}(\alpha^{\{e\}}_{\{e_{1},e_{2}\}}(\lambda^{(\{e\})}(F))) \\ \nonumber
 & = \omega^{(\{e_{1},e_{2}\})}(\lambda^{(\{e_{1},e_{2}\})}(\varsigma^{\{e\}}_{\{e_{1},e_{2}\}}(F))) \\ \nonumber
 & = \int_{G_{e_{2}}\times G_{e_{1}}}dg_{e_{2}}dg_{e_{1}}\!\ (F_{T_{\{e_{1},e_{2}\}}}\ast \varsigma^{\{e\}}_{\{e_{1},e_{2}\}}(F))(1_{e_{2}},g_{e_{2}};1_{e_{1}},g_{e_{1}}) \\ \nonumber
 & = \int_{G_{e_{2}}\times G_{e_{1}}}dg_{e_{2}}dg_{e_{1}}\!\ \int_{G_{e_{2}}\times G_{e_{1}}}dh_{e_{2}}dh_{e_{1}}\!\ F_{T_{\{e_{1},e_{2}\}}}(h_{e_{2}},g_{e_{2}};h_{e_{1}},g_{e_{1}})\delta_{1_{e_{1}}}(h^{-1}_{e_{1}})F(h^{-1}_{e_{2}},h^{-1}_{e_{2}}g_{e_{2}}h^{-1}_{e_{1}}g_{e_{1}}) \\ \nonumber
 & = \int_{G_{e}}dg_{e}\!\ \int_{G_{e}}dh_{e}\!\ \underbrace{\left(\int_{G_{e_{1}}}dg_{e_{1}}\!\ F_{T_{\{e_{1},e_{2}\}}}(h_{e},g_{e}g^{-1}_{e_{1}};1_{e_{1}},g_{e_{1}})\right)}_{=F_{T_{\{e\}}}(h_{e},g_{e})}F(h^{-1}_{e},h^{-1}_{e}g_{e}) \\ \nonumber
 & = \int_{G_{e}}dg_{e}\!\ (F_{T_{\{e\}}}\ast F)(1_{e},g_{e}) \\ \nonumber
 & = \omega^{(\{e\})}(\lambda^{(\{e\})}(F)), \\[0.25cm]
 \label{eq:elementarystatecoherence2}
 & \omega^{(\{e_{1},e_{2}\})}(\alpha^{\{e_{2}\}}_{\{e_{1},e_{2}\}}(\lambda^{(\{e_{2}\})}(F))) \\ \nonumber
 & = \omega^{(\{e_{1},e_{2}\})}(\lambda^{(\{e_{1},e_{2}\})}(\varsigma^{\{e_{2}\}}_{\{e_{1},e_{2}\}}(F))) \\ \nonumber
 & = \int_{G_{e_{2}}\times G_{e_{1}}}dg_{e_{2}}dg_{e_{1}}\!\ (F_{T_{\{e_{1},e_{2}\}}}\ast \varsigma^{\{e_{2}\}}_{\{e_{1},e_{2}\}}(F))(1_{e_{2}},g_{e_{2}};1_{e_{1}},g_{e_{1}}) \\ \nonumber
 & = \int_{G_{e_{2}}\times G_{e_{1}}}dg_{e_{2}}dg_{e_{1}}\!\ \int_{G_{e_{2}}\times G_{e_{1}}}dh_{e_{2}}dh_{e_{1}}\!\ F_{T_{\{e_{1},e_{2}\}}}(h_{e_{2}},g_{e_{2}};h_{e_{1}},g_{e_{1}})\delta_{1_{e_{1}}}(h^{-1}_{e_{1}})F(h^{-1}_{e_{2}},h^{-1}_{e_{2}}g_{e_{2}}) \\ \nonumber
 & = \int_{G_{e_{2}}}dg_{e_{2}}\!\ \int_{G_{e_{2}}}dh_{e_{2}}\!\ \underbrace{\left(\int_{G_{e_{1}}}dg_{e_{1}}\!\ F_{T_{\{e_{1},e_{2}\}}}(h_{e_{2}},g_{e_{2}};1_{e_{1}},g_{e_{1}})\right)}_{=F_{T_{\{e_{2}\}}}(h_{e_{2}},g_{e_{2}})}F(h^{-1}_{e_{2}},h^{-1}_{e_{2}}g_{e_{2}}) \\ \nonumber
 & = \int_{G_{e_{2}}}dg_{e_{2}}\!\ (F_{T_{\{e_{2}\}}}\ast F)(1_{e_{2}},g_{e_{2}}) \\ \nonumber
 & = \omega^{(\{e_{2}\})}(\lambda^{(\{e_{2}\})}(F)),
\end{align}
for $F\in C_{c}(\Pi_{e})$ resp.~$F\in C_{c}(\Pi_{e_{2}})$. Thus, the consistency conditions become simple integral equations:
\begin{align}
\label{eq:integralstatecoherence}
\int_{G_{e_{1}}}dg_{e_{1}}\!\ F_{T_{\{e_{1},e_{2}\}}}(h_{e},g_{e}g^{-1}_{e_{1}};1_{e_{1}},g_{e_{1}}) & = F_{T_{\{e\}}}(h_{e},g_{e}), \\ \nonumber
\int_{G_{e_{1}}}dg_{e_{1}}\!\ F_{T_{\{e_{1},e_{2}\}}}(h_{e_{2}},g_{e_{2}};1_{e_{1}},g_{e_{1}}) & = F_{T_{\{e_{2}\}}}(h_{e_{2}},g_{e_{2}}).
\end{align}
Moreover, the operation of edge inversion \eqref{eq:algorient} gives:
\begin{align}
\label{eq:stateinversion}
& \omega^{(\{e^{-1}\})}(\alpha^{\{e\}}_{\{e^{-1}\}}(\lambda^{(\{e\})}(F))) \\ \nonumber
 & = \omega^{(\{e^{-1}\})}(\lambda^{(\{e^{-1}\})}(\varsigma^{\{e\}}_{\{e^{-1}\}}(F))) \\ \nonumber
 & = \int_{G_{e^{-1}}}dg_{e^{-1}}\!\ (F_{T_{\{e^{-1}\}}}\ast \varsigma^{\{e\}}_{\{e^{-1}\}}(F))(1_{e^{-1}},g_{e^{-1}}) \\ \nonumber
 & = \int_{G_{e^{-1}}}dg_{e^{-1}}\!\ \int_{G_{e^{-1}}}dh_{e^{-1}}\!\ F_{T_{\{e^{-1}\}}}(h_{e^{-1}},g_{e^{-1}})F(\alpha_{(h^{-1}_{e^{-1}}g_{e^{-1}})^{-1}}(h^{-1}_{e^{-1}})^{-1},(h^{-1}_{e^{-1}}g_{e^{-1}})^{-1}) \\ \nonumber
 & = \int_{G_{e^{-1}}}dg_{e^{-1}}\!\ \int_{G_{e^{-1}}}dh_{e^{-1}}\!\ F_{T_{\{e^{-1}\}}}(h_{e^{-1}},g_{e^{-1}})F(\alpha_{g^{-1}_{e^{-1}}}(h_{e^{-1}}),g^{-1}_{e^{-1}}h_{e^{-1}}) \\ \nonumber
 & = \int_{G_{e^{-1}}}dg_{e^{-1}}\!\ \int_{G_{e^{-1}}}dh_{e^{-1}}\!\ F_{T_{\{e^{-1}\}}}(\alpha_{g_{e^{-1}}}(h_{e^{-1}}),g_{e^{-1}})F(h_{e^{-1}},h_{e^{-1}}g^{-1}_{e^{-1}}) \\ \nonumber
 & = \int_{G_{e}}dg_{e}\!\ \int_{G_{e}}dh_{e}\!\ \underbrace{F_{T_{\{e^{-1}\}}}(\alpha_{g^{-1}_{e}}(h_{e})^{-1},g^{-1}_{e})}_{=\varsigma^{\{e^{-1}\}}_{\{e\}}(F_{T_{\{e^{-1}\}}})(h_{e},g_{e})=F_{T_{\{e\}}}(h_{e},g_{e})}F(h^{-1}_{e},h^{-1}_{e}g_{e}) \\ \nonumber
 & = \omega^{(\{e\})}(\lambda^{(\{e\})}(F)).
\end{align}
Therefore, consistency w.r.t.~the inversion of a single edge requires:
\begin{align}
\label{eq:integralstateinversion}
F_{T_{\{e^{-1}\}}}(\alpha_{g^{-1}_{e}}(h_{e})^{-1},g^{-1}_{e}) & = \varsigma^{\{e^{-1}\}}_{\{e\}}(F_{T_{\{e^{-1}\}}})(h_{e},g_{e}) = F_{T_{\{e\}}}(h_{e},g_{e}).
\end{align}
But, this latter condition needs to be treated carefully, because the reversal of edge orientations is not implemented by the binary relation $\subset_{\uL}$ of the directed set $\Gamma_{\D,L}$ (cf.~\cite{StottmeisterCoherentStatesQuantumIII}).

\subsubsection{The compact case}
\label{sec:compact}
In the final part of this section, we specialize to the important case when $G$ is compact including the physically important cases $G = \Z_{2}, U(1), SU(2), SU(3)$. Because of this specialization, we have $C_{c}(G) = C_{0}(G) = C(G)$, and w.l.o.g.~we may fix the total mass of the Haar measure to unity, $m_{G}(G) = \vol(G) = 1$. 
A rephrasing of the latter is the existence of the distinguished unit vector,
\begin{align}
\label{eq:compacthilbertvect}
1_{\gamma} & \in L^{2}(\scrA^{E(\gamma)}),
\end{align}
for $\gamma\in\Gamma_{\D,L}$. The (normal) states,
\begin{align}
\label{eq:compactstate}
\omega^{(\gamma)}(a) & = \Tr(p_{1_{\gamma}}a) , & a & \in B(L^{2}(\scrA^{E(\gamma)})),
\end{align}
where $p_{1_{\gamma}}$ is the projection associated with \eqref{eq:compacthilbertvect}, form a coherent family \eqref{eq:statecoherence}. For convolution kernels \eqref{eq:convkernel}, $F\in C(\Pi^{E(\gamma)})$, we have the explicit formula (cp.~\eqref{eq:integralstatecoherence}),
\begin{align}
\label{eq:compactstateint}
\omega^{(\gamma)}(\lambda^{(\gamma)}(F)) & = \int_{G^{E(\gamma)}}dh\!\ \int_{\scrA^{E(\gamma)}}dg\!\ F(h,g),
\end{align}
involving the Haar measures $m^{(\gamma)}_{G}$, which we abbreviate by $dg$ and $dh$ on $\scrA^{E(\gamma)}$ resp.~$G^{E(\gamma)}$. As the computation is instructive for subsequent sections, we use \eqref{eq:compactstateint} to deduce the form of the Gel'fand ideal, $\fI_{\gamma}$, of $\omega^{(\gamma)}$:
\begin{align}
\label{eq:compactstateideal}
\omega^{(\gamma)}(\lambda^{(\gamma)}(F)^{*}\lambda^{(\gamma)}(F)) & = \int_{\scrA^{E(\gamma)}}dg\!\ \int_{G^{E(\gamma)}}dh\!\ (F^{*}\ast F)(h,g) \\ \nonumber
 &  = \int_{\scrA^{E(\gamma)}}dg\!\ \int_{G^{E(\gamma)}}dh\!\ \int_{G^{E(\gamma)}}dh'\!\ \overline{F((h')^{-1},(h')^{-1}g)} F((h')^{-1}h,(h')^{-1}g) \\ \nonumber
 &  = \int_{\scrA^{E(\gamma)}}dg\!\ \left(\int_{G^{E(\gamma)}}dh'\!\ \overline{F(h',g)} \right)\left(\int_{G^{E(\gamma)}}dh\!\ F(h,g) \right) \\ \nonumber
 &  = \int_{\scrA^{E(\gamma)}}dg\!\ |\hat{F}^{1}(\pi_{\utriv},g)|^{2},
\end{align}
where $\hat{F}^{1}$ denotes the Fourier transform of $F$ w.r.t.~the momenta, $G^{E(\gamma)}$, and $\pi_{\utriv}$ is the trivial representation of $G$ (see \eqref{eq:fourier} below). Thus, by density of $C(\Pi^{E(\gamma)})$, $\fI_{\gamma}$ is obtained from those $F$ that have $\hat{F}^{1}(\pi_{\utriv})=0$, and we have, as expected from \eqref{eq:compactstate}, $\cH_{\gamma} = L^{2}(\scrA^{E(\gamma)})$. In this case, the elementary refining isometries (cp.~\eqref{eq:hilbertsys}) are:
\begin{align}
\label{eq:compacthilbertsys}
(R^{\{e\}}_{\{e_{1},e_{2}\}}\phi)(g_{e_{2}},g_{e_{1}}) & = \phi(g_{e_{2}}g_{e_{1}}) = (p^{\{e_{1},e_{2}\}}_{\{e\}})^{*}\phi(g_{e_{2}},g_{e_{1}}), \\ \nonumber
(R^{\emptyset}_{\{e\}}z)(g_{e}) & = z = (p^{\{e\}}_{\emptyset})^{*}z(g),
\end{align}
with $\phi\in C(G_{e})$ and $z\in C(\{1\})\cong\C$. Furthermore, compactness of $G$ implies that $\overline{\scrA}$ (see corollary \ref{cor:groupsys}) is a compact Hausdorff space, and the family of Haar measures $\{m^{(\gamma)}_{G}\}_{\gamma\in\Gamma_{\D,L}}$ forms a cylindrical measure that gives an actual measure $m_{\overline{\scrA}}$ by the Riez-Markov-Kakutani theorem (cf.~\cite{ThiemannModernCanonicalQuantum}). The latter leads to a realization of the inductive limit Hilbert space as an $L^{2}$-space,
\begin{align}
\label{eq:compactL2limit}
\cH_{\D,L} & = L^{2}(\overline{\scrA}),
\end{align}
and we may deduce from our results in the subsequent section that
\begin{align}
\label{eq:compactVNA}
\scrM_{\D,L} & = B(L^{2}(\overline{\scrA})).
\end{align}
The action $\underline{L}:\underline{G}\rightarrow\Homeo(\overline{\scrA})$ preserves the measure $m_{\overline{\scrA}}$ because the unit vector \eqref{eq:compacthilbertvect} is obviously invariant for the action of $G^{E(\gamma)}$, which is equivalent to the natural adjointness relations for the gauge field momentum operators,
\begin{align}
\label{eq:compactmomentum}
(\underline{L}_{\underline{h}})^{*} & = \underline{L}_{\underline{h}^{-1}}, & \underline{h}&\in\underline{G}.
\end{align}
Moreover, because the limit state, $\omega_{\D,L}$, is invariant under the automorphic action $\alpha_{\overline{\tau}}$ (see corollary \ref{cor:algsys}), the latter is unitarily implemented,
\begin{align}
\label{eq:compactgaugeaction}
U_{\overline{\tau}} : \overline{G} & \longrightarrow \cU(L^{2}(\overline{\scrA})),
\end{align}
and the underlying transformation group, $\overline{\tau}:\overline{G}\rightarrow\Homeo(\overline{\scrA})$, preserves the measure $m_{\overline{\scrA}}$.

\begin{remark}
\label{rem:compactstate}
In view of remark \ref{rem:gaugealg}, we note that $\omega^{(\gamma)}$ induces the unique (up to unitary equivalence) irreducible representation of $K(L^{2}(\scrA^{E(\gamma)}))$ in the sense of generalize Stone-von Neumann theorem (cf.~\cite{WilliamsCrossedProductsOf, GrundlingQCDOnAn}). The inductive limit Hilbert space $L^{2}(\overline{\scrA})$ is essentially what is known as the Ashtekar-Isham-Lewandowski Hilbert space in the context of loop quantum gravity (cf.~\cite{ThiemannModernCanonicalQuantum}). The associated partial isometries $R^{\gamma}_{\gamma'}$, $\gamma\subset_{\uL}\gamma'$, \eqref{eq:hilbertsys} are the refinement maps invoked in \cite{AriciQuantumLatticeGauge} (cp.~also remark \ref{rem:localalg}). The cyclic vectors $\Omega_{\gamma}$, $\gamma\in\Gamma_{\D,L}$, are known as the strong-coupling vacua in lattice gauge theory \cite{KogutTheLatticeGauge, KogutAnIntroductionTo, RotheLatticeGaugeTheories}.
\end{remark}

We conclude our general considerations with the observation that any normalized class-function, $f\in C(G)$\footnote{Presumably, the restriction to continuous functions is unnecessary. But, it is sufficient for our purposes in the following.}, with positive Fourier matrix coefficients, i.e.
\begin{align}
\label{eq:simplecoherenceconditions}
f(1_{G}) & = 1, & \forall g,h\in G: f(\alpha_{g}(h)) & = f(h), & \forall \pi\in\hat{G} : \hat{f}(\pi) & \geq 0,
\end{align}
where $\hat{G}$ is the unitary dual of $G$, gives rise to a coherent family of states (cp.~\eqref{eq:integralstatecoherence} \& \eqref{eq:integralstateinversion}) via:
\begin{align}
\label{eq:simpletensorstates}
T_{\gamma} & = \lambda^{(\gamma)}(F_{\gamma})\in B(L^{2}(\scrA^{E(\gamma)}))_{*}, & F_{\gamma} & = \bigotimes_{e\in E(\gamma)}(f \otimes 1)_{e}\in C(\Pi^{E(\gamma)}).
\end{align}
We consider particular examples of such coherent families of states in section \ref{sec:can11}.

\begin{remark}
\label{rem:inhomogeneoustensorstates}
We note that the coherent families of states defined by \eqref{eq:simpletensorstates} allow for an immediate generalization by replacing the single function $f$ of \eqref{eq:simplecoherenceconditions} with a family of functions adapted to the cofinal sequence of lattices, $\{\gamma_{N}\}_{N\in\N_{0}}$, see figure \ref{fig:cofinal}. Specific examples of this more general construction of coherent families of states are given in section \ref{sec:hkinhom}.
\end{remark}

\section{Yang-Mills theory in 1+1 dimensions}
\label{sec:ym11}
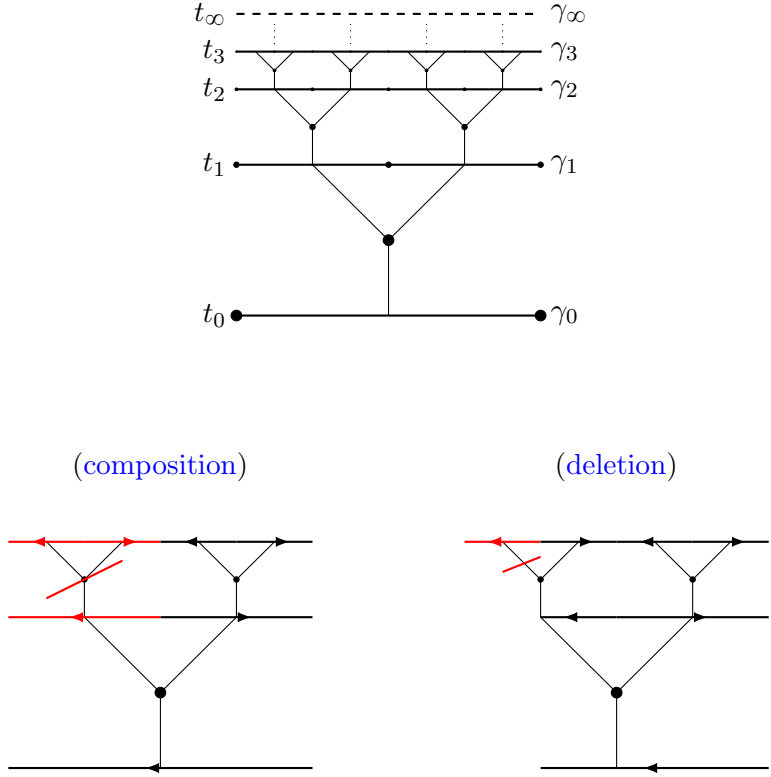
\begin{figure}[t]
	\begin{tikzpicture}
	
	
	\draw[thick, dashed] (3,10) to (7,10);
	\draw[thick] (3,9.5) to (7,9.5) (3,9) to (7,9) (3,8) to (7,8) (3,6) to (7,6);
	
	\draw[dotted] (3.5,9.6) to (3.5,9.9) (4.5,9.6) to (4.5,9.9) (5.5,9.6) to (5.5,9.9) (6.5,9.6) to (6.5,9.9);
	
	\draw (5,6) to (5,7) (5,7) to (4,8) (5,7) to (6,8) (4,8) to (4,8.5) (6,8) to (6,8.5) (4,8.5) to (3.5,9) (4,8.5) to (4.5,9) (6,8.5) to (5.5,9) (6,8.5) to (6.5,9) (3.5,9) to (3.5,9.25) (4.5,9) to (4.5,9.25) (5.5,9) to (5.5,9.25) (6.5,9) to (6.5,9.25) (3.5,9.25) to (3.25,9.5) (3.5,9.25) to (3.75,9.5) (4.5,9.25) to (4.25,9.5) (4.5,9.25) to (4.75,9.5) (5.5,9.25) to (5.25,9.5) (5.5,9.25) to (5.75,9.5) (6.5,9.25) to (6.25,9.5) (6.5,9.25) to (6.75,9.5);
	
	\draw (3,6) node[left]{$t_{0}$} (3,8) node[left]{$t_{1}$} (3,9) node[left]{$t_{2}$} (3,9.5) node[left]{$t_{3}$} (3,10) node[left]{$t_{\infty}$};
	
	\draw (7,6) node[right]{$\gamma_{0}$} (7,8) node[right]{$\gamma_{1}$} (7,9) node[right]{$\gamma_{2}$} (7,9.5) node[right]{$\gamma_{3}$} (7,10) node[right]{$\gamma_{\infty}$};
	
	\filldraw (5,7) circle (2pt) (4,8.5) circle (1pt) (6,8.5) circle (1pt) (3.5,9.25) circle (0.5pt) (4.5,9.25) circle (0.5pt) (5.5,9.25) circle (0.5pt) (6.5,9.25) circle (0.5pt);
	
	\filldraw (3,6) circle (2pt) (7,6) circle (2pt) (3,8) circle (1pt) (5,8) circle (1pt) (7,8) circle (1pt) (3,9) circle (0.5pt) (4,9) circle (0.5pt) (5,9) circle (0.5pt) (6,9) circle (0.5pt) (7,9) circle (0.5pt) (3,9.5) circle (0.25pt) (3.5,9.5) circle (0.25pt) (4,9.5) circle (0.25pt) (4.5,9.5) circle (0.25pt) (5,9.5) circle (0.25pt) (5.5,9.5) circle (0.25pt) (6,9.5) circle (0.25pt) (6.5,9.5) circle (0.25pt) (7,9.5) circle (0.25pt);
	
	
	
	\draw (2,4) node{(\textb{composition})};

	\draw (2,0) to (2,1) (2,1) to (1,2) (2,1) to (3,2) (1,2) to (1,2.5) (3,2) to (3,2.5) (1,2.5) to (0.5,3) (1,2.5) to (1.5,3) (3,2.5) to (2.5,3) (3,2.5) to (3.5,3);
	
	
	\filldraw (2,1) circle (2pt) (1,2.5) circle (1pt) (3,2.5) circle (1pt);
	
	
	\draw[thick, red] (0.5,2.25) to (1.5,2.75);
	
	\draw[->-=0.55, thick] (4,0) to (0,0); 
	\draw[->-=0.6, thick] (2,2) to (4,2); 
	\draw[->-=0.6, thick, red] (2,2) to (0,2);
	\draw[->-=0.7, thick] (3,3) to (4,3);
	\draw[->-=0.7, thick] (3,3) to (2,3);
	\draw[->-=0.7, thick, red] (1,3) to (2,3);
	\draw[->-=0.7, thick, red] (1,3) to (0,3);
	
	
	\draw (8,4) node{(\textb{deletion})};
	
	\draw (8,0) to (8,1) (8,1) to (7,2) (8,1) to (9,2) (7,2) to (7,2.5) (9,2) to (9,2.5) (7,2.5) to (6.5,3) (7,2.5) to (7.5,3) (9,2.5) to (8.5,3) (9,2.5) to (9.5,3);
	
	
	\filldraw (8,1) circle (2pt) (7,2.5) circle (1pt) (9,2.5) circle (1pt);
	
	
	\draw[thick, red] (6.5,2.6) to (7,2.8);
	
	\draw[->-=0.55, thick] (10,0) to (7,0); 
	\draw[->-=0.6, thick] (8,2) to (10,2); 
	\draw[->-=0.7, thick] (8,2) to (7,2);
	\draw[->-=0.7, thick] (9,3) to (10,3);
	\draw[->-=0.7, thick] (9,3) to (8,3);
	\draw[->-=0.7, thick] (7,3) to (8,3);
	\draw[->-=0.7, thick, red] (7,3) to (6,3);

	\end{tikzpicture}
\caption{Illustration of the correspondence between dyadic partitions and binary rooted trees (top, edge orientations are suppressed), and the basic coarsening operations (bottom).}
\label{fig:treecut}
\end{figure}

Although parts of what follows immediately generalize to arbitrary dimensions, we now specialize to 1+1 dimensional case (referred to as YM$_{1+1}$), i.e.~we have $d=1$ and unoriented dyadic lattices $|\Gamma_{\D,L}|$ correspond to dyadic partitions of the finite interval $[0,L]$ or the circle $\T^{1}_{L}$ and, thus, to binary rooted trees $\fT$ (figure \ref{fig:treecut}, top). Moreover, since a choice of cofinal sequence $\{\gamma_{N}\}_{N\in\N_{0}}$ according to remark \ref{rem:cofinal} is in one-to-one correspondence with the sequence of unoriented lattices $|\Gamma_{\D,L}|$, each $\gamma_{N}$ corresponds to a unique dyadic partition which in turn can be identified with a binary rooted tree $t\in\fT$. In the context of this identification, we have two natural bijections: the first is between the edges, $E(\gamma_{N})$, of a lattice $\gamma_{N}$ and the leaves, $\ell(t)$, of a binary rooted tree $t\in\fT$, and the second is between the vertices $V(\gamma_{N})$ and the partition $\cP_{t}=\{0=\sigma_{0}<\sigma_{1}<...<\sigma_{n(t)}=L\}\subset L\D$, where $|V(\gamma_{N})|=n(t)+1$. Moreover, the basic coarsening operations given in definition \ref{def:edgeop} correspond to cutting operations on a tree $t\in\fT$, i.e.~composition of two edges is given by cutting at one of the outer (binary) nodes of $t$ while removal of an edge is given by cutting one of the outer leaves of $t$ (figure \ref{fig:treecut}, bottom). Clearly, the cutting of an outer node corresponds to the removal of the elementary binary tree with two leaves, $\uY$, and, thus, successive compositions of edges are identified with the removal of binary rooted forests, $f\in\cF$. To allow for direct comparison with our companion article \cite{BrothierConstructionsOfConformal}, we note the additional identification of the directed subset $\overleftarrow{\Gamma}_{\D,L}\subset\Gamma_{\D,L}$ of lattices with orientations fixed to the left, i.e.~from $L$ towards $0$, with the directed set of binary rooted trees $\fT$.\\[0.1cm]
Restricting to the directed subset $\overleftarrow{\Gamma}_{\D,L}$ in certain situations is justified by the construction of a $*$-isomorphism between the $C^{*}$-inductive limit algebras associated with the latter and the cofinal sequence $\{\gamma_{N}\}_{N\in\N_{0}}$, see remark \ref{rem:cofinal}. To this end, we observe that $\overleftarrow{\Gamma}_{\D,L}$ is sequential w.r.t.~to the level and totally ordered with respect to $\subset_{\uL}$ (or $\subset_{\uR}$). Therefore, we denote its elements by $\cev{\gamma}_{N}$, $N\in\N_{0}$.

\begin{prop}
\label{prop:cofinaliso}
There exists an isomorphism of $C^{*}$-algebra,
\begin{align*}
\zeta : \varinjlim_{\{\gamma_{N}\}_{N\in\N_{0}}}B(L^{2}(\scrA^{E(\gamma_{N})})) = \fA_{\D,L} & \longrightarrow \overleftarrow{\fA}_{\D,L} := \varinjlim_{\{\cev{\gamma}_{N}\}_{N\in\N_{0}}}B(L^{2}(\scrA^{E(\gamma)})).
\end{align*}
\begin{proof}
Without loss of generality we assume the orientation of the initial element, $\gamma_{0}$, of the cofinal sequence to be to the left. The elementary morphisms that define the inductive-limit algebras are, cp.~\eqref{eq:algsys} to \eqref{eq:orientunitary}, 
\begin{align*}
\alpha^{\gamma_{0}}_{\gamma_{1}} : B(L^{2}(G_{e})) & \longrightarrow B(L^{2}(G_{e_{2}}\times G_{e_{1}})), \\
a & \longmapsto U^{*}_{\uL}(a\otimes\mathds{1}_{e_{1}})U_{\uL}, \\[0.1cm]
\alpha^{\cev{\gamma}_{0}}_{\cev{\gamma}_{1}} : B(L^{2}(G_{e})) & \longrightarrow B(L^{2}(G_{e_{2}}\times G_{e_{1}})), \\
a & \longmapsto U_{\uL}(a\otimes\mathds{1}_{e_{1}})U^{*}_{\uL}.
\end{align*}
This implies that we have a simple commutative diagram involving an orientation reversal on the edge $e_{1}$ because $\Ad_{\mathds{1}_{e_{2}}\otimes U_{\iota}}(U^{*}_{\uL}) = U_{\uL}$:
\begin{align*}
\xymatrix{
B(L^{2}(G_{e})) \ar[dd]_{\id} \ar[rr]^{\alpha^{\gamma_{0}}_{\gamma_{1}}} & & B(L^{2}(G_{e_{2}}\times G_{e_{1}})) \ar[dd]^{\Ad_{\mathds{1}_{e_{2}}\otimes U_{\iota}}} \\
\\
B(L^{2}(G_{e})) \ar[rr]_{\alpha^{\cev{\gamma}_{0}}_{\cev{\gamma}_{1}}} & & B(L^{2}(G_{e_{2}}\times G_{e_{1}})).
}
\end{align*}

\begin{figure}[h]
	\begin{tikzpicture}
	
	\draw[->-=0.55, thick] (2,3) to (0,3);
	\filldraw (0,3) circle (2pt) (2,3) circle (2pt);
	\draw (3,3) node{$\subset_{\uL}$};
	\draw[->-=0.6, thick] (5,3) to (4,3);
	\filldraw (4,3) circle (2pt) (5,3) circle (2pt) (6,3) circle (2pt);
	\draw[->-=0.6, thick] (5,3) to (6,3);
	
	\draw (1,2) node{$\id\downarrow$};
	\draw (5,2) node{$\downarrow\id\times\iota$};
	
	\draw[->-=0.55, thick] (2,1) to (0,1);
	\filldraw (0,1) circle (2pt) (2,1) circle (2pt);
	\draw (3,1) node{$\subset_{\uL}$};
	\draw[->-=0.6, thick] (5,1) to (4,1);
	\filldraw (4,1) circle (2pt) (5,1) circle (2pt) (6,1) circle (2pt);
	\draw[->-=0.6, thick] (6,1) to (5,1);
	
	\end{tikzpicture}
	\caption{Geometrical origin of the isomorphism of proposition \ref{prop:cofinaliso}}
\label{fig:cofinaliso}
\end{figure}
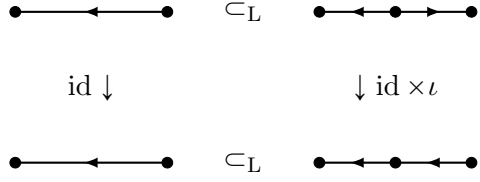

Clearly, this diagram has a simple geometric origin (see figure \ref{fig:cofinaliso}), and proceeding inductively we obtain a sequence $\{\zeta_{N}\}_{N\in\N_{0}}$ with initial elements $\zeta_{0} = \id$, $\zeta_{1}=\Ad_{\mathds{1}_{e_{2}}\otimes U_{\iota}}$. More explicitly, we assume that $\zeta_{N} = \Ad_{U_{N}}$ has been constructed up to some $N\in\N_{0}$ with a unitary $U_{N}\in\cU(L^{2}(\scrA^{E(\gamma_{N})}))$. Then, we define $U_{N+1}$ by
\begin{align*}
\Ad_{U_{N+1}}\circ\!\ \alpha^{\gamma_{N}}_{\gamma_{N+1}} & = \alpha^{\cev{\gamma}_{N}}_{\cev{\gamma}_{N+1}}\circ\Ad_{U_{N}} = \Ad_{\alpha^{\cev{\gamma}_{N}}_{\cev{\gamma}_{N+1}}(U_{N})}\circ\!\ \alpha^{\cev{\gamma}_{N}}_{\cev{\gamma}_{N+1}},
\end{align*}
which is possible because both $\alpha^{\gamma_{N}}_{\gamma_{N+1}}$ and $\alpha^{\cev{\gamma}_{N}}_{\cev{\gamma}_{N+1}}$ are unitarily conjugate to one another:
\begin{align*}
\alpha^{\cev{\gamma}_{N}}_{\cev{\gamma}_{N+1}} & = \Ad_{(1\otimes U_{\iota})^{\otimes 2^{N}}}\circ\!\ \alpha^{\gamma_{N}}_{\gamma_{N+1}}.
\end{align*}
Thus, an admissible choice for $U_{N+1}$ is:
\begin{align*}
U_{N+1} & = \alpha^{\cev{\gamma}_{N}}_{\cev{\gamma}_{N+1}}(U_{N})(1\otimes U_{\iota})^{\otimes 2^{N}}.
\end{align*}
From this we infer the existence of the inductive limit $\varinjlim_{N\in\N_{0}}\zeta_{N} = \zeta$, cf.~\cite{EvansQuantumSymmetriesOn}.
\end{proof}
\end{prop}

\begin{remark}
\label{rem:cofinaliso}
It is important to note, that the isomorphism $\zeta:\fA_{\D,L}\rightarrow\overleftarrow{\fA}_{\D,L}$ is only an isomorphism of $C^{*}$-algebras. If we supply both algebras with states, $\omega_{\D,L}$ and $\cev{\omega}_{\D,L}$, for example, by choosing two coherent families of states \eqref{eq:statecoherence}, there is a priori no guarantee that the corresponding von Neumann algebras will be equivalent, because
\begin{align*}
\cev{\omega}_{\D,L}\circ\zeta & = \omega_{\D,L}
\end{align*}
may not hold.
\end{remark}

\begin{remark}
\label{rem:subiso}
The existence of the isomorphism $\zeta:\fA_{\D,L}\rightarrow\overleftarrow{\fA}_{\D,L}$ should be considered in the light of another natural injective $*$-morphism that results from the inductive system $\{B(L^{2}(\scrA^{E(\gamma)})),\alpha^{\gamma'}_{\gamma}\}_{\gamma\subset_{\uL}\gamma'\in\Gamma_{\D,L}}$. Namely, we have $\cev{\gamma}_{N}\subset_{\uL}\gamma_{N}$ for all $N\in\N_{0}$ which results in the commutative diagram:
\begin{align*}
\xymatrix{
... \ar[rr]^{\alpha^{\cev{\gamma}_{N-1}}_{\cev{\gamma}_{N}}} & & B(L^{2}(\scrA^{E(\cev{\gamma}_{N})})) \ar[dd]_{\alpha^{\cev{\gamma}_{N}}_{\gamma_{N+1}}} \ar[rr]^{\alpha^{\cev{\gamma}_{N}}_{\cev{\gamma}_{N+1}}} & & B(L^{2}(\scrA^{E(\cev{\gamma}_{N+1})})) \ar[dd]^{\alpha^{\cev{\gamma}_{N+1}}_{\gamma_{N+2}}} \ar[rr]^{\alpha^{\cev{\gamma}_{N-1}}_{\cev{\gamma}_{N}}} & & ... \\
\\
... \ar[rr]_{\alpha^{\gamma_{N}}_{\gamma_{N+1}}} & & B(L^{2}(\scrA^{E(\gamma_{N+1})})) \ar[rr]_{\alpha^{\gamma_{N+1}}_{\gamma_{N+2}}} & & B(L^{2}(\scrA^{E(\gamma_{N+2})})) \ar[rr]_{\alpha^{\gamma_{N+2}}_{\gamma_{N+3}}} & & ... ,
}
\end{align*}
and, thus, an injective $*$-morphism $\cev{\alpha}:\overleftarrow{\fA}_{\D,L}\rightarrow\fA_{\D,L}$. In this way, the local net associated with $\overleftarrow{\fA}_{\D,L}$ always constitutes a subnet of that corresponding to $\fA_{\D,L}$, cf.~corollary \ref{cor:localalgsys}.
\end{remark}

\begin{remark}[Inductive system of a cofinal sequence of oriented dyadic trees]
\label{rem:algsyscofinalbasic}
Regarding the cofinal sequence $\{\gamma_{N}\}_{N\in\N_{0}}$, where we choose the edge orientation of the initial element $\gamma_{0}$ to be to the left as in proposition \ref{prop:cofinaliso}, we observe that the inductive system $\{B(L^{2}(\scrA^{E(\gamma_{N})})),\alpha^{\gamma_{N}}_{\gamma_{N+1}}\}_{N\in\N_{0}}$ results from an iteration of the following two injective $*$-morphisms
\begin{align*}
\alpha_{\uL} : B(L^{2}(G_{e})) & \longrightarrow B(L^{2}(G_{e_{2}}\times G_{e_{1}})), \\
 a & \longmapsto U^{*}_{\uL}(a\otimes\mathds{1}_{e_{1}})U_{\uL}, \\
\alpha_{\uR} : B(L^{2}(G_{e})) & \longrightarrow B(L^{2}(G_{e_{2}}\times G_{e_{1}})), \\
 a & \longmapsto V^{*}_{\uR}(\mathds{1}_{e_{2}}\otimes a)V_{\uR},
\end{align*}
which correspond to the upper and lower parts of figure \ref{fig:leftrightrefine}. The unitary $V_{\uR}\in\cU(L^{2}(G\times G))$ is defined by
\begin{align*}
(V_{\uR}\psi)(g_{e_{2}},g_{e_{1}}) & = \psi(g_{e_{2}},g_{e_{1}}g_{e_{2}}), & \psi\in C_{c}(G_{e_{2}}\times G_{e_{1}}).
\end{align*}
and results from conjugation of $U_{\uL}$ with the exchange of (tensor) factors, $\flip:L^{2}(G_{e_{2}}\times G_{e_{1}})\rightarrow L^{2}(G_{e_{1}}\times G_{e_{2}})$:
\begin{align*}
\flip\circ\!\ U_{\uL}\circ\flip & = V_{\uR}.
\end{align*}

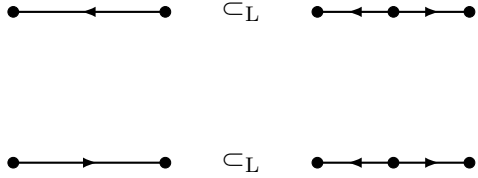
\begin{figure}[h]
	\begin{tikzpicture}
	
	\draw[->-=0.55, thick] (2,3) to (0,3);
	\filldraw (0,3) circle (2pt) (2,3) circle (2pt);
	\draw (3,3) node{$\subset_{\uL}$};
	\draw[->-=0.6, thick] (5,3) to (4,3);
	\filldraw (4,3) circle (2pt) (5,3) circle (2pt) (6,3) circle (2pt);
	\draw[->-=0.6, thick] (5,3) to (6,3);
	
	\draw[->-=0.55, thick] (0,1) to (2,1);
	\filldraw (0,1) circle (2pt) (2,1) circle (2pt);
	\draw (3,1) node{$\subset_{\uL}$};
	\draw[->-=0.6, thick] (5,1) to (4,1);
	\filldraw (4,1) circle (2pt) (5,1) circle (2pt) (6,1) circle (2pt);
	\draw[->-=0.6, thick] (5,1) to (6,1);
	
	\end{tikzpicture}
	\caption{Refinements of left- and right-oriented edges w.r.t.~$\subset_{\uL}$.}
\label{fig:leftrightrefine}
\end{figure}

\end{remark}

\subsection{Jones' actions of Thompson's groups}
\label{sec:thompson}
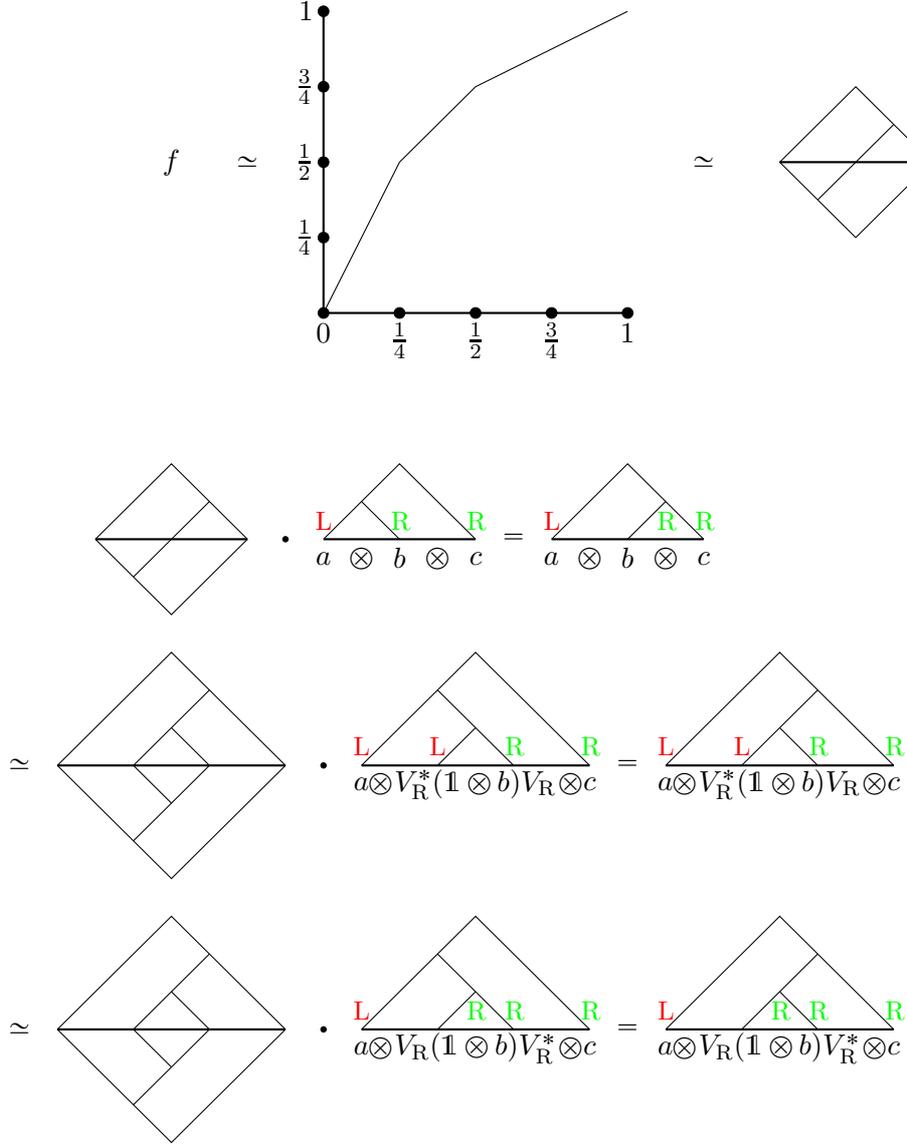
\begin{figure}[h]
	\begin{tikzpicture}
	
	
	\draw (2,2) node{$f$};
	
	\draw (3,2) node{$\simeq$};
	
	\filldraw (5,0) circle (2pt) (6,0) circle (2pt) (7,0) circle (2pt) (8,0) circle (2pt);
	\draw (5,0) node[below]{$\frac{1}{4}$} (6,0) node[below]{$\frac{1}{2}$} (7,0) node[below]{$\frac{3}{4}$} (8,0) node[below]{$1$};
	\draw[thick] (4,0) to (8,0);
	\filldraw (4,0) circle (2pt) (4,1) circle (2pt) (4,2) circle (2pt) (4,3) circle (2pt) (4,4) circle (2pt);
	\draw (4,0) node[below]{$0$} (4,1) node[left]{$\frac{1}{4}$} (4,2) node[left]{$\frac{1}{2}$} (4,3) node[left]{$\frac{3}{4}$} (4,4) node[left]{$1$};
	\draw[thick] (4,0) to (4,4);
	
	\draw (4,0) to (5,2) to (6,3) to (8,4);
	
	\draw (9,2) node{$\simeq$};
	
	\draw[thick] (10,2) to (12,2);
	\draw (10,2) to (11,3) to (12,2);
	\draw (11.5,2.5) to (11,2);
	\draw (10,2) to (11,1) to (12,2);
	\draw (10.5,1.5) to (11,2);
	
	
	\draw[thick] (1,-3) to (3,-3);
	\draw (1,-3) to (2,-2) to (3,-3);
	\draw (2.5,-2.5) to (2,-3);
	\draw (1,-3) to (2,-4) to (3,-3);
	\draw (1.5,-3.5) to (2,-3);
	
	\filldraw (3.5,-3) circle (1pt);
	
	\draw[thick] (4,-3) to (6,-3);
	\draw (4,-3) to (5,-2) to (6,-3);
	\draw (4.5,-2.5) to (5,-3);
	\draw (5.75,-2.75) node[right]{\small\color{green}{$\uR$}};
	\draw (4.25,-2.75) node[left]{\small\color{red}{$\uL$}};
	\draw (4.75,-2.75) node[right]{\small\color{green}{$\uR$}};
	\draw (4,-3.25) node{$a$} (4.5,-3.25) node{$\otimes$} (5,-3.25) node{$b$} (5.5,-3.25) node{$\otimes$} (6,-3.25) node{$c$};
	
	\draw (6.5,-3) node{$=$};
	
	\draw[thick] (7,-3) to (9,-3);
	\draw (7,-3) to (8,-2) to (9,-3);
	\draw (8.5,-2.5) to (8,-3);
	\draw (8.25,-2.75) node[right]{\small\color{green}{$\uR$}};
	\draw (8.75,-2.75) node[right]{\small\color{green}{$\uR$}};
	\draw (7.25,-2.75) node[left]{\small\color{red}{$\uL$}};
	\draw (7,-3.25) node{$a$} (7.5,-3.25) node{$\otimes$} (8,-3.25) node{$b$} (8.5,-3.25) node{$\otimes$} (9,-3.25) node{$c$};
	
	
	\draw (0,-6) node{$\simeq$};
	
	\draw[thick] (0.5,-6) to (3.5,-6);
	\draw (0.5,-6) to (2,-4.5) to (3.5,-6);
	\draw (2.5,-5) to (1.5,-6);
	\draw (2,-5.5) to (2.5,-6);
	\draw (0.5,-6) to (2,-7.5) to (3.5,-6);
	\draw (1.5,-7) to (2.5,-6);
	\draw (2,-6.5) to (1.5,-6);
	
	\filldraw (4,-6) circle (1pt);
	
	\draw[thick] (4.5,-6) to (7.5,-6);
	\draw (4.5,-6) to (6,-4.5) to (7.5,-6);
	\draw (6,-5.5) to (5.5,-6);
	\draw (5.5,-5) to (6.5,-6);
	\draw (4.75,-5.75) node[left]{\small\color{red}{$\uL$}};
	\draw (5.75,-5.75) node[left]{\small\color{red}{$\uL$}};
	\draw (6.25,-5.75) node[right]{\small\color{green}{$\uR$}};
	\draw (7.25,-5.75) node[right]{\small\color{green}{$\uR$}};
	\draw (4.5,-6.25) node{$a$} (4.75,-6.25) node{$\otimes$} (6,-6.25) node{$V^{*}_{\uR}(\mathds{1}\otimes b)V_{\uR}$} (7.25,-6.25) node{$\otimes$} (7.5,-6.25) node{$c$};
	
	\draw (8,-6) node{$=$};
	
	\draw[thick] (8.5,-6) to (11.5,-6);
	\draw (8.5,-6) to (10,-4.5) to (11.5,-6);
	\draw (10.5,-5) to (9.5,-6);
	\draw (10,-5.5) to (10.5,-6);
	\draw (8.75,-5.75) node[left]{\small\color{red}{$\uL$}};
	\draw (9.75,-5.75) node[left]{\small\color{red}{$\uL$}};
	\draw (10.25,-5.75) node[right]{\small\color{green}{$\uR$}};
	\draw (11.25,-5.75) node[right]{\small\color{green}{$\uR$}};
	\draw (8.5,-6.25) node{$a$} (8.75,-6.25) node{$\otimes$} (10,-6.25) node{$V^{*}_{\uR}(\mathds{1}\otimes b)V_{\uR}$} (11.25,-6.25) node{$\otimes$} (11.5,-6.25) node{$c$};
	
	
	\draw (0,-9.5) node{$\simeq$};
	
	\draw[thick] (0.5,-9.5) to (3.5,-9.5);
	\draw (0.5,-9.5) to (2,-8) to (3.5,-9.5);
	\draw (2.5,-8.5) to (1.5,-9.5);
	\draw (2,-9) to (2.5,-9.5);
	\draw (0.5,-9.5) to (2,-11) to (3.5,-9.5);
	\draw (1.5,-10.5) to (2.5,-9.5);
	\draw (2,-10) to (1.5,-9.5);
	
	\filldraw (4,-9.5) circle (1pt);
	
	\draw[thick] (4.5,-9.5) to (7.5,-9.5);
	\draw (4.5,-9.5) to (6,-8) to (7.5,-9.5);
	\draw (6,-9) to (5.5,-9.5);
	\draw (5.5,-8.5) to (6.5,-9.5);
	\draw (4.75,-9.25) node[left]{\small\color{red}{$\uL$}};
	\draw (5.75,-9.25) node[right]{\small\color{green}{$\uR$}};
	\draw (6.25,-9.25) node[right]{\small\color{green}{$\uR$}};
	\draw (7.25,-9.25) node[right]{\small\color{green}{$\uR$}};
	\draw (4.5,-9.75) node{$a$} (4.75,-9.75) node{$\otimes$} (6,-9.75) node{$V_{\uR}(\mathds{1}\otimes b)V^{*}_{\uR}$} (7.25,-9.75) node{$\otimes$} (7.5,-9.75) node{$c$};
	
	\draw (8,-9.5) node{$=$};
	
	\draw[thick] (8.5,-9.5) to (11.5,-9.5);
	\draw (8.5,-9.5) to (10,-8) to (11.5,-9.5);
	\draw (10.5,-8.5) to (9.5,-9.5);
	\draw (10,-9) to (10.5,-9.5);
	\draw (8.75,-9.25) node[left]{\small\color{red}{$\uL$}};
	\draw (9.75,-9.25) node[right]{\small\color{green}{$\uR$}};
	\draw (10.25,-9.25) node[right]{\small\color{green}{$\uR$}};
	\draw (11.25,-9.25) node[right]{\small\color{green}{$\uR$}};
	\draw (8.5,-9.75) node{$a$} (8.75,-9.75) node{$\otimes$} (10,-9.75) node{$V_{\uR}(\mathds{1}\otimes b)V^{*}_{\uR}$} (11.25,-9.75) node{$\otimes$} (11.5,-9.75) node{$c$};
		
	\end{tikzpicture}
	\caption{Exemplary action of an $f\in F$ on $\fA_{\D,L}$ in graphical notation. {\color{red}{$\uL$}} and {\color{green}{$\uR$}} refer to the orientation of the edges in the partition corresponding to a binary rooted tree. The last and the next-to-last line illustrate the two possible refinements of the operator on the middle edge, $b$, compatible with $\subset_{\uL}$.}
\label{fig:orientedtreeaction}
\end{figure}

In view of our companion article \cite{BrothierConstructionsOfConformal}, we explain in this subsection how the construction of the semi-continuum field algebra $\fA_{\D,L}$ in 1+1-dimensions via the inductive system $\{B(L^{2}(\scrA^{E(\gamma)})),\alpha^{\gamma}_{\gamma'}\}_{\gamma\subset_{\uL}\gamma'\in\Gamma_{\D,L}}$ leads to an action of Thompson's groups $F\subset T\subset V$ on $\fA_{\D, L}$ by automorphisms \cite{JonesSomeUnitaryRepresentations, JonesANoGo}. In contrast with \cite{BrothierConstructionsOfConformal}, we illustrate this action in a manner familiar from loop quantum gravity-type models, e.g.~\cite{ThiemannModernCanonicalQuantum}, which gives us the advantage that the extension of Jones' action, initially given on unoriented dyadic lattices\footnote{Unoriented dyadic lattices directly correspond to binary rooted trees, and Jones' action is constructed by exploiting the categorical structure of the direct set of binary rooted trees. i.e.~Thompson's group $F$ arises as the group of fractions of said category \cite{CannonIntroductoryNotesOn, JonesSomeUnitaryRepresentations}.}, to oriented dyadic lattices is almost self-evident. For this purpose, we consider Thompson's group $F$ as the group of piecewise linear homeomorphims of $f:[0,1]\rightarrow[0,1]$ that are differentiable everywhere except from a finite set of dyadic rationals $\D_{f}\subset\D$ and the differentials are multiplications by powers of $2$ where they exist \cite{CannonIntroductoryNotesOn}. Positivity of the differentials that are defined almost everywhere implies that elements of $F$ are orientation preserving. Therefore, it is clear that $F$ acts (asymptotically) on the family of oriented dyadic lattices $\Gamma_{\D,L}$ in dimension 1 by conjugating with the scaling of the interval $[0,L]$ to $[0,1]$ und using suitable refining operations. The action of $F\act\fA_{\D,L}$ is compatible with $\subset_{\uL/\uR}$, i.e.~$f(\gamma)\subset_{\uL/\uR}f(\gamma')$ for $\gamma\subset_{\uL/\uR}\gamma'$ and $f\in F$, and it can be visualized in the following sense, cp.~figure \ref{fig:orientedtreeaction}:\\[0.1cm]
Consider the dense $*$-subalgebra $\fA^{(0)}_{\D,L}\subset\fA_{\D,L}$ given by the algebraic inductive limit. Then, any element $a\in\fA^{(0)}_{\D,L}$ is uniquely determined by an element $a_{\gamma_{\min}}\in B(L^{2}(\scrA^{E(\gamma_{\min})}))$ for a minimal $\gamma_{\min}\in\Gamma_{\D,L}$. Given $f\in F$, we assign a new element $(f\cdot a)_{f(\gamma_{\min})}\in B(L^{2}(\scrA^{E(f(\gamma_{\min}))}))$. We initially assume that $a_{\gamma_{\min}} = \otimes_{e\in E(\gamma_{\min})}a_{e}$ is an elementary tensor and $\gamma_{\min}$ corresponds to an $f$-adapted partition of $[0,1]$, i.e.~the break points of $f$ form a subset of the vertices of $\gamma_{\min}$ after rescaling, $L\D_{f}\subset V(\gamma_{\min})$. Therefore, we know that $f(\gamma_{\min})$ results from taking the images of edges in $E(\gamma_{\min})$ and applying the necessary refinements (see proposition \ref{prop:groupsys}) to obtain an element of $\Gamma_{\D,L}$. This allows us to define $(f\cdot a)_{f(\gamma_{\min})}$ by subjecting $\otimes_{e'=f(e): e\in E(f(\gamma_{\min}))}a_{f^{-1}(e')}$ to the necessary refinements (see proposition \ref{prop:algsys}) according to the associated identification of tensor factors. Since this assignment is isometric, we may extend it uniquely to all of $B(L^{2}(\scrA^{E(\gamma_{\min})}))$ by the density of elementary tensors. Next, we assume that $\gamma_{\min}$ is not $f$-adapted. Since $\Gamma_{\D,L}$ is directed, we now that there is an $f$-adapted $\gamma'_{\min}$ s.t. $\gamma_{\min}\subset_{\uL}\gamma'_{\min}$, and we may apply the previous formula to $\alpha^{\gamma_{\min}}_{\gamma'_{\min}}(a_{\gamma_{\min}})$. Thus, we obtain a map $f\cdot\!\ .\!\ :\fA^{(0)}_{\D,L}\rightarrow\fA^{(0)}_{\D,L}$ which extends uniquely to $\fA_{\D,L}$. It is easily checked that this defines an automorphic action $F\act\fA_{\D,L}$ which coincides with the categorical construction of Jones when lattices are unoriented and correspond directly to binary rooted trees. The extension of this action to $T$ and $V$ is straightforward. 

\begin{remark}[Jones' actions and the locality structure]
\label{rem:localV}
Although we have actions of Thompson's groups $F\subset T\subset V$ on the semi-continuum field algebra $\fA_{\D,L}$, only the actions of $F$ and $T$ are compatible with the locality structure of $\fA_{\D,L}$ given in corollary \ref{cor:localalgsys} as the concept of star domain is not invariant w.r.t.~$V$.
\end{remark}

\begin{remark}[Extensions to von Neumann algebras]
\label{rem:thompsonVNA}
Initially, the actions of $F\subset T\subset V$ are defined on $\fA_{\D,L}$ and their extension to the von Neumann algebra $\scrM_{\D,L}$ (form some state $\omega_{\D,L}$) is a nontrivial matter. A sufficient a priori reason can be given by the (quasi-)invariance of $\omega_{\D,L}$ w.r.t.~one of the actions or its restriction to a suitable subgroup, e.g.~the dyadic rotations in $T$, see \cite{BrothierConstructionsOfConformal}. 
\end{remark}

\begin{remark}
\label{rem:thompsonrestrict}
It follows from the construction of the Jones actions of $F\subset T\subset V$ on $\fA_{\D,L}$ and remark \ref{rem:subiso} that their restrictions to $\overleftarrow{\fA}_{\D,L}$ are well-defined.
\end{remark}

\subsection{Fourier transform, the holonomy map, and duality}
\label{sec:duality}
As it is well-know from the general setting of lattice models, notions of duality are extremely useful and provide valuable insight into the structure of models \cite{KogutTheLatticeGauge, KogutAnIntroductionTo, CreutzQuarksGluonsLattices, ItzyksonStatisticalFieldTheory1}. Therefore, we introduce two consistent systems of transformations w.r.t.~to our projective and inductive systems that elucidate especially well an instance of duality in the $d=1$ case.

\subsubsection{Fourier transform}
\label{sec:fourier}
For the first system of transformations, we restrict to the case when $G$ is a compact group\footnote{A similar construction can be done in the context of quantum groups.}, and make use of the Fourier transform (cf.~\cite{RuzhanskyPseudoDifferentialOperators}), 
\begin{align}
\label{eq:fourier}
\scrF : L^{2}(G) & \longrightarrow L^{2}(\hat{G}), & \scrF[\phi](\pi)_{nm} & = \hat{\phi}(\pi)_{nm} = \int_{G}dg\!\ \overline{\pi_{mn}(g)}\phi(g) = (\pi_{mn},\phi)_{L^{2}(G)}, \\ \nonumber
\scrF^{-1} : L^{2}(\hat{G}) & \longrightarrow L^{2}(G), & \scrF^{-1}[\varphi](g) & = \check{\varphi}(g) = \sum_{[\pi]\in\hat{G}}d_{\pi}\Tr_{V_{\pi}}(\pi(g)\varphi(\pi)),
\end{align}
between $G$ and its unitary dual $\hat{G}$, and the coherence of the latter w.r.t.~the projective and inductive systems given in corollary \ref{cor:groupsys}. Here, $\pi:G\rightarrow\cU(V_{\pi})$ is a irreducible, unitary representation defining an equivalence class $[\pi]\in\hat{G}$. $\{\pi_{mn}\}_{m,n=1}^{d_{\pi}}$ are the matrix coefficient w.r.t.~to an orthonormal basis $\{v^{\pi}_{k}\}^{d_{\pi}}_{k=1}$ of $V_{\pi}$, and $d_{\pi} = \dim(V_{\pi})$ is the (complex) dimension of the representation. Thus, the Fourier transform requires a fixed choice of representatives for every equivalence class in $\hat{G}$, e.g.~a suitable subrepresentation of the left regular representation $\lambda:G\rightarrow\cU(L^{2}(G))$. In \eqref{eq:fourier} and the following we allow for a slight abuse of notation and identify $\hat{G}$ with a fixed collection of matrix representatives. More precisely, we consider $\hat{G}$ together with its counting measure as in \eqref{eq:fourier} to be given by a specific realization of the group von Neumann algebra $\cR_{\lambda}(G)$ \cite{PedersenCAlgebrasAnd}:
\begin{align}
\label{eq:matrixdual}
L^{\infty}(\hat{G}) & \cong \cR_{\lambda}(G) \cong \bigoplus^{\li}_{[\pi]\in\hat{G}}M_{d_{\pi}}(\C),
\end{align}
where we consider the $\li$-closure of the (algebraic) direct sum. We remark that it appears to be convenient to interpret $L^{\infty}(\hat{G})$ as von Neumann-algebraic representation of the discrete quantum group $\hat{G}$, e.g.~\cite{KustermansLocallyCompactQuantum, TimmermannAnInvitationTo}. By means of the Fourier transform and the duality pairing,
\begin{align}
\label{eq:fourierduality}
\hat{G}\times G & \longrightarrow \C, \\ \nonumber
(\pi_{mn}, g) & \longmapsto \pi_{mn}(g),
\end{align}
we have basic dual field operators in $B(L^{2}(\hat{G}))$ (cp.~\eqref{eq:basickernels}),
\begin{align}
\label{eq:dualfieldop}
\hat{\lambda}(\pi_{mn}) & = \scrF \circ M(\pi_{mn})\circ \scrF^{-1}, \\ \nonumber
\hat{M}(\pi_{mn}) & = \scrF \circ \lambda(\pi_{mn}\otimes 1)\circ \scrF^{-1},
\end{align}
for all matrix coefficients $\pi_{mn}\in\hat{G}$. The (infinite) matrix representations of which are:
\begin{align}
\label{eq:dualfieldopmatrix}
\hat{\lambda}(\pi_{mn})_{\pi'_{m'n'}}^{\pi''_{m''n''}} & = \int_{G}dg\!\ \overline{\pi''_{m'n'}(g)}\pi_{mn}(g)\pi'_{m''n''}(g), \\ \nonumber
\hat{M}(\pi_{mn})_{\pi'_{m'n'}}^{\pi''_{m''n''}} & = d^{-2}_{\pi}\delta_{\pi,\pi'}\delta_{\pi,\pi''}\delta_{m,m'}\delta_{n,m''}\delta_{n',n''}.
\end{align}
Thus, we deduce that $\hat{\lambda}(\pi_{mn})$ acts by tensor convolution w.r.t.~to the representation $\pi$ (Clebsch-Gordan coefficients) on $L^{2}(\hat{G})$ while $\hat{M}(\pi_{mn})$ acts by right multiplication with the transpose of the suitably normalized matrix unit in the direct summand corresponding to $\pi$:
\begin{align}
\label{eq:fouriermatrixunits}
(E(\pi)_{mn})(\pi')_{n'm'} = d^{-1}_{\pi}\delta_{\pi,\pi'}\delta_{m,m'}\delta_{n,n'}.
\end{align}
The analogue of the canonical commutation relations \eqref{eq:Gccr} reads,
\begin{align}
\label{eq:fourierGccr}
\left(\hat{M}(\pi'_{m'n'})\hat{\lambda}(\pi_{mn})\hat{M}(\pi''_{m''n''})\right)_{\sigma'_{k'l'}}^{\sigma''_{k''l''}} & = d^{-1}_{\pi'}d^{-1}_{\pi''}\delta_{\pi',\sigma'}\delta_{\pi'',\sigma''}\delta_{m',k'}\delta_{n'',k''}\hat{\lambda}(\pi_{mn})_{\pi'_{n',l'}}^{\pi''_{m'',l''}}.
\end{align}
Now, we define the following (dual) maps corresponding to the elementary operations of edge composition and deletion.

\begin{definition}
\label{def:fouriergroupsys}
Let $\hat{G}$ be the unitary dual of $G$ as given by \eqref{eq:matrixdual}. Then, we consider the maps:
\begin{align*}
\hat{p}^{\{e_{1},e_{2}\}}_{\{e\}} : \hat{G}_{e} & \longrightarrow \hat{G}_{e_{2}}\otimes\hat{G}_{e_{1}}, & \hat{j}^{\{e\}}_{\{e_{1},e_{2}\}} : \hat{G}_{e_{2}}\otimes\hat{G}_{e_{1}} & \longrightarrow \hat{G}_{e}\cong\hat{G}_{e}\otimes\{\mathds{1}_{\hat{G}}\}, \\
\pi_{mn} & \longmapsto \sum^{d_{\pi}}_{k=1}\pi_{mk}\otimes\pi_{kn}, & \pi_{mn}\otimes\pi'_{m'n'} & \longmapsto \pi_{mn}\sim\pi_{mn}\otimes\mathds{1}_{\hat{G}}, \\[0.25cm]
\hat{p}^{\{e\}}_{\emptyset} : \{\hat{1}\} & \longrightarrow \hat{G}_{e}, & \hat{j}^{\emptyset}_{\{e\}} : \hat{G}_{e} & \longrightarrow \{\hat{1}\}, \\
\hat{1} & \longmapsto \pi_{\utriv}, & \pi_{mn} & \longmapsto \hat{1}.
\end{align*}
\end{definition}

The following result is an immediate consequence of the preceding definition.

\begin{corollary}
\label{cor:fouriergroupsys}
The maps of definition \ref{def:fouriergroupsys} correspond to those of \ref{def:groupop} via the Fourier transform \eqref{eq:fourier}, i.e.~$\hat{p}^{\{e_{1},e_{2}\}}_{\{e\}}$ and $\hat{p}^{\{e\}}_{\emptyset}$ induce maps between the corresponding $L^{2}$-spaces via pullback such that:
\begin{align*}
(\hat{p}^{\{e_{1},e_{2}\}}_{\{e\}})^{*}\scrF[(p^{\{e_{1},e_{2}\}}_{\{e\}})^{*}\scrF^{-1}[\varphi]](\pi)_{nm} & = \varphi(\pi)_{nm}, \\
(\hat{p}^{\{e\}}_{\emptyset})^{*}\scrF[(p^{\{e\}}_{\emptyset})^{*}\scrF^{-1}[w]](\hat{1}) & = w(\hat{1}),
\end{align*}
for $\varphi\in L^{2}(\hat{G})$ and $w\in L^{2}(\{\hat{1}\})\cong\C$, and we have the following relations among the dual field operators:
\begin{align*}
\alpha^{\{e\}}_{\{e_{1},e_{2}\}}(\hat{\lambda}^{(\{e\})}(\pi_{mn})) & = (\hat{\lambda}^{(\{e_{2}\})}\otimes\hat{\lambda}^{(\{e_{1}\})})(\hat{p}^{\{e_{1},e_{2}\}}_{\{e\}}(\pi_{mn})), \\ \nonumber
\alpha^{\{e\}}_{\{e_{1},e_{2}\}}(\hat{M}^{(\{e\})}(\hat{j}^{\{e\}}_{\{e_{1},e_{2}\}}(\pi_{mn}\otimes\pi'_{m'n'}))) & = \hat{M}^{(\{e_{2}\})}(\pi_{mn})\otimes\hat{M}^{(\{e_{1}\})})(\mathds{1}_{\hat{G}}), \\[0.25cm] \nonumber
\alpha^{\emptyset}_{\{e\}}(\hat{\lambda}^{(\emptyset)}(\hat{1})) & = \hat{\lambda}^{(\{e_{2}\})}(\hat{p}^{\{e_{1}\}}_{\emptyset}(\hat{1})), \\ \nonumber
\alpha^{\emptyset}_{\{e\}}(\hat{M}^{(\emptyset)}(\hat{j}^{\emptyset}_{\{e\}}(\pi_{mn}))) & = \hat{M}^{(\{e_{2}\})}(\mathds{1}_{\hat{G}}),
\end{align*}
where $B(L^{2}(\hat{G}))\cong B(L^{2}(G))$ via conjugation with the Fourier transform is implicitly understood. Moreover, we slightly abuse notation and denote the Fourier transform for several copies of $G$ and the trivial group $\{1\}$ by the same symbol $\scrF$.\end{corollary}
The corollary implies that we can equally well base our construction of the semi-continuum field algebra $\fA_{\D,L}$ on $B(L^{2}(\hat{G}))$ together with the dual projective and inductive systems coming from definition \ref{def:fouriergroupsys}. Nevertheless, it is important to note that the compatible dual inductive limit structure of the Hilbert space $\hat{\cH}_{\D,L}\cong L^{2}(\overline{\scrA})$ is obtained from the right inverses of $(\hat{p}^{\{e_{1},e_{2}\}}_{\{e\}})^{*}$ and $(\hat{p}^{\{e\}}_{\emptyset})^{*}$ given in the preceding corollary:
\begin{align}
\label{eq:fouriercompacthilbertsys}
(\hat{R}^{\{e\}}_{\{e_{1},e_{2}\}}\varphi)(\pi,\pi')_{nm,n'm'} & = d^{-1}_{\pi} \delta_{\pi,\pi'} \delta_{n,m'}\varphi(\pi)_{n'm} = (((\hat{p}^{\{e_{1},e_{2}\}}_{\{e\}})^{*})^{-1}\varphi)(\pi,\pi')_{nm,n'm'}, \\ \nonumber
(\hat{R}^{\emptyset}_{\{e\}}w)(\pi)_{nm} & = \delta_{\pi,\pi_{\utriv}}\delta_{m,1}\delta_{n,1}w(\hat{1}) = ((\hat{p}^{\{e\}}_{\emptyset})^{*})^{-1}w)(\pi)_{nm}.
\end{align}
These maps are well-defined because $\hat{G}$ is discrete, and the GNS vector of $\hat{\cH}_{\D,L}$ is given by the indicator function of the trivial representation, $\delta_{\pi_{\utriv}}$, on every edge. By standard reasoning \cite{EvansQuantumSymmetriesOn}, the coherent system of transformations given by the Fourier transform provides a unitary transformation,
\begin{align}
\label{eq:limitfourier}
\overline{\scrF} : L^{2}(\overline{\scrA})=\cH_{\D,L} & \longrightarrow \hat{\cH}_{\D,L} = \ell^{2}(\hat{G}_{\fr}),
\end{align}
where,
\begin{align}
\label{eq:fouriercompactL2limit}
\hat{\cH}_{\D,L} & = \varinjlim_{\gamma\in\Gamma_{\D,L}}\ell^{2}(\hat{G}^{E(\gamma)}),
\end{align}
is the dual inductive-limit Hilbert space with
\begin{align}
\label{eq:fouriercompacspacelimit}
\hat{G}_{\fr} & = \varinjlim_{\gamma\in\Gamma_{\D,L}}\hat{G}^{E(\gamma)}, & \hat{G}^{E(\gamma)} & = \bigotimes_{e\in E(\gamma)}\hat{G}_{e}.
\end{align}
For obvious reasons, we refer to $\overline{\scrF}$ as the Fourier transform for $\overline{\scrA}$. 

\begin{remark}
\label{rem:looptransform}
We note that our definition of the Fourier transform, $\overline{\scrF}$, for $\overline{\scrA}$ is nothing else but a basis-free version of the spin-network transform used in loop quantum gravity \cite{ThiemannModernCanonicalQuantum, AbbatiInductiveConstructionOf, ThiemannAnAccountOf}. But, the values taken by a function $\varphi^{(\gamma)}\in L^{2}(\hat{G}^{E(\gamma)})$ should not be confused with the coefficient w.r.t.~the spin-network basis due to the presence of non-trivial Mandelstam identities.
\end{remark}

\subsubsection{The holonomy map}
\label{sec:discretehol}
The second system of transformations arises from a realization of $\overline{\scrA}$ as affine, based dyadic-path group $W_{\D,L}(G)$ via a discrete analogue of the holonomy map (cp.~\cite{MouraoPhysicalPropertiesOf, DriverYangMillsTheory}). For simplicity, we consider $\overline{\scrA}$ as projective limit w.r.t.~$\overleftarrow{\Gamma}_{\D,L}$, but the generalization to $\Gamma_{\D,L}$ via a cofinal sequence $\{\gamma_{N}\}_{N\in\N_{0}}$ is straightforward.\\
\\

\begin{definition}
\label{def:discretehol}
We consider the countably infinite product,
\begin{align*}
W_{\D,L}(G) & := \{h:L\D\rightarrow G\ |\ h_{0}=1_{G}\} \cong \bigtimes_{d\in L\D_{\neq 0}}G,
\end{align*}
and define a map
\begin{align*}
\h : \overline{\scrA} & \longrightarrow W_{\D,L}(G) \\
\bar{g} & \longmapsto \{\h(\bar{g})_{\tau}\}_{\tau\in L\D},
\end{align*}
by the following prescription:
\begin{align*}
\h(\bar{g})_{\tau} & = \prod^{\leftarrow}_{\tau\geq\sigma\in\cP_{t}}(g_{t})_{\sigma}, & \h(\bar{g})_{0} & = 1_{G},
\end{align*}
with $g_{t} = p_{\gamma(t)}(\bar{g})$ s.t. $\tau\in\{0=\sigma_{0}<\sigma_{1}<...<\sigma_{n(t)}=L\}=\cP_{t}$, where the latter is the dyadic partition 
\begin{align*}
(0,L] & = \bigcup^{n}_{i=1}\underbrace{(\sigma_{i-1},\sigma_{i}]}_{=:I_{\sigma_{i}}}
\end{align*}
defined by $t\in\fT$, and $\gamma(t)\in\overleftarrow{\Gamma}_{\D,L}$ is the associated lattice. $\prod^{\leftarrow}_{\sigma\in\cP_{t}}$ denotes the ordered product of elements of $G$ according to the partition $\cP_{t}$ (indices decreasing to the left) and $(g_{t})_{\sigma} = g_{I_{\sigma}}$, $(g_{t})_{\sigma_{0}}=1_{G}$. We call $\h(\bar{g})\in W_{\D,L}(G)$ the (dyadic) holonomy of $\bar{g}\in\overline{\scrA}$ based at $0$.
\end{definition}

Naturally, $\overline{\scrA}$ inherits only an affine group structure from $W_{\D,L}(G)$ because the former can be interpreted as the groupoid homomorphism of the dyadic-path groupoid into the structure group, $\hom(\mathscr{P}_{\D,L},G)$, i.e.~two elements of $\overline{\scrA}$ differ by an element of $W_{\D,L}(G)$. The properties of the holonomy and the aforesaid realization of $\overline{\scrA}$ are summarized in the following proposition.

\begin{prop}
\label{prop:discretehol}
The holonomy map $\h:\overline{\scrA}\longrightarrow W_{\D,L}(G)$ is well-defined and a homeomorphism, if we endow $W_{\D,L}(G)\cong\bigtimes_{d\in \D_{\neq 0}}G$ with the Tychonoff topology\footnote{Recall that the Tychonoff topology is the initial topology for the natural projective structure of $W_{\D,L}(G)$}. Moreover, the push-forward of the measure $m_{\overline{\scrA}}$ satisfies $\h_{*}m_{\overline{\scrA}} = \times_{d\in \D_{\neq 0}}m_{G}$ for the corresponding Borel $\sigma$-algebras.
\begin{proof}
We only show that $\h$ is well-defined and give a formula for the inverse, $\h^{-1}$, because it is instructive to see how the computation works. For a complete proof we refer to proposition 2.10 of our companion paper \cite{BrothierConstructionsOfConformal}.\\
\\
Given $\bar{g}\in\overline{\scrA}$ and $\tau\in\D$, we consider $g_{t} = p_{\gamma(t)}(\bar{g})$, $g_{t'}=p_{\gamma(t')}(\bar{g})$ for some $t,t'\in\fT$ such that $\tau\in\cP_{t}\cap\cP_{t'}$. Then, we can always find $t''\in\fT$ s.t. $t,t'\leq t''$ as $\fT$ is directed, i.e.~$ft = t'', f't' = t''$ for $f,f'\in\cF$. Therefore, we have by the definition of $\bar{g}$:
\begin{align*}
\h(\bar{g})_{\tau} & = \prod^{\leftarrow}_{\tau\geq\sigma\in\cP_{t}}(g_{t})_{\sigma} = \prod^{\leftarrow}_{\tau\geq\sigma\in\cP_{t}}\underbrace{\left(\prod^{\leftarrow}_{\sigma\geq\sigma''\in\cP_{t''}(I_{\sigma})}(g_{t''})_{\sigma''}\right)}_{=p^{\gamma(t'')}_{\gamma(t)}(g_{t''})_{\sigma}} = \prod^{\leftarrow}_{\tau\geq\sigma''\in\cP_{t''}}(g_{t''})_{\sigma''}, \\
\h(\bar{g})_{\tau} & = \prod^{\leftarrow}_{\tau\geq\sigma'\in\cP_{t'}}(g_{t'})_{\sigma'} = \prod^{\leftarrow}_{\tau\geq\sigma'\in\cP_{t'}}\underbrace{\left(\prod^{\leftarrow}_{\sigma'\geq\sigma''\in\cP_{t''}(I_{\sigma'})}(g_{t''})_{\sigma''}\right)}_{=p^{\gamma(t'')}_{\gamma(t')}(g_{t''})_{\sigma'}} = \prod^{\leftarrow}_{\tau\geq\sigma''\in\cP_{t''}}(g_{t''})_{\sigma''},
\end{align*}
where $,\cP_{t''}(I_{\sigma'}),\cP_{t''}(I_{\sigma'})$ are the subpartitions of $I_{\sigma}, I_{\sigma'}\subset (0,L]$ defined by $t''$. Thus, $\h(\bar{g})_{\tau}$ is well-defined for $\bar{g}\in\overline{\scrA}, \tau\in \D$. A natural inverse,
\begin{align*}
\h^{-1} : W_{\D,L}(G) & \longrightarrow \overline{\scrA},
\end{align*}
is given by:
\begin{align*}
\h^{-1}(h)_{t} & = (h_{\sigma_{1}},h_{\sigma_{1}}^{-1}h_{\sigma_{2}},...,h_{\sigma_{n(t)-1}}^{-1}h_{\sigma_{n(t)}}),
\end{align*}
for $t\in\fT$ and $\cP_{t}=\{0=\sigma_{0}<\sigma_{1}<...<\sigma_{n(t)}=1\}$. Assuming $t\leq t' = ft$, we infer
\begin{align*}
p^{\gamma(t')}_{\gamma(t)}(\h^{-1}(h)_{t'})_{\sigma} & = \prod^{\leftarrow}_{\sigma\geq\sigma'\in\cP_{t'}(I_{\sigma})}(\h^{-1}(h)_{t'})_{\sigma'} = (\h^{-1}(h)_{t})_{\sigma}.
\end{align*}
Thus, $\h^{-1}$ is well-defined. By construction, we have:
\begin{align*}
 \h^{-1}(\h(\bar{g}))_{t}& = (\h(\bar{g})_{\sigma_{1}},\h(\bar{g})_{\sigma_{1}}^{-1}\h(\bar{g})_{\sigma_{2}},...,\h(\bar{g})_{\sigma_{n(t)-1}}^{-1}\h(\bar{g})_{\sigma_{n(t)}}) \\ 
 & = ((g_{t})_{\sigma_{1}},(g_{t})_{\sigma_{1}}^{-1}(g_{t})_{\sigma_{1}}(g_{t})_{\sigma_{2}},...,((g_{t})_{\sigma_{1}}...(g_{t})_{\sigma_{n(t)-1}})^{-1}(g_{t})_{\sigma_{1}}...(g_{t})_{\sigma_{n(t)}}) \\
 & = ((g_{t})_{\sigma_{1}},...,(g_{t})_{\sigma_{n(t)}}) \\
 & = g_{t},\\
 \h(\h^{-1}(h))_{\tau} & = \prod^{\leftarrow}_{\tau\geq\sigma\in\cP_{t}}(\h^{-1}(h)_{t})_{\sigma} \\
 & = (\h^{-1}(h)_{t})_{\sigma_{1}}(\h^{-1}(h)_{t})_{\sigma_{2}}...(\h^{-1}(h)_{t})_{\sigma}...(\h^{-1}(h)_{t})_{\sigma_{n(t)}} \\
 & = h_{\sigma_{1}}h_{\sigma_{1}}^{-1}...h_{\sigma}h_{\sigma}^{-1}...h_{\tau} \\
 & = h_{\tau},
\end{align*}
showing that $\h^{-1}\circ\h = \id = \h\circ\h^{-1}$.
\end{proof}
\end{prop}

As $\h$ is a homeomorphism, we use it to transfer the (continuous) actions $\underline{L}$ and $\overline{\tau}$ to $W_{\D,L}(G)$ (see corollary \ref{cor:groupsys} and proposition \ref{prop:gaugesys}).

\begin{corollary}
\label{cor:discretehol}
We obtain the actions,
\begin{align*}
L^{W} : G^{(L\D)}_{0} \times W_{\D,L}(G) & \longrightarrow W_{\D,L}(G),
\end{align*}
and,
\begin{align*}
\tau^{W} : G^{L\D} \times W_{\D,L}(G) & \longrightarrow W_{\D,L}(G),
\end{align*}
by conjugating $\underline{L}$ and $\overline{\tau}$ with the holonomy map. Additionally, we use the natural identifications,
\begin{align*}
\underline{G} & \cong \bigoplus_{d\in L\D}G = G^{(L\D)},& \underline{G}_{0} & \cong \bigoplus_{d\in L\D_{\neq 0}}G = G^{(L\D)}_{0},
\end{align*}
where we associate an element $\underline{g}\in\underline{G}$ with the trivial extension of its minimal tree representation, $j^{\gamma(t_{0})}(g_{\sigma_{1}},...,g_{\sigma_{n(t_{0})}}) = \underline{g}$ for $t_{0}\in\fT$, and,
\begin{align*}
\overline{G} & \cong \bigtimes_{d\in L\D}G = G^{L\D}, & \overline{G}_{0} & \cong \bigtimes_{d\in L\D_{\neq 0}}G = W_{\D,L}(G).
\end{align*}
Here, $G^{(L\D)}_{0}$ is the direct-sum subgroup of $W_{\D,L}(G)$ with subspace topology coming from the box topology. The actions take the explicit forms:
\begin{align*}
L^{W}_{\underline{g}}(h)_{\tau} & = g_{\sigma_{1}}h_{\sigma_{1}}g_{\sigma_{2}}h_{\sigma_{1}}^{-1}h_{\sigma_{2}}...h_{\sigma_{j-1}}g_{\sigma_{j}}h_{\sigma_{j-1}}^{-1}...h_{\tau} \\
 & \!\!\!\!\stackrel{\tau=\sigma_{m}}{=} \left(\prod_{j=1}^{m}\alpha_{h_{\sigma_{j-1}}}(g_{\sigma_{j}})\right)h_{\tau}, \\
\tau^{W}_{\overline{g}}(h)_{\tau} & = g_{0}h_{\tau}g^{-1}_{\tau}.
\end{align*}
Thus, the holonomy map intertwines the action of the gauge group, $\overline{\tau}:G^{L\D}\act\overline{\scrA}$, with the action of $G^{L\D}$ on its normal subgroup $G^{L\D}_{0}=W_{\D,L}(G)$ by (right) group multiplication up to an additional factor of $G$ acting from the left (at $0\in L\D$).
\end{corollary}

\begin{remark}
\label{rem:discretehol}
The action of the gauge group, $\tau^{W}:G^{L\D}\act W_{\D,L}(G)$, makes the implications of different boundary conditions especially transparent: free boundary conditions correspond to the action of the (full) dyadic-path group $G^{L\D}$, periodic boundary conditions correspond to the restricted action of the dyadic-loop group $\scrL_{\D,L}(G) = \{h:L\D\rightarrow G\ |\ h_{L}=h_{0}\}\subset G^{L\D}$. Moreover, we can restrict the gauge group to the based dyadic-path group, $G^{L\D}_{0} = W_{\D,L}(G)$, or the based dyadic-loop group, $\scrL_{\D,L}(G)_{0}=W_{\D,L}(G)\cap\scrL_{\D,L}(G)$, to study only gauge transformations that vanish at $0$ respectively at $0$ and $L$.
\end{remark}

\subsubsection{Duality}
\label{sec:abelianduality}
For sake of clarity, we describe the duality that is implied by the Fourier transform and the holonomy map when $G$ is compact and abelian. To this end, we observe that the explicit formula for the action $L^{W}:\underline{G}\act W_{\D,L}(G)$ given in corollary \ref{cor:discretehol} can be further simplified for abelian $G$:
\begin{align}
\label{eq:finiterankaction}
L^{W}_{\underline{g}}(h)_{\tau} & \!\!\!\!\stackrel{\tau=\sigma_{m}}{=} \left(\prod_{i=1}^{m}\alpha_{h_{\sigma_{i-1}}}(g_{\sigma_{i}})\right)h_{\tau} = \underbrace{\left(\prod^{\leftarrow}_{\tau\geq\sigma\in\cP_{t}}g_{\sigma}\right)}_{\h(\underline{g})_{\tau}}h_{\tau} \\ \nonumber
 & = L_{\h(\underline{g})}(h)_{\tau},
\end{align}
where we consider $\underline{G}\subset\overline{\scrA}$ as a dense, continuously embedded subset because of proposition \ref{prop:groupsys} and corollary \ref{cor:groupsys}. We call the image of $\h:\underline{G}\rightarrow W_{\D,L}(G)$ the finite rank dyadic-path group (cp.~\eqref{eq:fouriercompacspacelimit}):
\begin{align}
\label{eq:finiterankgroup}
G_{\fr} & = \{h : L\D \rightarrow G\ |\ h_{0}=1_{G}\ \wedge\ \exists t\in\fT:\forall\sigma\in\cP_{t}:h_{I_{\sigma}\cap L\D}=\textup{const.}\}.
\end{align}
By definition, the holonomy map intertwines the action $\underline{G}\act\overline{\scrA}$ with the action $G_{\fr}\act W_{\D,L}(G)$ by (left) multiplication.\\
\\
Next, we observe that Fourier duality allows us to transfer our construction from the action $\underline{G}\act\overline{\scrA}$ to the action $W_{\D,L}(\hat{G})\tca\hat{G}_{\fr}$ because, on the one hand, $\hat{G}$ is an abelian, discrete group, and, on the other hand, the projective and inductive systems based on definition \ref{def:fouriergroupsys} lead precisely to the latter pair of limit objects. Following this, we invoke the inverse of the holonomy map to pass from the pair $W_{\D,L}(\hat{G})\tca\hat{G}_{\fr}$ to the pair $\overline{\hat{\scrA}}\tca\underline{\hat{G}}$. To summarize, we have:

\begin{theo}
\label{th:abelianduality}
Given an abelian, compact group $G$ and its abelian, discrete, unitary dual $\hat{G}$ (Pontryagin dual), the Fourier transform, $\scrF$, and the holonomy map, $\h$, provide an isomorphism of topological actions:
\begin{align*}
\h^{-1}\circ\scrF : \underline{G}\act\overline{\scrA} & \longrightarrow \overline{\hat{\scrA}}\tca\underline{\hat{G}}.
\end{align*}
\end{theo}

\subsection{Canonical states}
\label{sec:can11}
\begin{figure}[t]
	\begin{tikzpicture}

	
	\foreach \x in {0} 
		\foreach \y in {0,1}
		\foreach \v in {0,1} 	
		{
			\draw[->-=0.55, thick] (1+2*\x,4+2*\y) to (3+2*\x,4+2*\y);
			\draw[->-=0.55, thick] (1+2*\y,4+2*\x) to (1+2*\y,6+2*\x);
			\filldraw (1+2*\y,4+2*\v) circle (2pt);
		}
	
	\draw (3.25,5) node{$e_{1}$};
	\draw (2,6.25) node{$e_{2}$};
	\draw (0.75,5) node{$e_{3}$};	
	\draw (2,3.75) node{$e_{4}$};
	\draw (2,5) node{$f$};

\end{tikzpicture}
\caption{The edges, $\partial f = \{e_{4},e^{-1}_{3},e^{-1}_{2},e_{1}\}\subset E(\gamma)$, around a face, $f\in F(\gamma)$, of a lattice $\gamma\in\Gamma_{\D,L}$. $g_{f} = g_{e_{4}}g^{-1}_{e_{3}}g^{-1}_{e_{2}}g_{e_{1}}$ is the associated holonomy.}
\label{fig:facehol}
\end{figure}
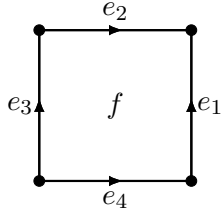

So far, we have mainly been concerned with the kinematical aspects of Hamiltonian lattice gauge theory, and, thus, it has been possible to work in a rather abstract operator-algebraic setting. But, as we intend to discuss dynamical aspects of the theory as well, we would like to introduce specific representations of the semi-continuum field algebra, $\pi : \fA_{\D,L}\rightarrow \cH$, that allow for the introduction of a one-parameter, unitary time evolution group,
\begin{align}
\label{eq:timeev}
U : \R & \longrightarrow \cU(\cH), 
\end{align}
which is associated with a Hamiltonian affiliated, $H\!\ \eta\!\ \pi(\fA_{\D,L})''$, to the semi-continuum von Neumann field algebra,
\begin{align}
\label{eq:hamiltonian}
U_{t} & = e^{-itH}, & t&\in\R.
\end{align}
As we construct our field algebra from a scale of lattice theories, it is natural to use suitable lattice approximations for the Hamiltonian, e.g.~the Kogut-Susskind Hamiltonian for Yang-Mills theory \cite{KogutHamiltonianFormulation}.\\[0.1cm]
Therefore, we introduce in the following canonical Gibbs states for YM$_{1+1}$, which are based on the Kogut-Susskind Hamiltonian and its finite-temperature canonical ensemble on each lattice $\gamma\in\Gamma_{\D,L}$. We call these states the heat-kernel states for Yang-Mills theory for reasons that become apparent in the next subsection. Additionally, we use the duality discussed in the previous subsection, especially theorem \ref{th:abelianduality}, to define the dual heat-kernel states, and we show that the latter have a natural interpretation on their own.\\[0.1cm]
Before we define the heat-kernel states for YM$_{1+1}$, we recall some general facts about the Yang-Mills Hamiltonian in a lattice approximation in arbitrary dimensions. Initially, we assume that $G$ is a compact Lie group. Then, the Kogut-Susskind Hamiltonian on a lattice, $\gamma_{N}\in\Gamma_{\D,L}$ of level $N\in\N_{0}$, is formally given by:
\begin{align}
\label{eq:KShamiltonian}
H^{(N)}_{\pi} & = \frac{g_{N}^{2}}{2 a_{\gamma_{N}}}\sum_{e\in E(\gamma)}\big(-\Delta_{G_{e}}\big)+\frac{2}{a_{\gamma_{N}}g_{N}^{2}}\sum_{f\in F(\gamma)}\big(d_{\pi}-\Re(M(\Tr_{V_{\pi}}(\pi_{f})))\big),
\end{align}
for some $\pi\in\hat{G}$ and the dimensionless coupling constant $g_{N}\in\R_{>0}$. Here, $\Delta_{G_{e}}$ denotes the bi-invariant Laplace (or quadratic Casimir) operator on $C^{\infty}(G_{e})$ associated with the negative of the Killing form, $\langle\ .\ \rangle_{\fg_{e}}$, and $M(\Tr_{V_{\pi}}(\pi_{f}))$ is the multiplication associated with the trace of the holonomy around a face, i.e.~a Wilson loop, in the representation $\pi$ (see figure \ref{fig:facehol}):
\begin{align}
\label{eq:KScomponents}
(\Delta_{G_{e}}\phi)(g_{e}) & = \sum^{\dim\fg_{e}}_{i=1}(R^{2}_{X_{e,i}}\phi)(g_{e}), \\ \nonumber
(M(\Tr_{V_{\pi}}(\pi_{f}))\psi)(g_{e_{4}},g_{e_{3}},g_{e_{2}},g_{e_{1}}) & = \Tr_{V_{\pi}}(\pi(g_{e_{4}}g^{-1}_{e_{3}}g^{-1}_{e_{2}}g_{e_{1}}))\psi(g_{e_{4}},g_{e_{3}},g_{e_{2}},g_{e_{1}}),
\end{align}
for $\phi\in C^{\infty}(G_{e})$, $\psi\in C^{\infty}(G_{e_{4}}\times G_{e_{3}}\times G_{e_{2}}\times G_{e_{1}})$, and $\partial f = \{e_{4},e^{-1}_{3},e^{-1}_{2},e_{1}\}$. $\{X_{e,i}\}^{\dim\fg_{e}}_{i=1}$ is an orthonormal basis of $\fg_{e}$ w.r.t.~$\langle\ .\ \rangle_{\fg_{e}}$, and,
\begin{align}
\label{eq:rightinvariantvf}
(R_{X_{e}}\phi)(g_{e}) & = \frac{\ud}{\ud t}_{|t=0}(\lambda_{\exp_{G_{e}}(tX_{e})}\phi)(g_{e}) = \frac{\ud}{\ud t}_{|t=0}f(\exp_{G_{e}}(tX_{e})g_{e}), & X_{e}&\in\fg_{e},
\end{align}
is a right-invariant vector field. We note that by definition $H^{(N)}_{\pi}$ is invariant under edge inversion \eqref{eq:algorient} and, therefore, it only depends on the level $N$. Moreover, $H^{(N)}_{\pi}$ is gauge-invariant on $C^{\infty}(\scrA^{E(\gamma_{N})})$, i.e.
\begin{align}
\label{eq:KSinvariance}
\Ad_{U_{\tau^{\gamma_{N}}}(g)}(H^{(N)}_{\pi}) & = H^{(N)}_{\pi}, & g&\in G^{V(\gamma_{N})}.
\end{align}

Since Yang-Mills theory has only a single (running) coupling constant, $g_{N}$, we have two distinguished, gauge-invariant, formal limits of $H^{(N)}_{\pi}$: the strong-coupling limit $g_{N}>>1$,
\begin{align}
\label{eq:KShamiltonianstrong}
H^{(N)}_{s} & = \frac{g_{N}^{2}}{2 a_{\gamma_{N}}}\sum_{e\in E(\gamma)}\big(-\Delta_{G_{e}}\big),
\end{align}
and the weak-coupling limit $g_{N}<<1$,
\begin{align}
\label{eq:KShamiltonianweak}
H^{(N)}_{\pi,w} & = \frac{2}{a_{\gamma_{N}}g_{N}^{2}}\sum_{f\in F(\gamma)}\big(d_{\pi}-\Re(\Tr_{V_{\pi}}(M(\pi_{f})))\big).
\end{align}
It is evident from \eqref{eq:KShamiltonian} that $H^{(N)}_{\pi}$ and the formal limits $H^{(N)}_{s}$, $H^{(N)}_{\pi,w}$ are symmetric operators on $C^{\infty}(\scrA^{E(\gamma_{N})})$. It is well-known that $H^{(N)}_{s}$ is essentially self-adjoint with domain $D(H^{(N)}_{s})=C^{\infty}(\scrA^{E(\gamma_{N})})$. Moreover, for a compact Lie group $G$, the weak-coupling limit $H^{(N)}_{\pi,w}$ is a multiplication operator with a bounded function in $C^{\infty}(\scrA^{E(\gamma_{N})})$. Therefore, $H^{(N)}_{\pi,w}$ is an infinitesimal perturbation of the strong-coupling limit $H^{(N)}_{s}$ in the sense of the Kato-Rellich theorem \cite{KatoPerturbationTheoryFor, ReedMethodsOfModern2}, and $H^{(N)}_{\pi}$ is essentially self-adjoint on $D(H^{(N)}_{s})=C^{\infty}(\scrA^{E(\gamma_{N})})$.\\[0.25cm]
Because of the identities,
\begin{align}
\label{eq:stronglimitconsistency}
\tfrac{1}{2}\big(\Delta_{G_{e_{2}}}\otimes\mathds{1}_{e_{1}}+\mathds{1}_{e_{2}}\otimes\Delta_{G_{e_{1}}}\big)R^{\{e\}}_{\{e_{1},e_{2}\}} & = R^{\{e\}}_{\{e_{1},e_{2}\}}\Delta_{G_{e}}, \\ \nonumber
\big(\Delta_{G_{e_{2}}}\otimes\mathds{1}_{e_{1}}+\mathds{1}_{e_{2}}\otimes\Delta_{G_{e_{1}}}\big)R^{\{e_{2}\}}_{\{e_{1},e_{2}\}} & = R^{\{e_{2}\}}_{\{e_{1},e_{2}\}}\Delta_{G_{e_{2}}},
\end{align}
and the stability of the domains w.r.t.~the inductive system of isometries associated with the representation of the semi-continuum field algebra on the Hilbert space $L^{2}(\overline{\scrA})$ (see section \ref{sec:compact} and use \eqref{eq:compacthilbertsys}),
\begin{align}
\label{eq:domainconsistency}
R^{\gamma}_{\gamma'}C^{\infty}(\scrA^{E(\gamma)}) & \subset C^{\infty}(\scrA^{E(\gamma')}), & \gamma & \subset_{\uL}\gamma',
\end{align}
we have:

\begin{theo}
\label{th:KShamiltonianstrong}
For $g_{N+1} = 2^{-1}g_{N}$, the strong-coupling limits of the Kogut-Susskind Hamiltonians, $\{H^{(N)}_{s}\}_{\gamma\in\Gamma_{\D,L}}$, $N = N(\gamma)$, form a coherent family of essentially self-adjoint operators. Thus, there exists an essentially self-adjoint inductive-limit Hamiltonian:
\begin{align*}
\underline{H}_{s} & = \varinjlim_{\gamma\in\Gamma_{\D,L}}H^{(N)}_{s}, 
\end{align*}
with domain $D(\underline{H}_{s})=\bigcup_{\gamma\in\Gamma_{\D,L}}C^{\infty}(\scrA^{E(\gamma)})\subset L^{2}(\overline{\scrA})$, which is gauge-invariant:
\begin{align*}
Ad_{U_{\overline{\tau}}(g)}(\underline{H}_{s}) & = \underline{H}_{s}, & g&\in\overline{G}.
\end{align*}
\end{theo}

As a consequence of this theorem, we infer the following corollary from the fact that $d\pi(\Delta_{G}) = -c_{\pi}\mathds{1}_{V_{\pi}}$ with $c_{\pi}\geq0$ for any $\pi\in\hat{G}$, cf.~\cite{AshtekarCoherentStateTransforms}.

\begin{corollary}
\label{cor:KShamiltonianstrongfourier}
The Fourier transform, $\overline{\scrF}$, diagonalizes the strong-coupling Kogut-Susskind Hamiltonian, $\underline{H}_{s}$. The spin-network functions form a complete system of eigenvectors for $\underline{H}_{s}$ in $L^{2}(\overline{\scrA})$. In particular, the GNS vector, $\Omega_{\D,L}$, satisfies:
\begin{align*}
\underline{H}_{s}\Omega_{\D,L} & = 0,
\end{align*}
which justifies the name ``strong-coupling limit vacuum'', cf.~remark \ref{rem:compactstate}.
\end{corollary}

Based on the fact that the 1+1-dimensional Kogut-Susskind Hamiltonian \eqref{eq:KShamiltonian} is equivalent to its strong-coupling limit \eqref{eq:KShamiltonianstrong} because the set of faces of a lattice, $\gamma\in\Gamma_{\D,L}$, is empty, $F(\gamma)=\emptyset$, we obtain the vacuum or ground state sector of YM$_{1+1}$ from theorem \ref{th:KShamiltonianstrong}. In this case, the dimensionless coupling constant, $g_{N}$, is related to the dimensionful, bare coupling constant, $g_{\ub}$, by
\begin{align}
\label{eq:11dcoupling}
g_{N} & \sim a_{\gamma_{N}}g_{\ub},
\end{align}
where the latter enters the Yang-Mills action in the form, cp.~\cite{CreutzQuarksGluonsLattices, KogutTheLatticeGauge}:
\begin{align}
\label{eq:ymaction}
S[A] & = -\frac{1}{4g_{\ub}^{2}}\int_{\R}dt\!\!\int_{\Sigma}d^{d}x\!\ \eta^{ik}\eta^{jl}\langle F_{ij}, F_{kl}\rangle_{\fg}.
\end{align}
Here, $F = dA + A\wedge A$ is the curvature $2$-form of the gauge field $A$ (a connection $1$-form in a principal $G$-bundle $P\rightarrow\R\times\Sigma$), and $\eta$ is a Lorentzian spacetime metric on $\R\times\Sigma$. Incidentally, \eqref{eq:11dcoupling} together with the assumptions of theorem \ref{th:KShamiltonianstrong} implies 
\begin{align}
\frac{g_{N+1}}{a_{\gamma_{N+1}}} & = 2^{-1}\frac{g_{N}}{a_{\gamma_{N+1}}} = \frac{g_{N}}{a_{\gamma_{N}}},	
\end{align}
which is equivalent to a constant string tension $\tfrac{g^{2}_{N}}{a^{2}_{\gamma_{N}}}=\textup{const.}$, which leads to a linear potential of the $Q\bar{Q}$-state in the heavy-quark limit \cite{KogutTheLatticeGauge}, and, thus, a bare coupling, $g_{\ub}$, independent of the level $N$. Therefore, the coherent family of Kogut-Susskind Hamiltonians acquires the correct scaling, and the bare coupling constant, $g_{\ub}$, remains unrenormalized as observed before, e.g.~\cite{WittenOnQuantumGauge, AshtekarSU(N)QuantumYang}. We recover the well-known structure of the holonomy or Wilson loop observables \cite{WittenOnQuantumGauge, DimockCanonicalQuantizationOf, AshtekarSU(N)QuantumYang}, i.e.~denoting by $c_{\pi}, \pi\in\hat{G},$ the eigenvalues of $-\Delta_{G}$ as above and by $\chi_{\pi} = \Tr_{V_{\pi}}(\pi)\in C(G)\cong C(\scrA^{E(\gamma_{0})})$, we have in agreement with corollary \ref{cor:KShamiltonianstrongfourier}:
\begin{align}
\label{eq:ymholobs}
H^{(N)}_{s}\alpha^{\gamma_{0}}_{\gamma_{N}}(M(\chi_{\pi}))\Omega_{\gamma_{N}} & = H^{(N)}_{s}R^{\gamma_{0}}_{\gamma_{N}}(M(\chi_{\pi})\Omega_{\gamma_{0}}) = R^{\gamma_{0}}_{\gamma_{N}}(H^{(0)}_{s}M(\chi_{\pi})\Omega_{\gamma_{0}}) \\ \nonumber
 & = \tfrac{1}{2}c_{\pi}Lg^{2}_{b}R^{\gamma_{0}}_{\gamma_{N}}(M(\chi_{\pi})\Omega_{\gamma_{0}}) = \tfrac{1}{2}c_{\pi}Lg^{2}_{b}\alpha^{\gamma_{0}}_{\gamma_{N}}(M(\chi_{\pi}))\Omega_{\gamma_{N}},
\end{align}
for any $N\in\N_{0}$. This implies that the holonomy observables $\alpha^{\gamma_{0}}(M(\chi_{\pi}))\in\fA_{\D,L}$ generate eigenvectors of $\underline{H}_{s}$ from the strong-coupling limit vacuum $\Omega_{\D,L}$:
\begin{align}
\label{eq:ymholobslim}
\underline{H}_{s}\alpha^{\gamma_{0}}(M(\chi_{\pi}))\Omega_{\D,L} & = \tfrac{1}{2}c_{\pi}Lg^{2}_{b}\alpha^{\gamma_{0}}(M(\chi_{\pi}))\Omega_{\D,L},
\end{align}
where $\alpha^{\gamma_{0}}:B(L^{2}(\scrA^{E(\gamma_{0})}))\rightarrow\fA_{\D,L}$ is the canonical injective $*$-morphism. Moreover, if we restrict $\underline{H}_{s}$ to the gauge-invariant subspace $\cH^{\overline{G}}_{\D,L}\subset\cH_{\D,L}$, we will find that $\Omega_{\D,L}$ is cyclic for the holonomy observables, because any gauge-invariant spin-network function in one spatial dimension has only trivial 2-valent intertwiners and, thus, is proportional to $\alpha^{\gamma_{0}}(M(\chi_{\pi}))\Omega_{\D,L}\in L^{2}(\overline{\scrA})$ for some $\pi\in\hat{G}$, see e.g.~\cite{ThiemannModernCanonicalQuantum}.

\begin{remark}
\label{eq:11KShamiltonianstrong}
The existence of the 1+1-dimensional Kogut-Susskind Hamiltonian, $\underline{H}_{s}$, might seem to contradict the results in \cite{DriverYangMillsTheory}, but this is only apparent because the Hilbert space, $L^{2}(W_{L=1}(G),\tilde{\rho}_{s})$, on the Wiener or based path group, $W_{L=1}(G)=\{h\in C([0,1],G)\ |\ h(0)=e\}$, of $G$ (with Wiener measure of ``inverse mass'' $s>0$) constructed there differs from $L^{2}(\overline{\scrA})$. In section \ref{sec:hkdual}, we explain how the former can be obtained as a representation of the subalgebra $C(\overline{\scrA})$ of the semi-continuum field algebra $\fA_{\D,L}$ for a compact Lie group $G$. We also give an argument that said representation is related to the weak-coupling limit of the Kogut-Susskind Hamiltonian \eqref{eq:KShamiltonianweak}.
\end{remark}

\subsubsection{The heat-kernel state}
\label{sec:hk}
After these introductory remarks on the Kogut-Susskind Hamiltonian and, especially, its strong-coupling limit, we now define the heat-kernel states. For this purpose, we recall that the heat kernel, $\rho_{\beta}, \beta>0$, on a compact Lie group $G$ is the fundamental solution at $1_{G}$ to the heat equation, cf.~\cite{HallTheSegalBargmann} and references therein:
\begin{align}
\label{eq:hkequation}
\frac{\ud}{\ud\beta}\rho_{\beta} & = \frac{1}{2}\Delta_{G}\rho_{\beta}, & \lim_{\beta\rightarrow0+}\rho_{\beta} & = \delta_{1_{G}}.
\end{align}
It is known that $\rho_{\beta}$ is a smooth, real, strictly positive class-function with locally uniformly convergent Fourier series:
\begin{align}
\label{eq:hkseries}
\rho_{\beta} & = \sum_{\pi\in\hat{G}}d_{\pi}\!\ e^{-\frac{\beta}{2}c_{\pi}}\chi_{\pi}.
\end{align}
From the Fourier series of $\rho_{\beta}$, we deduce that it has unit mass w.r.t.~the normalized Haar measure on $G$:
\begin{align}
\label{eq:hkint}
\int_{G}dg\!\ \rho_{\beta}(g) & = 1.
\end{align}
Because of \eqref{eq:hkequation}, we can use the integrated left regular representation to have an explicit expression for the Gibbs states on $B(L^{2}(G))$ associated with the Hamiltonian $H = -\tfrac{1}{2}\Delta_{G}$:
\begin{align}
\label{eq:hkstatesingle}
\omega_{\beta}(a) & = Z(\beta)^{-1}\Tr_{L^{2}}(\lambda(\rho_{\beta}\otimes 1)a) = Z(\beta)^{-1}\Tr_{L^{2}}(e^{-\beta H}a), \\ \nonumber
Z(\beta) & = \Tr_{L^{2}}(\lambda(\rho_{\beta}\otimes 1)) = \Tr_{L^{2}}(e^{-\beta H}) = \rho_{\beta}(1_{G}),
\end{align}
for $a\in B(L^{2}(G))$ and any $\beta>0$. Moreover, due to our general observations in \ref{sec:compact} and, notably, \eqref{eq:simplecoherenceconditions} \& \eqref{eq:simpletensorstates}, we have a coherent family of states for any $\beta>0$,
\begin{align}
\label{eq:hkstates}
\omega^{(\gamma)}_{\beta}(a_{\gamma}) & = (Z^{(\gamma)}(\beta))^{-1}\Tr_{L^{2}}(\lambda^{(\gamma)}(F_{\gamma}(\beta))a_{\gamma}), & \gamma\in\Gamma_{\D,L}
\end{align}
where $F_{\gamma}(\beta) = \bigotimes_{e\in E(\gamma)}(\rho_{\beta}\otimes 1)_{e}$, $Z^{(\gamma)}(\beta)=\rho_{\beta}(1_{G})^{|E(\gamma)|}$, and $a_{\gamma}\in B(L^{2}(\scrA^{E(\gamma)}))$. Therefore, we have a (projective) limit state,
\begin{align}
\label{eq:hkstatelim}
\omega_{\beta,\D,L} & = \varprojlim_{\gamma\in\Gamma_{\D,L}}\omega^{(\gamma)}_{\beta},
\end{align}
on the semi-continuum field algebra $\fA_{\D,L}$, which we call the heat-kernel state (at $\beta>0$). By construction $\omega_{\beta,\D,L}$ is gauge-invariant,
\begin{align}
\label{eq:hkstategaugeinv}
\omega_{\beta,\D,L}\circ\alpha_{\overline{\tau}}(g) & = \omega_{\beta,\D,L}, & g&\in\overline{G},
\end{align}
because $\rho_{\beta}$ is a class function, and, thus, the action of the gauge group $\overline{G}$ is unitarily implemented on the semi-continuum von-Neumann field algebra $\scrM_{\beta,\D,L}$, cf.~proposition \ref{prop:gaugesys}.

\begin{remark}
\label{rem:hkinfty}
We infer from \eqref{eq:hkseries} that
\begin{align*}
\lim_{\beta\rightarrow\infty}\rho_{\beta} & = 1
\end{align*}
in $C(G)$, which entails
\begin{align*}
\lim_{\beta\rightarrow\infty}\sum_{\pi\in\hat{G}\setminus\{\pi_{\utriv}\}}d_{\pi}^{2}\!\ e^{-\frac{\beta}{2}c_{\pi}} & = 0.
\end{align*}
This, in turn, is equivalent to
\begin{align*}
\lim_{\beta\rightarrow\infty}\Tr_{L^{2}}(|e^{-\beta H}-p_{1}|)=0
\end{align*}
where $p_{1}$ is the projection onto $1\in L^{2}(G)$. By generalization to an arbitrary finite number of copies of $G$, we have the existence of the weak limits,
\begin{align*}
\lim_{\beta\rightarrow\infty}\omega^{(\gamma)}_{\beta} & = \omega^{(\gamma)}, & \gamma&\in\Gamma_{\D,L},
\end{align*}
with $\omega^{(\gamma)}$ given by \eqref{eq:compactstate} -- the strong-coupling vacuum of $\gamma$. Thus, $\omega_{\beta,\D,L}$ converges weakly to $\omega_{\D,L}$ on the dense $*$-subalgebra $\bigcup_{\gamma\in\Gamma_{\D,L}}\alpha^{\gamma}(B(L^{2}(\scrA^{E(\gamma)})))\subset\fA_{\D,L}$, which implies weak convergence everywhere.
\end{remark}

By extending the relation expressed in \eqref{eq:hkstatesingle} to the product of an arbitrary finite number of copies of $G$, we identify the states $\omega^{(\gamma)}_{\beta}, \gamma\in\Gamma_{\D,L}$ as the Gibbs states of the Kogut-Susskind Hamiltonian in the strong-coupling limit:
\begin{align}
\label{eq:hkstatesKSstronglimit}
\lambda^{(\gamma)}(F_{\gamma}(\tfrac{\beta_{N}g_{N}^{2}}{a_{\gamma_{N}}})) & = e^{-\beta_{N} H^{(N)}_{s}},
\end{align}
where we allow for an explicit dependence of the inverse temperature $\beta_{N}$ on the level $N\in\N_{0}$.\\[0.1cm]
In 1+1 dimensions, where the Kogut-Susskind Hamiltonian coincides with its strong-coupling limit, this implies that the finite-temperature Gibbs states of Yang-Mills theory on the directed family of lattices $\Gamma_{\D,L}$ form a coherent family for:
\begin{align}
\label{eq:hkhomogeneousren}
\frac{\beta_{N}g_{N}^{2}}{a_{\gamma_{N}}} & = \textup{const. as a function of }N.
\end{align}
If we adopt the same scaling (or renormalization) of $g_{N}$ as in the strong-coupling vacuum \eqref{eq:11dcoupling}, we will have:
\begin{align}
\label{eq:hktemperatureren}
\beta_{N} & \sim a_{\gamma_{N}}^{-1}g^{-2}_{\ub}.
\end{align}
This indicates that (formally) $\omega_{\beta,\D,L}$ resembles a ground state, $\lim_{N\rightarrow\infty}\beta_{N} = \infty$, of the strong-coupling Kogut-Susskind Hamiltonian, $\underline{H}_{s}$, for any initial choice of $\beta_{0}$ -- a point of view that is further justified by the results presented in \cite{BrothierConstructionsOfConformal} on the type of $\scrM_{\beta,\D,L}$, cp.~also section \ref{sec:timezero}.

\subsubsection{Inhomogeneous heat-kernel states}
\label{sec:hkinhom}

There is an immediate generalization of the heat-kernel states defined in the previous section which we call inhomogeneous heat-kernel states. While the former are associated with constant $\beta_{N}>0$ at every level $N\in\N_{0}$, the latter are given by potentially nonconstant functions $\beta_{N}:E(\gamma_{N})\rightarrow\R_{>0}$. A coherent family of states $\{\omega^{(\gamma_{N})}_{\beta_{N}}\}_{N\in\N_{0}}$ can be constructed in the following way:\\[0.1cm]
We consider to initial values $\beta_{0,\uL}, \beta_{1,\uR}>0$, and we set $\beta_{0}=\beta_{0,\uL}$ respectively $\beta_{0}=\beta_{1,\uR}$ depending on whether we choose the orientation of the single edge $e\in E(\gamma_{0})$ to be oriented to the left or to the right. We realize the Gibbs state $\omega^{(\gamma_{0})}_{\beta_{0}}$ accordingly. Next, we set $\beta_{1}(e) = \beta_{0,\uL}a_{\gamma_{1}}g^{-2}_{1}\textup{ or }\beta_{1,\uR}a_{\gamma_{1}}g^{-2}_{1}$ for $e\in E(\gamma_{1})$ oriented to the left or to the right respectively. Defining $\omega^{(\gamma_{1})}_{\beta_{1}} = \otimes_{e\in E(\gamma_{1})}\omega_{\beta_{1}(e)}$ gives $\omega^{(\gamma_{1})}_{\beta_{1}}\circ\!\ \alpha^{\gamma_{0}}_{\gamma_{1}} = \omega^{(\gamma_{0})}_{\beta_{0}}$. Clearly, we may define $\omega^{(\gamma_{2})}_{\beta_{2}}$ by repeating the refinement from $\omega^{(\gamma_{0})}_{\beta_{0}}$ to $\omega^{(\gamma_{1})}_{\beta_{1}}$ for the tensor factors $\omega_{\beta_{1}(e)}$, $e\in E(\gamma_{2})$, separately. This involves a choice of two additional values $\beta_{\frac{1}{2},\uL}, \beta_{\frac{1}{2},\uR}>0$ associated with the edges containing the vertex $\tfrac{L}{2}\in V(\gamma_{2})$, and results in a function $\beta_{2}:E(\gamma_{2})\rightarrow\R_{>0}$. We define $\omega^{(\gamma_{2})}_{\beta_{2}} = \otimes_{e\in E(\gamma_{2})}\omega_{\beta_{2}(e)}$. Iterating this procedure, we realize that the functions $\beta_{N}:E(\gamma_{N})\rightarrow\R_{>0}$ are obtained from two freely specified sequences $\{\beta_{d,\uL}\}_{d\in\D}$ and $\{\beta_{d,\uR}\}_{d\in\D}$ with the convention $\beta_{0,\uR} = \beta_{1,\uR}$. The (projective) limit state,
\begin{align}
\label{eq:hkinhomstatelim}
\omega_{(\beta_{\uL},\beta_{\uR}),\D,L} & = \varprojlim_{N\in\N_{0}}\omega^{(\gamma_{N})}_{\beta_{N}},
\end{align}
on $\fA_{\D,L}$ is the inhomogeneous heat-kernel state (at $(\beta_{\uL},\beta_{\uR})$). Again, because $\rho_{\beta}$ is a class function, $\omega_{(\beta_{\uL},\beta_{\uR}),\D,L}$ is gauge-invariant, and we have unitary action of the gauge group $\overline{G}$ on $\scrM_{(\beta_{\uL},\beta_{\uR}),\D,L}$.
\\[0.1cm]
It is shown in our companion paper \cite{BrothierConstructionsOfConformal} that the properties of the semi-continuum von-Neumann field algebra $\scrM_{\beta,\D,L}$ and the extendibility of Jones' actions of Thompson's groups (subsection \ref{sec:thompson}), may depend sensitively on the properties of the sequences $\{\beta_{d,\uL}\}_{d\in\D}$ and $\{\beta_{d,\uR}\}_{d\in\D}$, e.g.~their $p$-summability. Therefore, we infer that the nature of a full continuum limit is affected by these properties as well, cp.~section \ref{sec:rgp}.

\subsubsection{The dual heat-kernel state}
\label{sec:hkdual}
As another state intimately connected to the Kogut-Susskind Hamiltonian and its weak-coupling limit \eqref{eq:KShamiltonianweak} (see below and also \ref{sec:can21}), we introduce the dual heat-kernel states. For simplicity, we first consider $G=U(1)$ and its unitary dual $\hat{G}=\Z$. Associated with the latter is the pair of inductive/projective systems of topological groups resp. spaces $(\underline{\Z},\overline{\scrA}_{\Z})$, see corollary \ref{cor:groupsys}.\\[0.1cm]
We consider the state $\omega_{\beta}:B(\ell^{2}(\Z))\rightarrow\C$ defined by the trace-class operator
\begin{align}
\label{eq:hkdualkernel}
\lambda(F_{\beta}) & = \sum_{m'\in\Z}F_{\beta}(m')\lambda_{m'},
\end{align}
where $F_{\beta}(m') = \delta_{0,m'}\rho^{(\Z)}_{\beta}\in\ell^{1}(\Z)$. $\rho^{(\Z)}_{\beta}$ is the (dual) heat kernel on $\Z$ w.r.t.~the discrete Laplacian $(\Delta^{(\Z)}\phi)(m) = \sum_{n:|m-n|=1}(\phi(m)-\phi(n))$, i.e.:
\begin{align}
\label{eq:hkdualdef}
\frac{\ud}{\ud\beta}\rho^{(\Z)}_{\beta} & = -\frac{1}{2}\Delta^{(\Z)}\rho^{(\Z)}_{\beta}, \\
\lim_{\beta\rightarrow0+}\rho^{(\Z)}_{\beta} & = \delta_{0}.
\end{align}
Clearly, $\lambda(F_{\beta}) = M(\rho^{(\Z)}_{\beta})$ is positive, $\rho^{(\Z)}_{\beta}\geq0$, self-adjoint, $(\rho^{(\Z)}_{\beta})^{*}=\rho^{(\Z)}_{\beta}$, and normalized:
\begin{align}
\label{eq:hkdualnorm}
\Tr_{\ell^{2}}(\lambda(F_{\beta})) & = \sum_{m\in\Z}\rho^{(\Z)}_{\beta}(m) = 1.
\end{align}
Because $B(\ltwo(\Z))\cong\li(\Z)\rtimes\Z$ as von Neumann algebras, we find that $\omega_{\beta}$ is the canonical crossed-product extension \cite{KadisonFundamentalsOfThe2} of the measure associated with $\rho^{(Z)}_{\beta}$ by
\begin{align}
\label{eq:hkdualcom}
\li(\Z)\ni f & \longmapsto \rho^{(\Z)}_{\beta}(f) = \sum_{m\in\Z}\rho^{(\Z)}_{\beta}(m)f(m)\in\C.
\end{align}
Namely, for $a = \sum_{m'\in\Z}F_{a}(m')\lambda_{m'}\in B(\ltwo(\Z))$, we have:
\begin{align}
\label{eq:hkdualcrossedprod}
\omega_{\beta}(a) & = \rho^{(\Z)}_{\beta}(F_{a}(0)). 
\end{align}
An explicit expression for $\rho^{(\Z)}_{\beta}$ is given in terms of modified Bessel functions of the first kind $I_{m}(\beta)$, $m\in\Z$ via Fourier transform:
\begin{align}
\label{eq:hkdualfourier}
\rho^{(\Z)}_{\beta}(m) & = \int_{\fS}\frac{d\varphi}{2\pi}e^{-\frac{\beta}{2}\lambda(\varphi)}\chi_{\varphi}(m) = e^{-\beta}I_{m}(\beta), \\ \nonumber
(\Delta^{(\Z)}\chi_{\varphi})(m) & = \underbrace{2(1-\cos(\varphi))}_{=\lambda(\varphi)}\chi_{\varphi}(m).
\end{align}
Using the notation introduced in subsection \ref{sec:discretehol}, we define a consistent family of tensor product states based on $\omega_{\beta}$ for the $C^{*}$-inductive limit $\fA_{\D,L}$ associated with the pair $(\underline{\Z},\overline{\scrA}_{\Z})$:
\begin{align}
\label{eq:hkdualstate}
\omega^{(\gamma(t))}_{\{\beta_{t}\}} & = \otimes_{\sigma\in\cP_{t}}\omega_{\beta_{t}(\sigma)},\ t\in\fT,
\end{align}
where $\beta_{t}:\cP_{t}\rightarrow\R_{>0}$. The family of states will be coherent if we choose $\beta_{t}(\sigma)=\beta|I_{\sigma}|$ for some fixed $\beta>0$ and $|I_{\sigma}|$ the length of the interval $I_{\sigma}$ for $\sigma\in\cP_{t}$. We call the resulting projective-limit state,
\begin{align}
\label{eq:hkdualstatelim}
\omega_{\beta,\D,L} & = \varprojlim_{t\in\fT}\omega^{(\gamma(t))}_{\{\beta_{t}\}},
\end{align}
on $\fA_{\D,L}$ the dual heat-kernel state as it is naturally associated with a discrete group instead of a compact group as in the case of the heat-kernel state. More specifically, while the heat-kernel state results in a crossed-product construction for the pair $(\Z_{\fr},W_{\D,L}(\Z))$ together with an infinite-product measure $m^{L\D}_{\beta}$ via Fourier duality with the pair $(\underline{U(1)},\overline{\scrA}_{U(1)})$ (see \cite{BrothierConstructionsOfConformal}), the dual heat-kernel state can be interpreted as a crossed-product construction for $(\Z_{\fr},W_{\D,L}(\Z))$ with a infinite-convolution measure $m^{\ast L\D}_{\beta}$ via the holonomy map (see especially theorem \ref{th:abelianduality}).\\[0.1cm]
It is clear from the construction that the action of an element $\tilde{k}\in\Z_{\fr}$ affects the convolution measure $m^{\ast L\D}_{\beta}$ only at the finite number of discontinuities of $\tilde{k}$. But, it is so far not clear whether $\tilde{k}_{*}m^{\ast L\D}_{\beta}$ and $m^{\ast L\D}_{\beta}$ are equivalent. Independent of the latter, we can mimic the action of the gauge group $\fS^{L\D}\act W_{\D,L}(\fS)$ in the Fourier domain by the point-wise group operation $\Z^{L\D}\act W_{\D,L}(\Z)$. Although, it should be noted that the use of $\fS^{L\D}$ resp. $\Z^{L\D}$ is a kind of maximal choice for the gauge transformations and by no means necessary. For example, we could also use $\fS^{(L\D)}$ resp. $\Z^{(L\D)}$ as a kind of minimal choice.\\[0.1cm]
An important question about the dual heat-kernel state concerns its gauge invariance. Because of theorem \ref{th:abelianduality}, there are two possible notions of gauge transformations associated with the pair $(\underline{\Z},\overline{\scrA}_{\Z})$: we may either interpret the latter as a $\Z$-gauge theory or a $U(1)$-gauge theory. The first choice leads to the following gauge transformations for $a\in B(\ltwo(\scrA^{E(\gamma(t))}_{\Z}))$, $t\in\fT$:
\begin{align}
\label{eq:hkdualgauge1}
 & \alpha^{(\gamma(t))}_{k}(F_{a})(n_{\sigma_{1}},m_{\sigma_{1}};...;n_{\sigma_{n(t)}},m_{\sigma_{n(t)}}) &  &  \\ \nonumber
 & = F_{a}(n_{\sigma_{1}},-k_{\sigma_{0}}+m_{\sigma_{1}}+k_{\sigma_{1}};...;n_{\sigma_{n(t)}},-k_{\sigma_{n(t)-1}}+m_{\sigma_{n(t)}}+k_{\sigma_{n(t)}}), & k&\in\Z^{V(\gamma(t))},
\end{align}
with convolution kernel $F_{a}$ s.t. $\lambda(F_{a})=a$. The second choice results in:
\begin{align}
\label{eq:hkdualgauge2}
 & \alpha^{(\gamma(t))}_{g}(F_{a})(n_{\sigma_{1}},m_{\sigma_{1}};...;n_{\sigma_{n(t)}},m_{\sigma_{n(t)}}) &  &  \\ \nonumber
 & = (g^{-1}_{\sigma_{0}}g_{\sigma_{n(t)}})^{n_{\sigma_{1}}}...(g^{-1}_{\sigma_{n(t)-1}}g_{\sigma_{n(t)}})^{n_{\sigma_{n(t)}}}F_{a}(n_{\sigma_{1}},m_{\sigma_{1}};...;n_{\sigma_{n(t)}},m_{\sigma_{n(t)}}), & g&\in U(1)^{V(\gamma(t))}.
\end{align}
A direct calculation shows that the family \eqref{eq:hkdualstate} is invariant under the latter transformations but not the former. Thus, the dual heat-kernel state is gauge-invariant for the action of $\underline{U(1)}$ resulting from \eqref{eq:hkdualgauge2} but not for the action of $\underline{\Z}$ resulting from \eqref{eq:hkdualgauge1}. Nevertheless, we explain below that its is possible to find a substitute for the action of $\underline{\Z}$ w.r.t.~the representation induced by the dual heat-kernel state.\\[0.1cm]
We note that the construction of the dual heat-kernel state extends partially to the compact pair $(\underline{U(1)},\overline{\scrA}_{U(1)})$, and more generally to any pair $(\underline{G},\overline{\scrA})$ with $G$ a compact Lie group. More precisely, we define a state on the $C^{*}$-subalgebra $C(\overline{\scrA})\subset\fA_{\D,L}$ (i.e.~dropping the left action by $\underline{G}$) by a consistent family as in \eqref{eq:hkdualstate} using $\rho^{(G)}_{\beta} = \sum_{\pi\in\hat{G}}e^{-\frac{\beta}{2}\lambda_{\pi}}\chi_{\pi}$ instead of $\rho^{(\Z)}_{\beta}$. Unfortunately, this state does not easily extend to $\fA_{\D,L}$ by means of an explicit formula because $M(\rho^{(G)}_{\beta})$ is not trace-class in $B(L^{2}(G))$. Nevertheless, by the Hahn-Banach theorem the existence of an extension is ensured. Moreover, the dual heat-kernel state is connected to the weak-coupling limit of the Kogut-Susskind Hamiltonian in the following sense:\\[0.1cm]
For $H^{(N)}_{\pi,w}$ as in \eqref{eq:KShamiltonianweak}, we have an asymptotic equality (up to some normalization constant),
\begin{align}
\label{eq:KSweakasymp}
e^{-\beta_{N}H^{(N)}_{\pi,w}} & \sim \prod_{f\in F(\gamma)}\rho^{(G)}_{\frac{a_{\gamma_{N}}g^{2}_{N}}{\beta_{N}}}(g_{f}), & \tfrac{\beta_{N}}{a_{\gamma_{N}}g^{2}_{N}} & \rightarrow \infty
\end{align}
as functions on $\scrA^{E(\gamma)}$. In the context of Euclidean quantum field theory, these two expressions are known as the Wilson action and generalized Villain action respectively \cite{VillainTheoryOfOne, ItzyksonStatisticalFieldTheory1}. Now, naively these expressions vanish in dimension $d=1$ because there are no 2-dimensional faces in this case, i.e.~$F(\gamma)=\emptyset$. But, if we think of a face, $f\in F(\gamma)$, as being bounded by edges, $e_{+},e_{-}\in E(\gamma)$, $\partial_{\pm}e_{\pm} = f$, we may identify faces with vertices, $F(\gamma)\equiv V(\gamma)$. Therefore, a possible interpretation of the weak-coupling limit in dimension $d=1$ is given by:
\begin{align}
\label{eq:KSweakasymp1}
e^{-\beta_{N} H^{(N)}_{\pi,w}} & \sim \prod_{e\in E(\gamma)}\rho^{(G)}_{\frac{a_{\gamma_{N}}g^{2}_{N}}{\beta_{N}}}(g_{e}),
\end{align}
where a holonomy around a face $f$ (see figure \ref{fig:facehol}) is given by a holonomy along a bounding edge $e$, $\partial_{\pm}e_{\pm} = f$. Thus, the product over faces, $F(\gamma)$, is replaced with a product over edges, $e\in E(\gamma)$. Now, we observe that \eqref{eq:KSweakasymp1} is precisely the convolution kernel of \eqref{eq:hkdualstate}, and we obtain a consistent choice of states for
\begin{align}
\label{eq:hkdualren}
\frac{g^{2}_{N}}{\beta_{N}} & = \textup{const. as a function of }N,
\end{align}
which should be contrasted with the scaling relation in the strong-coupling limit \eqref{eq:hkhomogeneousren}.
\begin{remark}
\label{rem:o2vector}
Taking a closer look at equation \eqref{eq:hkdualfourier}, we note that the Fourier transform of the heat kernel on $\Z$ corresponds to the Boltzmann weight of the $O(2)$-vector model. More precisely, the Boltzmann weight $p_{\beta}$ for the relative angle $\phi$ between two neighboring vectors $n,n'\in\fS=U(1)$ is (up to a normalization):
\begin{align*}
p_{\beta}(\phi) & = e^{-\frac{\beta}{2}\lambda(\phi)}, & \lambda(\phi) & = 2(1-\cos(\phi)).
\end{align*}
We return to this observation and potentially interesting applications of the dual heat-kernel state to the related $O(2)$-quantum rotor model  \cite{SachdevQuantumPhaseTransitions} in section \ref{sec:o2rotor}.
\end{remark}
In view of section \ref{sec:rgp}, we note that the dual heat-kernel state yields a paradigm for the notion of a full continuum limit. By construction (using the holonomy map) this state, when restricted to $C(\overline{\scrA})$, is nothing but the Wiener measure $\mu^{W}$ on $C(W_{\D,L}(G))$ because $\D\subset[0,1)$ is dense (see section \ref{sec:timezero} for further details). Therefore, we can restrict the support of the state to $W_{L}(G)=\{h\in C(L\D,G)\ |\ h(0)=e\}$ since the latter has full measure. In view of the gauge-invariance of the dual heat-kernel state, we observe that an interpretation as $G$-gauge theory, in this case corresponding to the transformations \eqref{eq:hkdualgauge1}, is possible for the following reason: If we choose the gauge group $H_{e}(G)\subset G^{L\D}$, i.e.~the Sobolev-Lie group of finite-energy loops, we will obtain a unitarizable (left) action on $L^{2}(W_{L}(G),\mu^{W})$ because $\mu^{W}$ is quasi-invariant (cf.~\cite{DriverYangMillsTheory, AlbeverioNoncommutativeDistributions}). By a result of Albeverio et al.~\cite{AlbeverioFactorialRepresentationsOf} the corresponding representation of $H_{e}(G)$ in $\cU(L^{2}(W_{L}(G),\mu^{W}))$ is factorial for $G=SU(2)$\footnote{The factor will be of type III, if we also take a thermodynamic limit.}. For abelian $G$ this action of the gauge group $H_{e}(G)$ also allows us to reintroduce a notion of momenta (corresponding to the action of $\h(\underline{G})=G_{\fr}\subset W_{\D,L}(G)$) because of corollary \ref{cor:discretehol}, but for nonabelian $G$ the situation remains somewhat unclear.

\subsection{The global and local von-Neumann field algebras}
\label{sec:timezero}
Before we discuss the structure of the global and local von-Neumann field algebras in view of the results of our companion article \cite{BrothierConstructionsOfConformal}, we show the existence of a peculiar isomorphism of the semi-continuum field algebra $\fA_{\D,L}$ with the infinite $C^{*}$-tensor product $\fB_{\D,L}=\bigotimes^{\min}_{d\in\D}B(L^{2}(G))^{\otimes 2}$\footnote{The construction of this isomorphism easily extends to the $1+d$-dimensional case by invoking a cofinal sequence of lattices.}. This isomorphism provides a link with tensor network renormalization and the multi-scale entanglement renormalization ansatz (MERA) \cite{VidalAClassOf, EvenblyTNRMERA} as will be explained below. Moreover, its construction is similar to that of an isomorphism between two AF algebras with identical Bratteli diagrams \cite{EvansQuantumSymmetriesOn}, cp.~also proposition \ref{prop:cofinaliso}.

\begin{prop}[A natural equivalence with the infinite tensor product]
\label{prop:tensoriso}
There exists an isomorphism of $C^{*}$-algebras
\begin{align*}
\eta : \fA_{\D,L} & \longrightarrow \fB_{\D,L}.
\end{align*}
\begin{proof}
As in proposition \ref{prop:cofinaliso}, we think of $\fA_{\D,L}$ as the inductive limit along the cofinal sequence of lattices, $\varinjlim_{\{\gamma_{N}\}_{N\in\N_{0}}}B(L^{2}(\scrA^{E(\gamma_{N})})) = \fA_{\D,L}$. The inductive system $\{B(L^{2}(\scrA^{E(\gamma_{N})})),\alpha^{\gamma_{N}}_{\gamma_{N+1}}\}_{N\in\N_{0}}$ results from an iteration of the injective $*$-morphisms
\begin{align*}
\alpha_{\uL} : B(L^{2}(G_{e})) & \longrightarrow B(L^{2}(G_{e_{2}}\times G_{e_{1}})), \\
 a & \longmapsto U^{*}_{\uL}(a\otimes\mathds{1}_{e_{1}})U_{\uL}, \\
\alpha_{\uR} : B(L^{2}(G_{e})) & \longrightarrow B(L^{2}(G_{e_{2}}\times G_{e_{1}})), \\
 a & \longmapsto V^{*}_{R}(\mathds{1}_{e_{2}}\otimes a)V_{\uR},
\end{align*}
while $\fB_{\D,L}$ corresponds to the inductive system $\{B(L^{2}(\scrA^{E(\gamma_{N})})),(\alpha_{\utriv})^{\gamma_{N}}_{\gamma_{N+1}}\}_{N\in\N_{0}}$ resulting from an iteration of the left and right tensor-factor embeddings
\begin{align*}
(\alpha_{\utriv})_{\uL} : B(L^{2}(G_{e})) & \longrightarrow B(L^{2}(G_{e_{2}}\times G_{e_{1}})), \\
 a & \longmapsto a\otimes\mathds{1}_{e_{1}}, \\
(\alpha_{\utriv})_{\uR} : B(L^{2}(G_{e})) & \longrightarrow B(L^{2}(G_{e_{2}}\times G_{e_{1}})), \\
 a & \longmapsto \mathds{1}_{e_{2}}\otimes a.
\end{align*}
Without loss of generality we assume the orientation of the initial element, $\gamma_{0}$, of the cofinal sequence to be to the left. We inductively obtain a sequence $\{\eta_{N}\}_{N\in\N_{0}}$ with initial elements $\zeta_{0} = \id$, $\eta_{1}=\Ad_{U_{\uL}}$. Assuming that $\eta_{N} = \Ad_{U_{N}}$ has been constructed up to some $N\in\N_{0}$ with a unitary $U_{N}\in\cU(L^{2}(\scrA^{E(\gamma_{N})}))$. Then, we define $U_{N+1}$ by
\begin{align*}
\Ad_{U_{N+1}}\circ\!\ \alpha^{\gamma_{N}}_{\gamma_{N+1}} & = (\alpha_{\utriv})^{\gamma_{N}}_{\gamma_{N+1}}\circ\Ad_{U_{N}} = \Ad_{(\alpha_{\utriv})^{\gamma_{N}}_{\gamma_{N+1}}(U_{N})}\circ\!\ (\alpha_{\utriv})^{\gamma_{N}}_{\gamma_{N+1}},
\end{align*}
which is possible because both $\alpha^{\gamma_{N}}_{\gamma_{N+1}}$ and $(\alpha_{\utriv})^{\gamma_{N}}_{\gamma_{N+1}}$ are unitarily conjugate to one another by construction. From this we infer the existence of the inductive limit $\varinjlim_{N\in\N_{0}}\eta_{N} = \eta$, cf.~\cite{EvansQuantumSymmetriesOn}.
\end{proof}
\end{prop}

Now, the link with the MERA arises as follows: Given a state $\omega_{\D,L}$ on $\fA_{\D,L}$, the system of $*$-morphisms $\{(\alpha_{\utriv})^{\gamma_{N}}_{\gamma_{N+1}}\}_{N\in\N_{0}}$ descends via GNS construction to a system of MERA isometries while the intertwining unitaries between the former and the $*$-morphisms defining $\fA_{\D,L}$ are the MERA disentanglers. More generally, if we consider a MERA with basic refining operation at the level of Hilbert spaces,
\begin{align}
\label{eq:MERAisometry}
R : L^{2}(\scrA^{E(\gamma)}) & \longrightarrow L^{2}(\scrA^{E(\gamma')}), \\ \nonumber
\psi_{E(\gamma)} & \longmapsto U\left(\psi_{E(\gamma)}\otimes 1^{\otimes E(\gamma')\setminus E(\gamma)}\right),
\end{align}
for some $\gamma\subset_{\uL}\gamma'$ and an associated inclusion of edges $E(\gamma)\subset E(\gamma')$ together with a disentangler $U\in\cU(L^{2}(\scrA^{E(\gamma')}))$, 
we can form a compatible basic refining $*$-morphism, cp.~also \cite{MilstedQuantumYangMills}:
\begin{align}
\label{eq:MERAmorph}
\alpha_{R} : B(L^{2}(\scrA^{E(\gamma)})) & \longrightarrow B(L^{2}(\scrA^{E(\gamma')})), \\ \nonumber
a_{E(\gamma)} & \longmapsto U\left(a_{E(\gamma)}\otimes\mathds{1}_{E(\gamma')\setminus E(\gamma)}\right)U^{*},
\end{align}
such that
\begin{align}
\label{eq:MERAcomp}
\alpha_{R}(a_{E(\gamma)})R(\Psi_{E(\gamma)}) & = R(a_{E(\gamma)}\Psi_{E(\gamma)}).
\end{align}
It is important to distinguish $\alpha_{R}$ from the adjoint action, $\Ad_{R}(\!\ .\!\ ) = R(\!\ .\!\ )R^{*}$, which is also compatible in the above sense, cp.~remark \ref{rem:localalg}:
\begin{align}
\label{eq:MERAadjoint}
\Ad_{R}(a_{E(\gamma)})R(\Psi_{E(\gamma)}) & = R(a_{E(\gamma)}\Psi_{E(\gamma)}).
\end{align}
We return to the discussion of the difference between $\alpha_{R}$ and $\Ad_{R}$ in section \ref{sec:rgp}.

\begin{remark}[Local algebras]
\label{rem:yangmillslocal}
Although, $\eta$ provides a natural isomorphism between the $C^{*}$-algebras $\fA_{\D,L}$ and $\fB_{\D,L}$ (it is a natural equivalence of the associated tensor functors, cf.~\cite{BrothierConstructionsOfConformal}), it does not preserve their locality structures, cp.~corollary \ref{cor:localalgsys}. Phrased in another way, this means that the underlying inductive systems corresponds to different systems of local fields. To see this, we consider exemplarily the local algebras $\fA_{\D,L}([0,\tfrac{L}{2}+\varepsilon)\cup(L-\varepsilon,L])$ and $\fB_{\D,L}([0,\tfrac{L}{2}+\varepsilon)\cup(L-\varepsilon,L])$ for some $0<\varepsilon<\tfrac{L}{2}$ and periodic boundary conditions ($\Sigma = \T_{L}^{d}$). Clearly, $\fB_{\D,L}([0,\tfrac{L}{2}+\varepsilon)\cup(L-\varepsilon,L])$ contains the equivalence class generated by elements $M(f)$, $\textup{const.}\neq f\in C(\scrA^{E(\gamma_{0})})$, because $\alpha^{\gamma_{0}}_{\gamma_{1}}(a) = a\otimes\mathds{1}$ for all $a\in B(L^{2}(\scrA^{E(\gamma_{0})}))$. But, the equivalence class generated by $\eta^{-1}_{1}(M(f)\otimes\mathds{1}) = U^{*}_{\uL}(M(f)\otimes\mathds{1})U_{\uL} = M(f\circ p^{\gamma_{1}}_{\gamma_{0}})$ is not in $\fA_{\D,L}([0,\tfrac{L}{2}+\varepsilon)\cup(L-\varepsilon,L])$ by proposition \ref{prop:groupsys}.
\end{remark}

\begin{remark}[Non-equivariance]
\label{rem:yangmillsnonequi}
It is important to note, that the action of Thompson's group $F$ (see section \ref{sec:thompson}) is not equivariant w.r.t.~the natural isomorphism $\eta : \fA_{\D,L} \longrightarrow \fB_{\D,L}$. Consider, for example, an elementary tensor $a\ot b\in B(L^{2}(\scrA^{E(\gamma_{1})}))$ and $f\in F$ with
\begin{equation*}
f \simeq
\resizebox{!}{!}{
\begin{tikzpicture}[baseline={(0,-3)}]
	
	\draw[thick] (1,-3) to (3,-3);
	\draw (1,-3) to (2,-2) to (3,-3);
	\draw (2.5,-2.5) to (2,-3);
	\draw (1,-3) to (2,-4) to (3,-3);
	\draw (1.5,-3.5) to (2,-3);
	
\end{tikzpicture}
}
\end{equation*}
which results in
\begin{equation*}
f\cdot(a\otimes b)\simeq
\resizebox{!}{!}{
\begin{tikzpicture}[baseline={(0,-3)}]
	
	\draw[thick] (1,-3) to (3,-3);
	\draw (1,-3) to (2,-2) to (3,-3);
	\draw (2.5,-2.5) to (2,-3);
	\draw (1,-3) to (2,-4) to (3,-3);
	\draw (1.5,-3.5) to (2,-3);
	
	\filldraw (3.5,-3) circle (1pt);
	
	\draw[thick] (4,-3) to (6,-3);
	\draw (4,-3) to (5,-2) to (6,-3);
	\draw (4.5,-2.5) to (5,-3);
	\draw (5.75,-2.75) node[right]{\small\textg{R}};
	\draw (4.25,-2.75) node[left]{\small\textr{L}};
	\draw (4.75,-2.75) node[right]{\small\textg{R}};
	\draw (4,-3.25) node{$a$} (4.5,-3.25) node{$\otimes$} (5,-3.25) node{$\mathds{1}$} (5.5,-3.25) node{$\otimes$} (6,-3.25) node{$b$};
	
	\draw (6.5,-3) node{$\simeq$};
	
	\draw[thick] (7,-3) to (9,-3);
	\draw (7,-3) to (8,-2) to (9,-3);
	\draw (8.5,-2.5) to (8,-3);
	\draw (8.75,-2.75) node[right]{\small\textg{R}};
	\draw (7.25,-2.75) node[left]{\small\textr{L}};
	\draw (7.75,-2.75) node[right]{\small\textg{R}};
	\draw (7,-3.25) node{$a$} (7.5,-3.25) node{$\otimes$} (8,-3.25) node{$\mathds{1}$} (8.5,-3.25) node{$\otimes$} (9,-3.25) node{$b$};
	
\end{tikzpicture}
}
\end{equation*}
for the action on $\fB_{\D,L}$. We compare this with the action on $\fA_{\D,L}$ by finding the corresponding expression for
\begin{equation*}
\eta_{
 \resizebox{!}{0.3cm}{\begin{tikzpicture}[baseline={(0,-3)}]
	
	\draw[thick] (1,-3) to (3,-3);
	\draw (1,-3) to (2,-2) to (3,-3);
	\draw (2.5,-2.5) to (2,-3);
	
	\draw (1,-3) node[below]{\LARGE\textr{L}} (2,-3) node[below]{\LARGE\textg{R}} (3,-3) node[below]{\LARGE\textg{R}};
	
\end{tikzpicture}}
}
(f\cdot\eta_{1}^{-1}(a\otimes b)),
\end{equation*}
where we use the extension of $\eta$ to (intermediate) incomplete dyadic trees respectively lattices, i.e.
\begin{equation*}
\eta_{
 \resizebox{!}{0.3cm}{\begin{tikzpicture}[baseline={(0,-3)}]
	
	\draw[thick] (1,-3) to (3,-3);
	\draw (1,-3) to (2,-2) to (3,-3);
	\draw (2.5,-2.5) to (2,-3);
	
	\draw (1,-3) node[below]{\LARGE\textr{L}} (2,-3) node[below]{\LARGE\textg{R}} (3,-3) node[below]{\LARGE\textg{R}};
	
\end{tikzpicture}}
} : B(L^{2}(\scrA^{E(\gamma)})) \longrightarrow B(L^{2}(\scrA^{E(\gamma)}))
\end{equation*}
is the restriction of $\eta$ to the algebras associated with $\gamma_{1}\subset_{\uL}\gamma\simeq\resizebox{!}{0.3cm}{\begin{tikzpicture}[baseline={(0,-3)}]
	
	\draw[thick] (1,-3) to (3,-3);
	\draw (1,-3) to (2,-2) to (3,-3);
	\draw (2.5,-2.5) to (2,-3);
	
	\draw (1,-3) node[below]{\LARGE\textr{L}} (2,-3) node[below]{\LARGE\textg{R}} (3,-3) node[below]{\LARGE\textg{R}};
	
\end{tikzpicture}}\subset_{\uL}\gamma_{2}$. Using the defining relation of $\eta$ (see proposition \ref{prop:tensoriso}), we find:

\begin{equation*}
\eta_{
 \resizebox{!}{0.3cm}{\begin{tikzpicture}[baseline={(0,-3)}]
	
	\draw[thick] (1,-3) to (3,-3);
	\draw (1,-3) to (2,-2) to (3,-3);
	\draw (2.5,-2.5) to (2,-3);
	
	\draw (1,-3) node[below]{\LARGE\textr{L}} (2,-3) node[below]{\LARGE\textg{R}} (3,-3) node[below]{\LARGE\textg{R}};
	
\end{tikzpicture}}
}
(f\cdot\eta_{1}^{-1}(a\otimes b))\simeq
\resizebox{!}{!}{
\begin{tikzpicture}[baseline={(0,-3)}]
	
	\draw[thick] (7,-3) to (9,-3);
	\draw (7,-3) to (8,-2) to (9,-3);
	\draw (8.5,-2.5) to (8,-3);
	\draw (8.75,-2.75) node[right]{\small\textg{R}};
	\draw (7.25,-2.75) node[left]{\small\textr{L}};
	\draw (7.75,-2.75) node[right]{\small\textg{R}};
	\draw (7,-3.25) node{$U(a$} (7.625,-3.25) node{$\otimes$} (8,-3.25) node{$\mathds{1}$} (8.325,-3.25) node{$\otimes$} (9,-3.25) node{$b)U^{*}$};
	
\end{tikzpicture}
},
\end{equation*}
where the unitary
\begin{equation*}
U = U_{
 \resizebox{!}{0.3cm}{\begin{tikzpicture}[baseline={(0,-3)}]
	
	\draw[thick] (1,-3) to (3,-3);
	\draw (1,-3) to (2,-2) to (3,-3);
	\draw (2.5,-2.5) to (2,-3);
	
	\draw (1,-3) node[below]{\LARGE\textr{L}} (2,-3) node[below]{\LARGE\textg{R}} (3,-3) node[below]{\LARGE\textg{R}};
	
\end{tikzpicture}}
}
U^{*}_{
 \resizebox{!}{0.3cm}{\begin{tikzpicture}[baseline={(0,-3)}]
	
	\draw[thick] (1,-3) to (3,-3);
	\draw (1,-3) to (2,-2) to (3,-3);
	\draw (1.5,-2.5) to (2,-3);
	
	\draw (1,-3) node[below]{\LARGE\textr{L}} (2,-3) node[below]{\LARGE\textg{R}} (3,-3) node[below]{\LARGE\textg{R}};
	
\end{tikzpicture}}
}
\end{equation*}
is given in terms of the unitaries implementing $\eta$. A simple but tedious computation shows that:
\begin{align*}
(U\psi)(g_{e_{3}},g_{e_{2}},g_{e_{1}}) & = \psi(g_{e_{3}}\alpha_{g^{-1}_{e_{1}}}(g^{-1}_{e_{2}}),g_{e_{2}},g_{e_{1}}), & \psi&\in C_{c}(G_{e_{3}}\times G_{e_{2}}\times G_{e_{1}}),
\end{align*}
which implies the non-equivariance.
\end{remark}

Following these initial observations regarding the structure of the semi-continuum field algebra $\fA_{\D,L}$, we turn to the discussion of the semi-continuum von-Neumann field algebras. As we explained in section \ref{sec:can11}, there are three natural classes of states associated with the strong- and weak-coupling limits of the Kogut-Susskind Hamiltonian for YM$_{1+1}$.\\[0.25cm]
The first and in some sense most elementary class is given by the heat-kernel states, $\omega_{\beta,\D,L}$, $\beta\in\R_{>0}$ with potentially well-defined limiting cases $\omega_{0,\D,L}$ and $\omega_{\infty,\D,L}$. In our companion paper \cite[Theorem 3.12]{BrothierConstructionsOfConformal}, we give precise description of the semi-continuum von-Neumann algebra,
\begin{align}
\label{eq:leftvnahklimit}
\overleftarrow{\scrM}_{\beta,\D,L} & = \pi_{\omega_{\beta,\D,L}}(\overleftarrow{\fA}_{\D,L}),
\end{align}
when $G$ is compact, separable and abelian. Here, we treat $\overleftarrow{\fA}_{\D,L}$ as a $C^{*}$-subalgebra $\fA_{\D,L}$ because of the natural inclusion stated in remark \ref{rem:subiso}. In this case, we know from the analysis in section \ref{sec:fourier} that $\overleftarrow{\fA}_{\D,L}$ can described in terms of the (discrete) dual pair $(\hat{G}_{\fr},W_{\D,L}(\hat{G}))$ via Fourier duality. The heat-kernel state $\omega_{\beta,\D,L}$ is then determined by the Borel probability measure on $(\hat{G},\cB(\hat{G}))$,
\begin{align}
\label{eq:hkprob1}
m_{\beta}(A) & = \sum_{\pi_{mn}\in A\in\cB(\hat{G})}m_{\beta}(\{\pi_{mn}\}), & m_{\beta}(\{\pi_{mn}\}) & = \rho_{\beta}(1_{G})^{-1}d_{\pi}e^{-\frac{\beta}{2}c_{\pi}}\delta_{mn},
\end{align}
as this gives the Gibbs state $\omega_{\beta}$ according to
\begin{align}
\label{eq:hkprob2}
\omega_{\beta}(a) & = \Tr_{L^{2}}(\lambda(h_{m_{\beta}}\otimes 1)a), & h_{m_{\beta}} & = \sum_{\pi\in\hat{G}}\sum^{d_{\pi}}_{m,n=1}m_{\beta}(\{\pi_{mn}\})\pi_{mn},
\end{align}
for $a\in B(L^{2}(G))$. Conversely, any Borel probability measure $m$ determines a state $\omega_{m}$ via the trace-class operator $\lambda(h_{m}\otimes 1)$ and, thus, by remark \ref{rem:compactstate} a limit state $\omega_{m,\D,L}$. Clearly, a representation $\pi\in\hat{G}$ is one-dimensional for abelian $G$ such that we may drop the reference to an orthonormal basis of the representation space $V_{\pi}$. Nevertheless, we state the more complicated formulae \eqref{eq:hkprob1} and \eqref{eq:hkprob2} as they remain valid for nonabelian $G$.\\[0.1cm]
To restate \cite[Theorem 3.12]{BrothierConstructionsOfConformal} in a way adapted to notation of this article, we consider the subgroup $N\subset\hat{G}$ preserving $m$ under pushforward, i.e.~$k_{*}m=m$ for $k\in\hat{G}$, which induces the subgroup $N_{\fr}\subset G_{\fr}$. For a section $\sigma:\hat{G}_{\fr}/N_{\fr}\to \hat{G}_{\fr}$ we consider the cocycle 
\begin{align}
\label{eq:finiterankcocycle}
\kappa : \hat{G}_{\fr}\times\hat{G}_{\fr}/N_{\fr} & \longrightarrow N_{\fr}, & (k,[k']) & \longmapsto \sigma(k[k'])^{-1} k \sigma([k']),
\end{align}
and the group action 
\begin{align}
\label{eq:finiterankcocycleaction}
\hat{G}_{\fr} & \act (W_{\D,L}(\hat{G})\times\hat{G}_{\fr}/N_{\fr}), & k\cdot (z,[k']) & = (\kappa(k,[k'])\!\ z , k[k']).
\end{align}
Using the latter, we form the crossed-product von Neumann algebra
\begin{align}
\label{eq:hkcrossedprod}
\overleftarrow{\scrB}_{m,\D,L} & = L^{\infty}(W_{\D,L}(\hat{G})\times\hat{G}_{\fr}/N_{\fr},\otimes_{d\in L\D_{\neq0}}m\otimes \mu_{\uc})\rtimes\hat{G}_{\fr},
\end{align}
where $\mu_{\uc}$ is the counting measure on $\hat{G}_{\fr}/N_{\fr}$. There is a specific normal state on $\overleftarrow{\scrB}_{m,\D,L}$ given by
\begin{align}
\label{eq:hkcrossedprodstate}
\varpi_{\D,L}\left(\sum_{k\in\hat{G}_{\fr}} f_{k}\!\ u_{k}\right) & = \int_{W_{\D,L}(\hat{G})} f_{1_{\hat{G}_{\fr}}}(z,[1_{\hat{G}_{\fr}}])\!\ d\!\otimes_{d\in L\D_{\neq0}}m(z),
\end{align}
where, for $k\in\hat{G}_{\fr}$, $f_{k}\in L^\infty(W_{\D,L}(\hat{G})\times\hat{G}_{\fr}/N_{\fr},\otimes_{d\in L\D_{\neq0}}m\otimes\mu_{\uc})$ and $u_{k}$ denotes the unitary of $\overleftarrow{\scrB}_{m,\D,L}$ implementing the action. There is a natural action of Thompson's group $V\act L\D$ induced by its action on the unit interval and rescaling. This induces an action on $\hat{G}^{L\D}$ by shifting tensor indices which descends to the subgroups $W_{\D,L}(\hat{G})$ and $\hat{G}_{\fr}$ and passes to the quotient $\hat{G}_{\fr}/ N_{\fr}$. This way we obtain an action $V\act\overleftarrow{\scrB}_{m,\D,L}$:
\begin{align}
\label{eq:hkcrossedprodaction}
v\cdot\left(\sum_{k\in\hat{G}_{\fr}} f_{k}\!\ u_{k}\right) = \sum_{k\in\hat{G}_{\fr}} (v^{-1})^{*}f_{k}\!\ u_{vk},
\end{align}
for $v\in V$, $f_{k}\in L^{\infty}(W_{\D,L}(\hat{G})\times\hat{G}_{\fr}/N_{\fr},\otimes_{d\in L\D_{\neq0}}m\otimes \mu_{\uc})$. It is an immediate consequence that $\varpi_{\D,L}$ is invariant under this action. Now, we have:

\begin{theo}
\label{th:hklimit}
Let $G$ be a compact abelian separable group and $m$ a strictly positive Borel probability measure. We have a state-preserving isomorphism of von Neumann algebras 
\begin{align*}
(\overleftarrow{\scrM}_{m,\D,L},\omega_{m,\D,L}) & \tilde{\longrightarrow} (\overleftarrow{\scrB}_{m,\D,L},\varpi_{\D,L})
\end{align*}
and in particular $\overleftarrow{\scrM}_{m,\D,L}$ does not have any type III component.
Moreover, the Jones action of Thompson's group, $V\act\overleftarrow{\scrM}_{m,\D,L}$, preserves the state $\omega_{m,\D,L}$ and is conjugate to the Jones action $V\act\overleftarrow{\scrB}_{m,\D,L}$.\\[0.1cm]
If $\hat{G}$ is torsion free (e.g.~$G = U(1)$), then $(\overleftarrow{\scrM}_{m,\D,L},\omega_{m,\D,L})$ is isomorphic to
\begin{align*}
(L^{\infty}(W_{\D,L}(\hat{G}),\otimes_{d\in L\D_{\neq0}}m)\ovt B(\ell^2(\hat{G}_{\fr})),\otimes_{d\in L\D_{\neq0}}m\otimes\langle\delta_{e},\ .\ \delta_{e}\rangle),
\end{align*}
which is a type I$_\infty$ von Neumann algebra with a diffuse center and equipped with a non-faithful state.\\[0.1cm]
If $G$ is a finite group and $m$ is $\hat{G}$-invariant (i.e.~$m$ is the normalized Haar measure of $\hat{G}$), then $(\overleftarrow{\scrM}_{m,\D,L},\omega_{m,\D,L})$ is isomorphic to the hyperfinite type II$_1$ factor equipped with its trace.
\end{theo}

Remarkably, the theorem gives a very precise description of the limit algebra $\overleftarrow{\scrM}_{\beta,\D,L}$ as an explicitly realized crossed-product von Neumann algebra. In accordance with the discussion of the temperature scaling of heat-kernel state $\omega_{\beta,\D,L}$ (see \eqref{eq:hkhomogeneousren} and \eqref{eq:hktemperatureren}), the absence of any type III component implies that we should not interpret $\omega_{\beta,\D,L}$ as a finite-temperature equilibrium state as said property contradicts the KMS condition. Especially, considering the important cases $G=U(1)$ or $G=\Z_{n}\subset U(1)$, $n\in\N$, we infer that $\overleftarrow{\scrM}_{\beta,\D,L}$ is a type I$_{\infty}$ von Neumann algebra and the heat-kernel state is not faithful. The main reason behind the generic non-faithfulness of the heat-kernel state and the type I$_{\infty}$ situation is the singular nature of the action $\hat{G}_{\fr}\act(W_{\D,L}(\hat{G}),\otimes_{d\in L\D_{\neq0}}m)$ by Kakutani's theorem \cite{KakutaniOnEquivalenceOf}, see \cite[Theorem 3.11]{BrothierConstructionsOfConformal}. An exception to the generic situation may occur whenever $\hat{G}$ has torsion and, specifically, if $G$ is a finite group. In the latter case, if $m=m_{\hat{G}}$ is the normalized Haar measure, we will have a non-singular action $\hat{G}_{\fr}\act(W_{\D,L}(\hat{G}),\otimes_{d\in L\D_{\neq0}}m)$, and the heat-kernel state will correspond to the limiting case $\omega_{\beta=0,\D,L}$, i.e.~a tracial infinite-temperature state, leading to a type II$_{1}$ situation. For infinite $G$, the infinite-temperature state $\omega_{0,\D,L}$ is ill-defined because the non-normalized basic building block is only a tracial weight, i.e.~$\lim_{\beta\rightarrow0+}Z(\beta)\omega_{\beta} = \Tr_{L^{2}}$. \\[0.1cm]
Theorem \ref{th:hklimit} also leads to the conclusion that it is impossible to define a time-evolution group on $\overleftarrow{\scrM}_{\beta,\D,L}$ by means of Tomita-Takesaki theory. More precisely, even though the restrictions $\omega^{(\gamma_{N})}_{\beta}$, $N\in\N_{0}$, to finite levels are faithful states and induce time-evolution groups via Tomita-Takesaki theory with modular operators related to the Kogut-Susskind Hamiltonians, i.e.~$\log\Delta^{(N)}_{\beta}\propto H^{(N)}=H^{(N)}_{s}$, the limit state $\omega_{\beta,\D,L}$ falls short of this property. Also the restriction of $\omega_{\beta,\D,L}$ to its support, i.e.~$E\overleftarrow{\scrM}_{\beta,\D,L}E$ for some maximal family of orthogonal projections $\{E_{i}\}_{i\in I}\subset\scrM_{\beta,\D,L}$ such that
\begin{align}
\label{eq:hklimitsupport}
\omega_{\beta,\D,L}(\mathds{1}-E) & = \omega_{\beta,\D,L}(\sum_{\i\in I}E_{i}) = 0,
\end{align}
necessarily leads to a tracial state and, thus, a trivial modular operator $\Delta_{\beta,\D,L,E}$.\\[0.1cm]
Although, as stated above, theorem \ref{th:hklimit} only applies to the global von-Neumann field algebra $\overleftarrow{\scrM}_{\beta,\D,L}$, it easily extends to the local von-Neumann field algebras $\overleftarrow{\scrM}_{\beta,\D,L}(\cS)$ by restricting the set $L\D$ to its appropriately localized counterpart $L\D(\cS)$, cf.~\cite[Section 2.6]{BrothierConstructionsOfConformal} for details.
\begin{remark}
\label{rem:generalJones}
The Jones' action of $V\act\fA_{\D,L}$ preserves the heat-kernel states $\omega_{\beta,\D,L}$ also for nonabelian compact groups $G$, because the invariance in solely due to the assignment of some $\omega_{\beta}$ with constant $\beta>0$ to each ``edge algebra'' $B(L^{2}(G_{e}))$, $e\in E(\gamma_{N})$ at level $N\in\N_{0}$. This observation will apply to more general groups or other objects like quantum groups as well, if an analogue of the heat-kernel state is defined. 
\end{remark}
\begin{remark}
\label{rem:cofinalisostate}
So far, we have only explained how theorem \ref{th:hklimit} clarifies the structure of the von Neumann algebra associated with $(\overleftarrow{\fA}_{\D,L}, \omega_{\beta,\D,L})$, but it is evident from the structure of the proofs presented \cite{BrothierConstructionsOfConformal} that our analysis carries over to situation involving the full field algebra $(\fA_{\D,L},\omega_{\beta,\D,L})$ with minor adaptions mainly due to changing the sequence of lattices from $\{\cev{\gamma}_{N}\}_{N\in\N_{0}}$ to $\{\gamma_{N}\}_{N\in\N_{0}}$. But, it is important to note that $\omega_{\beta,\D,L}$ is not invariant under the isomorphism $\zeta:\fA_{\D,L}\rightarrow\overleftarrow{\fA}_{\D,L}$, see proposition \ref{prop:cofinaliso} and remark \ref{rem:cofinaliso}. Namely, restricting to level $N=2$, we have:
\begin{align*}
\omega^{\otimes 4}_{\beta}\circ\zeta_{2} & = \omega^{\otimes 4}_{\beta}\circ\Ad_{U_{2}},
\end{align*}
where $U_{2} = \mathds{1}\otimes\mathds{1}\otimes U_{\uL}(U_{\iota}\otimes\mathds{1})U^{*}_{\uL}$. But, because
\begin{align*}
(U_{\uL}(U_{\iota}\otimes\mathds{1})U^{*}_{\uL}\psi)(g_{2},g_{1}) & = \psi((g_{1}g_{2}g_{1})^{-1},g_{1}),
\end{align*}
for $\psi\in C(G^{\times 2})$, we find:
\begin{align*}
\omega^{\otimes 4}_{\beta}\circ\zeta_{2} & \neq \omega^{\otimes 4}_{\beta}.
\end{align*}
\end{remark}

As explained in remark \ref{rem:hkinfty}, it is possible to define as a limit case the heat-kernel state $\omega_{\beta=\infty,\D,L}$ which is equivalent to the strong-coupling vacuum introduced in section \ref{sec:compact}. This case is not explicitly covered by theorem \ref{th:hklimit}, but it is easy to see that $\omega_{\infty,\D,L}$ is a pure state, corresponding to the degenerate probability measure $m(\{\pi_{mn}\}) = \delta_{\pi,\pi_{\utriv}}\delta_{m,0}\delta_{n,0}$, which leads to an irreducible faithful representation of $\fA_{\D,L}$. Therefore, $\cM_{\infty,\D,L} = B(L^{2}(\overline{\scrA}))$ which is type I$_{\infty}$. Moreover, theorem \ref{th:KShamiltonianstrong} allows us to define the time evolution group $\Ad_{U_{-t}}$, $t\in\R$, on $\cM_{\infty,\D,L}$ with $U_{t}=e^{-it\underline{H}_{s}}$ given by the strong-coupling limit of the Kogut-Susskind Hamiltonian.\\[0.1cm]
Concerning the local von-Neumann field algebras, it is possible to introduce the local \textit{spacetime} von-Neumann field algebras, at least for the strong-coupling vacuum following standard reasoning, see e.g.~\cite{GlimmBosonQuantumField}:\\[0.1cm]
For any bounded open region of spacetime, $\cO\subset\R\times\Sigma$, with time slices $\cO(t) = \{x\ | (t,x)\in\cO\}\subset\Sigma$, $t\in\R$, that are open star domains, we define the local spacetime von-Neumann field algebra by:
\begin{align}
\label{eq:spacetimelocalalg}
\scrM_{\infty,\D,L}(\cO) & = \bigcup_{t\in\R}\Ad_{U_{-t}}\left(\scrM_{\infty,\D,L}(\cO(t))\right),
\end{align}
which satisfy Einstein causality because $\underline{H}_{s}$ has propagation speed $c=0$ (see remark \ref{rem:finiteprop} below). The reason for the vanishing propagation speed is the gauge invariance of the model, i.e.~YM$_{1+1}$ has no local but only global propagating degrees of freedom (see section \ref{sec:gaugeobs})

\begin{remark}
\label{rem:finiteprop}
In general, \eqref{eq:spacetimelocalalg} makes sense as definition of the local spacetime von-Neumann field algebras, whenever we are able to establish finite propagation speed $c<0$ for the time evolution $\Ad_{U_{-t}}$ associated with a pair $(\fA_{\D,L},\omega_{\D,L})$, i.e.
\begin{align*}
\Ad_{U_{-t}}\left(\scrM_{\D,L}(\cS)\right) & \subset \scrM_{\D,L}(\cS_{ct}),
\end{align*}
where $\cS_{ct}=\{x\in\Sigma\ |\ d_{\Sigma}(x,\cS)<c|t|\}$ and $d_{\Sigma}$ is the metric distance on $\Sigma$. Indeed, given a finite propagation speed $c$, spatial locality and isotony implies Einstein causality and spacetime isotony, cf.~corollary \ref{cor:localalgsys}.\\[0.1cm]
Interestingly, there is also the possibility to consider the counterparts of the local spacetime von-Neumann field algebras at finite level $N\in\N_{0}$:
\begin{align*}
B(L^{2}(\scrA^{E(\gamma_{N})}))(\cO) & = \bigcup_{t\in\R}\Ad_{U^{(N)}_{-t}}\left(B(L^{2}(\scrA^{E((\gamma_{N})_{\cO(t)})}))\right),
\end{align*}
where we use the notation of definition \ref{def:locallat}, and $U^{(N)}_{t} = e^{-it H^{(N)}}$ is time-evolution group of the presumed approximate Hamiltonian at level $N$. These algebras do not satisfy Einstein causality, but in favorable cases an approximate form of the latter may hold due to Lieb-Robinson bounds \cite{LiebTheFiniteGroup, NachtergaeleLiebRobinsonBounds, NachtergaeleLRBoundsHarmonic}. Strict Einstein causality might be recovered in suitable continuum limits, see \cite{OsborneContinuumLimitsOf} for an argument in this direction, and \cite{MorinelliScalingLimitsOf} for a proof in the case of scalar free fields on the lattice.
\end{remark}

The second class of semi-continuum von-Neumann field algebras results from the inhomogeneous heat-kernel states $\omega_{(\beta_{\uL},\beta_{\uR}),\D,L}$ parametrized by two sequences $\{\beta_{d,\uL}\}_{d\in\D}$ and $\{\beta_{d,\uR}\}_{d\in\D}$. Again, we restrict attention to the structure of the von Neumann subalgebra $\overleftarrow{\scrM}_{\beta_{\uL},\D,L}\subset\scrM_{(\beta_{\uL},\beta_{\uR}),\D,L}$ which only depends on the sequence $\{\beta_{d,\uL}\}_{d\in\D}$ as a consequence of remark \ref{rem:subiso}. Similarly to the first class, the analysis extends to the full global von-Neumann field algebra $\scrM_{(\beta_{\uL},\beta_{\uR}),\D,L}$ with only minor complications. For simplicity, we restrict the discussion of the second class to the case $G=U(1)$, but there appear to be no major obstacles to generalize the results to arbitrary separable, compact abelian groups. Using Fourier duality again, the major difference in comparison with the (homogeneous) heat-kernel state is the replacement of the measure space $(W_{\D,L}(\hat{G}),\otimes_{d\in L\D_{\neq0}}m_{\beta})$ by $(W_{\D,L}(\hat{G}),\otimes_{d\in L\D_{\neq0}}m_{\beta_{d,\uL}})$. Thus, the structure of $\overleftarrow{\scrM}_{\beta_{\uL},\D,L}$ is essentially determined by the action $\hat{G}_{\fr}\act (W_{\D,L}(\hat{G}),\otimes_{d\in L\D_{\neq0}}m_{\beta_{d,\uL}})$. The first indication that this entails a change in algebraic structure is given by the following proposition -- again as a result of Kakutani's theorem, cf.~\cite[Proposition 3.15]{BrothierConstructionsOfConformal}:

\begin{prop}
\label{prop:hkinhomnonsingular}
For $G=U(1)$ and $\hat{G}=\Z$, the action $\theta:\Z_{\fr}\act (W_{\D,L}(\Z),\otimes_{d\in L\D_{\neq0}}m_{\beta_{d,\uL}})$ is nonsingular if and only if $\beta_{\uL}\in\ell^{1}(\D)$, i.e.~$\sum_{d\in\D}\beta_{d,\uL}<\infty$.
\end{prop}

Subsequently, we conclude that $\overleftarrow{\scrM}_{\beta_{\uL},\D,L}$ has the structure of a crossed-product von Neumann algebra for the above action:

\begin{theo}
\label{th:inhomhklimit}
Assume $\beta_{\uL}\in\ell^{1}(\D)$, then there exists an isomorphism of von Neumann algebras, 
\begin{align*}
J : \overleftarrow{\scrM}_{\beta_{\uL},\D,L} & \longrightarrow L^\infty(W_{\D,L}(\Z),\otimes_{d\in L\D_{\neq0}}m_{\beta_{d,\uL}})\rtimes \Z_{\fr},
\end{align*}
satisfying
\begin{align*}
\omega_{(\beta_{\uL},\beta_{\uR}),\D,L}\circ J^{-1}\left(\sum_{g\in\Z_{\fr}} b_{g} u_{g}\right) = \int_{W_{\D,L}(\Z)} b_{1_{\Z_{\fr}}}(x)\!\ d\!\otimes_{d\in L\D_{\neq0}}\!\!m_{\beta_{d,\uL}}(x),
\end{align*}
where $u_{g}$, $g\in\Z_{\fr}$ implements the unitary action of $\Z_{\fr}$ on $L^\infty(W_{\D,L}(\Z),\otimes_{d\in L\D_{\neq0}}m_{\beta_{d,\uL}}).$
\end{theo}

Moreover, it is shown in \cite[Corollary 3.18]{BrothierConstructionsOfConformal} that the action, $\theta:\Z_{\fr}\act (W_{\D,L}(\Z),\otimes_{d\in L\D_{\neq0}}m_{\beta_{d,\uL}})$, is ergodic for a certain range of $\beta_{\uL}$ -- not excluding a possibly wider range:

\begin{corollary}
\label{cor:inhomhkergodic}
If $\beta_{\uL}:\D\longrightarrow\R_{>0}$ is $p$-summable for some $0<p<\tfrac{1}{2}$, then the action $\theta:\Z_{\fr}\act (W_{\D,L}(\Z),\otimes_{d\in L\D_{\neq0}}m_{\beta_{d,\uL}})$ is ergodic.
\end{corollary}

From general theorems \cite{KadisonFundamentalsOfThe2, TakesakiTheoryOfOperator3, BlackadarOperatorAlgebras}, we deduce that the crossed-product von Neumann algebra of theorem \ref{th:inhomhklimit} is of type III for said range of $\beta_{\uL}$:

\begin{theo}
\label{th:inhomhktype}
Consider $\beta_{\uL}\in\ell^{p}(\D)$ with $0<p<\tfrac{1}{2}$, then $\theta:\Z_{\fr}\act (W_{\D,L}(\Z),\otimes_{d\in L\D_{\neq0}}m_{\beta_{d,\uL}})$ is nonsingular, free, ergodic and of type III. In particular, the von Neumann algebra $\overleftarrow{\scrM}_{\beta_{\uL},\D,L}$ is a hyperfinite type III factor.
\end{theo}

A family of geometrically motivated examples of $\beta_{\uL}\in\ell^{p}(\D)$ with $0<p<\tfrac{1}{2}$ is given by
\begin{align}
\label{eq:geometricbeta}
\beta^{\tau}_{d,\uL} & = (d'-d)^{\tau}, & \tau & > 2,
\end{align}
where $[d,d')$ is the largest standard dyadic interval starting at $d\in\D$, i.e.~$\beta^{\tau}_{d,\uL}=2^{-\tau n}$ for $d=\tfrac{m}{2^{N}}$ with $N\in\N_{0}$ and $m=1,...,2^{N}-1$ and $m\mod 2 \equiv 1$. Thus, $\log\beta^{\tau}_{d,\uL}$ is proportional to the level $N$ at which $d$ appears first as a vertex of $\gamma_{N}$.\\[0.25cm]
In contrast with the heat-kernel states, the inhomogeneous heat-kernel states satisfying the assumptions of theorem \ref{th:inhomhktype} lead to a nontrivial time-evolution group which commutes with the action of the gauge group, i.e.~we consider the modular flow,
\begin{align}
\label{eq:modulartimeev}
\sigma:\R & \act \overleftarrow{\scrM}_{\beta_{\uL},\D,L}, & \sigma^{\beta_{\uL}}_{t}(a) & = \Delta_{\beta_{\uL}}^{it} a\Delta_{\beta_{\uL}}^{-it},
\end{align}
where $\Delta_{\beta_{\uL}}$ is the modular operator of the vector corresponding to $\omega_{\beta_{\uL},\D,L}$.\\[0.25cm]
In view of theorem \ref{th:hklimit}, the following result indicates that an implementation of the action of Thompson's group $F$ is not immediate for the inhomogeneous heat-kernel states (even for $\tau>1$):

\begin{prop}
\label{prop:inhomhksingularF}
Consider $\beta^{\tau}_{d,\uL}$ for $\tau>1$, and let $\otimes_{d\in L\D_{\neq0}}m_{\beta^{\tau}_{d,\uL}}$ be the associated measure on $W_{\D,L}(\Z)$. Then the generalized Bernoulli action of Thompson's group $F$
\begin{align*}
\kappa: F & \act W_{\D,L}(\Z), & \kappa_{f} (x)(d) & = x(f^{-1} d), & f\in F, x&\in W_{\D,L}(\Z), d\in \D,
\end{align*}
is singular w.r.t.~the measure $\otimes_{d\in L\D_{\neq0}}m_{\beta^{\tau}_{d,\uL}}$. In particular, the Jones action $F\act \overleftarrow{\fA}_{\D,L}$ does not extends to an action on the von Neumann algebra $\overleftarrow{\scrM}_{\beta^{\tau}_{\uL},\D,L}$.
\end{prop}

\begin{remark}
\label{rem:generalJonesinhom}
In view of the last proposition, remark \ref{rem:generalJones} and the geometric nature of Jones' action $V\act\fA_{\D,L}$, we expect that the latter cannot be easily implemented for the inhomogeneous heat-kernel states even when $G$ is an abelian compact group. 
We expect encountering similar problems when $G$ is replaced by a nonabelian compact group or a more general object like a quantum group or a subfactor.
\end{remark}

On the contrary, if we consider only the subgroup of dyadic rotations $\Rot$ of Thompson's group $T$, we can get an implementation for any $\beta_{\uL}$ of the form $\eqref{eq:geometricbeta}$. More generally, we have:

\begin{corollary}
\label{cor:inhomhkRot}
Define $\ell:\D\to \R, d\mapsto -\log_2(d'-d)$, and consider $\beta_{\uL}:\D\longrightarrow\R_{>0}$ such that $\beta_{d,\uL}$ only depends on $\ell(d)$, i.e.~$\beta_{\uL}=b\circ\ell$ for some function $b$. We restrict the action of $T\act (W_{\D,L}(\Z),\otimes_{d\in L\D_{\neq0}}m_{\beta_{d,\uL}})$ to the rotation subgroup $\Rot$. Then the action $\Rot\act (W_{\D,L}(\Z),\otimes_{d\in L\D_{\neq0}}m_{\beta_{d,\uL}})$ is nonsingular and, thus, the restricted Jones action $\Rot\act \overleftarrow{\fA}_{\D,L}$ extends to an action by automorphisms on the von Neumann algebra $\overleftarrow{\scrM}_{\beta_{\uL},\D,L}$. More generally, the action $\Rot\act (W_{\D,L}(\Z), \otimes_{d\in L\D_{\neq0}} \nu_{d})$ is nonsingular if $\{\nu_{d}\}_{d\in\D}$, is any family of probability measures that are all mutually equivalent and such that $\nu_{d} = \nu_{d'}$ if $\ell(d)=\ell(d')$.
\end{corollary}

Regarding the implementation of the rotation subgroup $\Rot$, we also recall the following observation from our companion paper \cite[Remark 3.26]{BrothierConstructionsOfConformal}.

\begin{remark}
\label{rem:inhomhkgauge}
Although, it is natural to have an action of the rotation subgroup $\Rot\subset T$ on the field algebra $\overleftarrow{\scrM}_{\beta_{\uL},\D,L}$ whenever $\beta_{\uL}$ is a geometric function on $\D$, we expect that it is not possible to define the generator of rotations as a strong limit,
\begin{align*}
s-\lim_{n\rightarrow\infty}\tfrac{1}{2^{n}}(U_{r_{n}}-1)\xi = L_{0}\xi,
\end{align*}
as a consequence of gauge invariance, cf.~\cite{LoeffelholzMathematicalStructureOf}, see also \cite{JonesANoGo, BrothierPythagoreanRepresentationsOf}. If we interpreted the (spatial) local net of von Neumann algebras, $\{\overleftarrow{\scrM}_{\beta,\D,L}(\cS)\}$, as an analogue of a conformal field theory, cf.~corollary \ref{cor:localalgsys}, $L_{0}$ would correspond to the conformal Hamiltonian as opposed to the Kogut-Susskind Hamiltonian which motivates our definition of the heat-kernel states.\\ 
In a general setting of theories with local gauge invariance, $L_{0}$ might exist on the level of observables, i.e.~the gauge invariant subsector of our model. But, in the case of YM$_{1+1}$ this result in a topological theory without local degrees of freedom as explained in the next section, see also \cite{DimockCanonicalQuantizationOf}.
\end{remark}

The third class of semi-continuum von-Neumann field algebras results from the dual heat-kernel states $\omega_{\beta,\D,L}$. As explained in section \ref{sec:hkdual}, these states are defined by coherent collections of states $\omega^{(\gamma(t))}_{\{\beta_{t}\}} = \otimes_{\sigma\in\cP_{t}}\omega_{\beta_{t}(\sigma)}$  parametrized by a families of functions $\beta_{t}:\cP_{t}\rightarrow\R_{>0}$, $t\in\fT$. The coherence condition necessary to define the projective-limit state $\omega_{\beta,\D,L}$ entails $\beta_{t}(\sigma)=\beta|I_{\sigma}|$ for some fixed $\beta>0$ and $|I_{\sigma}|$ the length of the interval $I_{\sigma}$ for $\sigma\in\cP_{t}$. We recall that, by construction, the linear functional $\omega^{(\gamma(t))}_{\{\beta_{t}\}}$ will only define a state on $B(L^{2}(\scrA^{n(t)}))$ if the construction is based on a discrete group $G$. In case $G$ is a compact group, we are forced to restrict attention to the commutative subalgebra $C(\scrA^{n(t)})$ (identified with multiplication operators on $L^{2}(\scrA^{n(t)})$) -- unless we intend to resort to an abstract extension via the Hahn-Banach theorem.\\[0.1cm]
For compact $G$, the explicit formula for the state on $C(\scrA^{n(t)})$ reads:
\begin{align}
\label{eq:hkdualformula}
\omega^{(\gamma(t))}_{\{\beta_{t}\}}(M^{(\gamma(t))}(f)) & = \int_{G^{n(t)}}\left(\prod^{n(t)}_{j=1}dg_{\sigma_{j}}\rho_{\beta|I_{\sigma_{j}}|}(g_{\sigma_{j}})\right)f(g_{\sigma_{1}},...,g_{\sigma_{n(t)}}) = m_{\beta_{t}}(f),
\end{align}
where $f\in C(\scrA^{n(t)})$ and $|I_{\sigma_{j}}|=\sigma_{j}-\sigma_{j-1}$, $j=1,...,n(t)$. Considering this expression for the inductive (sub)system $\{C(\scrA^{E(\cev{\gamma}_{N})})\}_{N\in\N_{0}}$ (cp.~proposition \ref{prop:cofinaliso} and the preceding discussion) and using the holonomy map (see proposition \ref{prop:discretehol}), we find that these states agree with the restriction of the Wiener measure $\mu^{W}_{L,\beta}$ on $W_{\D,L}(G)$ to cylindrical functions $\Cyl(W_{\D,L}(G))\subset C(W_{\D,L}(G))$, cf.~\cite{DriverYangMillsTheory, TaylorMeasureTheory}:
\begin{align}
\label{eq:cylindricalwienerint}
\int_{W_{L}(G)}d\mu^{W}_{L,\beta}\!\ (f\circ\h^{-1})_{\cP_{t}} & = \int_{G^{n(t)}}\left(\prod^{n(t)}_{j=1}dh_{\sigma_{j}}\rho_{\beta|I_{\sigma_{j}}|}(h^{-1}_{\sigma_{j-1}}h_{\sigma_{j}})\right)(f\circ\h^{-1})(h_{\sigma_{1}},...,h_{\sigma_{n(t)}}) \\ \nonumber
& = \int_{\scrA^{n(t)}}\left(\prod^{n(t)}_{j=1}dg_{\sigma_{j}}\rho_{\beta|I_{\sigma_{j}}|}(g_{\sigma_{j}})\right)f(g_{\sigma_{1}},...,g_{\sigma_{n(t)}}) \\ \nonumber
& = \omega^{(\gamma(t))}_{\{\beta_{t}\}}(M^{(\gamma(t))}(f)),
\end{align}
where $(f\circ\h^{-1})_{\cP_{t}}(h) = (f\circ\h^{-1})(h_{\sigma_{n(t)}},...,h_{\sigma_{1}}), h\in W_{L}(G), f\in C(\scrA^{n(t)})$.\\[0.1cm]
To be more precise, the projective-limit state $\omega_{\beta,\D,L}$ is defined on the $C^{*}$-algebra $C(\overline{\scrA})$ which is isomorphic to $C(W_{\D,L}(G))$ via the holonomy map , and, by the Riesz-Markov-Kakutani theorem \cite{FollandRealAnalysis}, it corresponds to a unique Radon measure $m^{(\beta)}_{\overline{\scrA}}$ on the compact space $\overline{\scrA}$ which has $\mu^{W}_{L,\beta}$ as pushforward to the compact space $W_{\D,L}(G)$.\\
For non-compact $G$, we use $L^{\infty}(\scrA^{n(t)})$ instead of $C(\scrA^{n(t)})$ and consider the weak closure of the inductive limit of the system $\{L^{\infty}(\scrA^{n(t)})\}_{t\in\fT}$ w.r.t.~the dual heat-kernel state, which is an abelian von Neumann algebra. By construction this abelian von Neumann algebra is $L^{\infty}(\overline{\scrA},m^{(\beta)}_{\overline{\scrA}})$ for the inductive limit of probability measure spaces $\varinjlim_{t\in\fT}(\scrA^{n(t)},m_{\beta_{t}}) = (\overline{\scrA},m^{(\beta)}_{\overline{\scrA}})$, since dual heat-kernel state agrees with the state induced by $m^{(\beta)}_{\overline{\scrA}}$ on a weakly-dense subalgebra.\\[0.1cm]
If $G$ is compact and non-discrete, the measures $m^{(\beta)}_{\overline{\scrA}}$, $\beta>0$, can be distinguished from the uniform measure $m_{\overline{\scrA}}$ (see proposition \ref{prop:discretehol}) by the following result about their supports, cf.~\cite{DriverYangMillsTheory, MouraoPhysicalPropertiesOf}:

\begin{prop}
\label{prop:hkdualsupport}
For $\beta>0$, the supports of the measures $m^{(\beta)}_{\overline{\scrA}}$ and $m_{\overline{\scrA}}$, are disjoint. Precisely, $\h_{*}m_{\overline{\scrA}}$ is supported on those $h\in W_{\D,L}(G)$ that are nowhere continuous on $\D$, and $\h_{*}m^{(\beta)}_{\overline{\scrA}} = \mu^{W}_{L,\beta}$ is supported on $W_{L}(G)=\{h\in C(L\D,G)\ |\ h(0)=e\}$, i.e.~those $h\in W_{\D,L}(G)$ that are everywhere continuous on $\D$.
\end{prop}

In summary, we find that the commutative subalgebra of the semi-continuum von-Neumann field algebra corresponding to gauge-field configurations for the dual heat-kernel state $\omega_{\beta,\D,L}$ -- after restricting by its support projection $E$ \eqref{eq:hklimitsupport} -- is given by:
\begin{align}
\label{eq:hkduallimitalgcom}
E\pi_{\omega_{\beta,\D,L}}(C(\overline{\scrA}))E & \cong L^{\infty}(W_{\D,L}(G),d\mu^{W}_{L,\beta}).
\end{align}
To consider the dual heat-kernel state beyond the commutative part of $\fA_{\D,L}$, we analyze the cases when $G$ is discrete or compact but not discrete separately. As before, we restrict attention to the subalgebra $\overleftarrow{\fA}_{\D,L}\subset\fA_{\D,L}$ for simplicity.
Again, the analysis carries over to the full field algebra with minor modification, although, the dual heat-kernel state $\omega_{\beta,\D,L}$ is not invariant under $\zeta:\fA_{\D,L}\rightarrow\overleftarrow{\fA}_{\D,L}$ by an analogous argument as in remark \ref{rem:cofinalisostate}.\\[0.1cm]
In the first case, when $G$ is discrete, e.g.~$G=\Z$ as in section \ref{sec:hkdual}, we can extend the dual heat-kernel state $\omega_{\beta,\D,L}$ to the whole field algebra $\fA_{\D,L}$ by construction, cp.~\eqref{eq:hkdualcrossedprod}.  At level $N\in\N_{0}$, the explicit formula for the restriction of the dual-heat kernel state is:
\begin{align}
\label{eq:hkduallevelN}
\omega^{(\cev{\gamma}_{N})}_{\beta_{t_{N}}}(a) & = \sum_{g\in G^{\times 2^{N}}}\rho^{\otimes 2^{N}}_{\beta_{t_{N}}}(g)F_{a}(0,g) = \rho^{\otimes 2^{N}}_{\{\beta_{t_{N}}\}}(F_{a}(0)),
\end{align}
where $t_{N}\in\fT$ is the complete rooted binary tree with $2^{N}$ leaves, and $a = \sum_{h\in G^{\times 2^{N}}}F_{a}(h)\lambda^{(\cev{\gamma}_{N})}_{h}\in B(L^{2}(\scrA^{\cev{\gamma}_{N}}))$. Invoking again the holonomy map, \eqref{eq:hkduallevelN} is equivalent to
\begin{align}
\label{eq:hkduallevelNconv}
\omega^{(\cev{\gamma}_{N})}_{\{\beta_{t_{N}}\}}(a) & = \sum_{g\in G^{\times 2^{N}}}\left(\prod^{2^{N}}_{n=1}\rho_{\beta_{t_{N}}}(g^{-1}_{\sigma_{n-1}}g_{\sigma_{n}})\right)\tilde{F}_{a}(0,g) = \rho^{\ast 2^{N}}_{\beta_{t_{N}}}(\tilde{F}_{a}(0)),
\end{align}
where $\tilde{F}_{a}$ corresponds to $F_{a}$. The important difference is that \eqref{eq:hkduallevelNconv} refers to the construction of $\overleftarrow{\fA}_{\D,L}$ in terms of the conjugate action $L^{W}:\underline{G}\act W_{\D,L}(G)$.
Thus, we are led to the question whether the action $L^{W}:\underline{G}\act (W_{\D,L}(G),\mu^{W}_{L,\beta})$ is measure-class preserving. If quasi-invariance of the measure held for a given $G$, the limit von Neumann algebra would be the crossed-product algebra with the canonical state induced by $\mu^{W}_{L,\beta}$, see \cite{BratteliOperatorAlgebrasAnd1, KadisonFundamentalsOfThe2} for the general construction of crossed products:
\begin{align}
\label{eq:hkdualvNAdisc}
\overleftarrow{\scrM}_{\beta,\D,L} & \stackrel{?}{\cong} L^{\infty}(W_{\D,L}(G),d\mu^{W}_{L,\beta})\rtimes_{L^{W}} \underline{G}.
\end{align}
In the second case, when $G$ is compact but not discrete, a simple extension of the dual heat-kernel state to $\overleftarrow{\fA}_{\D,L}$ respectively $\fA_{\D,L}$ fails because the analogue of \eqref{eq:hkduallevelN} does not define a state ($\delta_{1_{G}}$ is not normalizable). Nevertheless, the action of $L^{W}:\underline{G}\act (W_{\D,L}(G),\mu^{W}_{L,\beta})$ could be measure-class preserving. Then, if we had measure-class preserving action, we could still form an (implemented) crossed-product von Neumann algebra similar to \eqref{eq:hkdualvNAdisc} in the following way:
\begin{align}
\label{eq:hkdualvNAcpt1}
\overleftarrow{\scrM}_{\beta,\D,L} & \stackrel{?}{=} L^{\infty}(\overline{\scrA},dm^{(\beta)}_{\overline{A}})\vee U^{\underline{L}}(\underline{G}),
\end{align}
where $U^{\underline{L}}(\underline{G})$ is the von Neumann algebra generated by the unitarized action of $\underline{G}\act L^{2}(\overline{\scrA},dm^{(\beta)}_{\overline{\scrA}})$. Moreover, we might use the Wiener measure $\mu^{W}_{L,\beta}$ to define a dual heat-kernel weight by:
\begin{align}
\label{eq:hkdualweight}
\omega_{\beta,\D,L}(a) & = \mu^{W}_{L,\beta}(F_{a}(1_{\underline{G}})),  
\end{align}
for $a = \sum_{\underline{h}\in\underline{G}}F_{a}(\underline{h})(L^{W}_{\underline{h}})^{*}$ with only finitely many non-zero $F_{a}(\underline{h})\in L^{\infty}(W_{L}(G),d\mu^{W}_{L,\beta})$. But, by applying Kakutani's theorem in a similar way as in our companion paper \cite{BrothierConstructionsOfConformal}, we obtain the following negative result:
\begin{theo}
\label{th:hkdualkakutani} 
For compact Lie groups $G$ and $G=\Z_{2}$, the action
\begin{align*}
L^{W} : \underline{G} & \act (W_{\D,L}(G),\mu^{W}_{L,\beta})
\end{align*}
is singular.
\begin{proof}
Since $\overline{\scrA}$ is a closed subset of the compact Hausdorff space $\bigtimes_{N\in\N_{0}}\scrA^{E(\gamma_{N})}$ (cp.~\cite[Section 2.2]{BrothierConstructionsOfConformal}), the coherence condition satisfied by the states \eqref{eq:hkduallevelN} implies that the infinite-product probability measure,
\begin{align}
\label{eq:infprodhkmeasure}
m^{(\beta)} & = \otimes_{N\in\N_{0}}\!\ \rho^{\otimes 2^{N}}_{\beta_{t_{N}}},
\end{align}
has a well-defined restriction to $\overline{\scrA}$ (of unit mass). Now, we apply Kakutani's theorem \cite{KakutaniOnEquivalenceOf} to $m^{(\beta)}$ and its pushforward by some $\underline{h}\in\underline{G}$ -- possibly by dropping a finite number of factors from \eqref{eq:infprodhkmeasure} up to the minimal level, $N_{\underline{h}}$, of $\underline{h}$. At level $N\geq N_{\underline{h}}$, the Radon-Nikodym derivative of $(\underline{L}_{\underline{h}})_{*}m^{(\beta)}$ relative to $m^{(\beta)}$ is given by:
\begin{align}
\label{eq:}
\frac{d(\underline{L}_{\underline{h}})_{*}m^{(\beta)}}{dm^{(\beta)}}_{|\scrA^{E(\gamma_{N})}}(g_{\sigma_{1}},...,g_{\sigma_{2^{N}}}) & = \prod^{2^{N}}_{n=1}\frac{\rho_{\beta_{t_{N}}}(p_{N}(\underline{h})_{\sigma_{n}}g_{\sigma_{n}})}{\rho_{\beta_{t_{N}}}(g_{\sigma_{n}})},
\end{align}
where $p_{N}(\underline{h})\in G^{E(\gamma_{N})}$ is the projection of $\underline{h}\in\underline{G}$ to level $N$. The equivalence of the measures can decided by the Hellinger distance,
\begin{align}
\label{eq:hkdualhellinger}
d_{H}\left((\underline{L}_{\underline{h}})_{*}m^{(\beta)},m^{(\beta)}\right) & = \prod_{N\geq N_{\underline{h}}}\int_{\scrA^{E(\gamma_{N})}}\frac{d(\underline{L}_{\underline{h}})_{*}m^{(\beta)}}{dm^{(\beta)}}_{|\scrA^{E(\gamma_{N})}}\!\ d\rho^{\otimes 2^{N}}_{\beta_{t_{N}}} \\ \nonumber
 & = \prod_{N\geq N_{\underline{h}}}\int_{\scrA^{E(\gamma_{N})}}\prod^{2^{N}}_{n=1}\sqrt{\rho_{\beta_{t_{N}}}(p_{N}(\underline{h})_{\sigma_{n}}g_{\sigma_{n}})\rho_{\beta_{t_{N}}}(g_{\sigma_{n}})}dg_{\sigma_{n}},
\end{align}
between $(\underline{L}_{\underline{h}})_{*}m^{(\beta)}$ and $m^{(\beta)}$.\\[0.1cm]
For a compact Lie group $G$, we can estimate \eqref{eq:hkdualhellinger} for $\underline{h}\neq 1_{\underline{G}}$ directly by the Cauchy-Schwarz inequality:
\begin{align}
\label{eq:hkdualhellingercpt}
d_{H}\left((\underline{L}_{\underline{h}})_{*}m^{(\beta)},m^{(\beta)}\right) & = \prod_{N\geq N_{\underline{h}}}\int_{G^{2^{N}}}\frac{d(\underline{L}_{\underline{h}})_{*}m^{(\beta)}}{dm^{(\beta)}}_{|N}\!\ d\rho^{\otimes 2^{N}}_{\beta_{t_{N}}} \\ \nonumber
 & = \prod_{N\geq N_{\underline{h}}}\int_{G^{2^{N}}}\prod^{2^{N}}_{n=1}\sqrt{\rho_{\beta_{t_{N}}}(p_{N}(\underline{h})_{\sigma_{n}}g_{\sigma_{n}})\rho_{\beta_{t_{N}}}(g_{\sigma_{n}})}dg_{\sigma_{n}} \\ \nonumber
 & \leq \prod_{N\geq N_{\underline{h}}}\prod_{\sigma\in\supp(\underline{h})}\left(\int_{G}\rho_{\beta_{t_{N}}}(p_{N}(\underline{h})_{\sigma}g_{\sigma})\rho_{\beta_{t_{N}}}(g_{\sigma})dg_{\sigma}\right)^{\frac{1}{2}} \\ \nonumber
 & = \prod_{N\geq N_{\underline{h}}}\prod_{\sigma\in\supp(\underline{h})}\sqrt{\rho_{2\beta_{t_{N}}}(p_{N}(\underline{h})_{\sigma})} = 0,
\end{align}
because $\lim_{\beta\rightarrow0}\rho_{\beta}(h) = 0$ for $h\neq 1_{G}$ by Urakawa's Poisson summation formula for $\rho_{\beta}$ \cite{UrakawaTheHeatEquation}, see also \cite{HallPhaseSpaceBounds}. From the second to the third line, we use the heat-kernel convolution identity.\\[0.1cm]
For $G=\Z_{2}=\{\pm1\}$, we have $\rho^{(\Z_{2})}_{\beta}(\pm 1) = 1\pm e^{-\frac{\beta}{2}}$, such that
\begin{align}
\label{eq:hkdualhellingerZ2}
d_{H}\left((\underline{L}_{\underline{h}})_{*}m^{(\beta)},m^{(\beta)}\right) & = \prod_{N\geq N_{\underline{h}}}(1-e^{-2^{-N}\beta})^{\frac{1}{2}|\supp(\underline{h})|} = \delta_{\underline{h},0},
\end{align}
because $\lim_{N\rightarrow\infty}(1-e^{-2^{-N}\beta}) = 0$.
\end{proof}
\end{theo}
\begin{remark}
\label{rem:hkdualkakutani}
The above argument applies when $G$ is a Lie group of compact type, i.e.~there is an orthogonal isomorphism $G\cong K\times\R^{n}$ w.r.t. a choice of $\Ad_{G}$-invariant inner product on the Lie algebra $\fg$ \cite{HallGeometricQuantizationAnd, HelgasonDifferentialGeometryLie}, as well. But the definition of the space $\overline{\scrA}$ requires slight modification, see section \ref{sec:wiener}. In the case of $G=\R^{d}$, we can compute \eqref{eq:hkdualhellinger} explicitly:
\begin{align}
\label{eq:hkdualhellingerR}
d_{H}\left((\underline{L}_{\underline{x}})_{*}m^{(\beta)},m^{(\beta)}\right) & = \prod_{N\geq N_{\underline{x}}}e^{-\frac{1}{8\beta_{t_{N}}}\sum^{2^{N}}_{n=1}||p_{N}(\underline{x})_{\sigma_{n}}||_{\R^{d}}^{2}} \\ \nonumber
 & = \prod_{N\geq N_{\underline{x}}}e^{-\frac{2^{N}}{8\beta}\sum^{2^{N}}_{n=1}||p_{N}(\underline{x})_{\sigma_{n}}||_{\R^{d}}^{2}} \\ \nonumber
 & = e^{-\frac{1}{8\beta}||\underline{x}||_{\underline{R}^{d}}^{2}\sum_{N\geq N_{\underline{x}}}2^{N}} = \delta_{\underline{x},0},
\end{align}
which implies the inequivalence of the measure for any $\underline{x}\neq0$.
\end{remark}
Interestingly, there is another way to proceed in the infinite compact case when $G$ is abelian: Because of \eqref{eq:finiterankaction}, we can identify the action $L^{W}:\underline{G}\act W_{\D,L}(G)$ with the left multiplication $G_{\fr}\act W_{\D,L}(G)$. Then, next to the action $G_{\fr}\act (W_{\D,L}(G),\mu^{W}_{L,\beta})$, there is another measure-class preserving left multiplication $H_{L}(G)\act W_{L}(G)$, where $H_{L}(G)\subset W_{L}(G)$ is the subgroup of finite-energy path in $G$ \cite{JostLocalQuantumTheory, ArakiFactorizableRepresentationOf, StreaterInfinitelyDivisibleRepresentations, AlbeverioTheEnergyRepresentation, AlbeverioIrreducibilityAndReducibility, AlbeverioFactorialRepresentationsOf, MalliavinIntegrationOnLoop, AlbeverioNoncommutativeDistributions}. $H_{L}(G)$ is the analogue of the Cameron-Martin subspace for $(W_{L}(G),d\mu^{W}_{L,\beta})$ \cite{DriverACameronMartin}. The action $H_{L}(G)\act W_{L}(G)$ is unitarizable which suggests to define the semi-continuum von-Neumann field algebra of the dual heat-kernel state in this case as:
\begin{align}
\label{eq:hkdualvNAcpt2}
\overleftarrow{\scrM}_{\beta,\D,L} & = L^{\infty}(W_{L}(G),\mu^{W}_{L,\beta})\vee U^{\uL}(H_{L}(G)).
\end{align}
Again, $U^{\uL}(H_{L}(G))$ denotes the von Neumann algebra generated by the unitary representatives of the left multiplication $H_{L}(G)\act W_{L}(G)$. 
\begin{remark}
\label{rem:hkdualvNAnonab}
The algebra \eqref{eq:hkdualvNAcpt2} can be defined in the nonabelian case as well, but the correspondence of $L^{W}$ with the left multiplication $L$ is not as direct as in the abelian case. More precisely, the formula for $L^{W}$ given in corollary \ref{cor:discretehol},
\begin{align*}
L^{W}_{\underline{g}}(h)_{\tau} & \stackrel{\tau=\sigma_{m}}{=} \left(\prod_{j=1}^{m}\alpha_{h_{\sigma_{j-1}}}(g_{\sigma_{j}})\right)h_{\tau},
\end{align*}
shows that $\underline{g}\in\underline{G}$ acts on $h\in W_{\D,L}(G)$ by left multiplication only after parallel transporting the components of the former to the starting point of the latter by the respective partial holonomy. Thus, if we intended to translate this formula into a continuous setting and make sense of \eqref{eq:hkdualvNAcpt2} simultaneously, we would need to interpret the product \textb{$\prod_{\sigma\leq\tau}$} as a multiplicative integral, see e.g.~\cite{McKeanStochasticIntegrals}. Then, because the integrand should be conjugated with a partial holonomy depending on its evaluation point, we infer that formally $L^{W}:\underline{G}\act W_{\D,L}(G)$ would correspond to an action by a subgroup of $K(G)\subset W_{L}(G)$ s.t. $K(G)$ is normalized into $H(G)$ by $W_{L}(G)$.\\
If $G$ was a compact Lie group and $\underline{g} = (g_{\sigma_{1}},...,g_{\sigma_{m}})$ was given by as a collection of exponentials of Lie algebra elements (electric fluxes) $g_{\sigma_{j}}=\exp_{G}(E_{\sigma_{j}})$, $j=1,...,m$, we could rewrite the formula for $L^{W}$ as a path-ordered exponential:
\begin{align*}
L^{W}_{\underline{g}}(h)_{\tau} & = \cP\exp_{G}\left(\int^{\tau}_{0}d\sigma\Ad_{h_{\sigma}}(E_{\sigma})\right)h_{\tau}.
\end{align*}
But, it is important to keep in mind that $E$ is not interpreted as a gauge field (connection).
\end{remark}

At this point, an in-depth analysis of the algebras \eqref{eq:hkdualvNAdisc}, \eqref{eq:hkdualvNAcpt1} and \eqref{eq:hkdualvNAcpt2} remains open, but there are partial results for semisimple compact Lie groups $G$ \cite{AlbeverioFactorialRepresentationsOf, AlbeverioFactorialityOfRepresentations}:\\[0.1cm]
The construction of $\mu^{W}_{L,\beta}$ is meaningful for $\Sigma=\R$ (the thermodynamic limit), which we denote by $\mu^{W}_{\beta}$ with associated spaces $H(G)$ and $W(G)$. The vector $1\in L^{2}(W(G),d\mu^{W}_{\beta})$ is cyclic for $U^{\uL}(H(G))$. For $G=SU(n)$, $n\geq2$, $U^{\uL}(H(G))$ is a type III factor and its commutant is generated by the unitary representatives of the right multiplication $W(G)\curvearrowleft H(G)$, i.e.~$U^{\uL}(H(G))' = U^{\uR}(H(G))$. Similar results, besides the type III property, hold when $\mu^{W}_{\beta}$ is conditioned on path subgroups such that $\Sigma=\R_{\geq0}$ or $\Sigma=\fS$ (the latter refers the the subgroup, $\scrL_{L}(G)\subset W_{L}(G)$, of loops based at $1_{G}$).\\[0.1cm]
We conclude with a remark on the locality structures and symmetries associated with the algebras \eqref{eq:hkduallimitalgcom} and \eqref{eq:hkdualvNAcpt2}.
\begin{remark}
\label{rem:hkduallocality}
The definition \ref{def:discretehol} of the holonomy map shows that $h\in W_{\D,L}(G)$ is a nonlocal expression of $\overline{g}\in\overline{\scrA}$. Conversely, $g\in\overline{\scrA}$ is a nonlocal, quadratic function of $h\in W_{\D,L}(G)$. In other words, the holonomy map does not preserve the locality structures of $W_{\D,L}(G)$ and $\overline{\scrA}$ associated with their respective inductive-limit constructions. For, example, we consider an element $f\in L^{\infty}(\overline{\scrA},dm^{(\beta)}_{\overline{\scrA}})$ which is localized inside a standard dyadic interval $[d,d']$ and associated with an adapted partition $\cP$:
\begin{align*}
f(\overline{g}) & = f(\{g_{\sigma}\}_{d<\sigma\leq d'\in\cP}).
\end{align*}
Its counterpart $\tilde{f}\in L^{\infty}(W_{\D,L}(G),d\mu^{W}_{L,\beta})$ is:
\begin{align*}
\tilde{f}(h) & = f(\{h^{-1}_{\textup{pre}(\sigma)}h_{\sigma}\}_{d<\sigma\leq d'\in\cP}),
\end{align*}
where $\textup{pre}(\sigma)$ is the predecessor of $\sigma$ in $\cP$. This has some similarity with the (anti-)locality structures of the Pauli algebra and the Fermion algebra which are related by the Jordan-Wigner transform \cite{EvansQuantumSymmetriesOn}.\\[0.1cm]
If $G$ is compact but not discrete, we know from proposition \ref{prop:hkdualsupport} that $\mu^{W}_{\D,L}$ is supported on continuous path from $h:[0,L]\rightarrow G$ with $h_{0}=1_{G}$. Therefore, the we may consider parallel transports associated with arbitrary subintervals $(\tau,\tau']\subset(0,L]$:
\begin{align*}
g_{(\tau,\tau']} & = h^{-1}_{\tau}h_{\tau'}.
\end{align*}
In this way, it is possible to describe operators $f\in L^{\infty}(\overline{\scrA},dm^{(\beta)}_{\overline{\scrA}})$ with arbitrary localization, i.e.~localized in $(\tau,\tau']\subset(0,L]$, directly.\\[0.1cm]
This observation indicates that we expect to have a strongly continuous unitary representation of the subgroup of dyadic rotations $\Rot\subset T$ w.r.t.~to the topology inherited from the inclusion $\Rot\subset\fS$. This is also intuitively justified because of the coherence condition required by the construction of the dual heat-kernel state:
\begin{align*}
\beta_{t_{N}} & = 2\beta_{t_{N+1}},
\end{align*}
which is rotation invariant at each level $N\in\N_{0}$. For general partitions $\cP$ this is equivalent to:
\begin{align*}
\beta : \cP & \longrightarrow \R_{>0}, \\ 
\sigma & \longmapsto \beta|I_{\sigma}|. 
\end{align*}
We return to the question of rotational invariance for the dual heat-kernel state in section \ref{sec:wiener}, where we argue that rotational invariance is also present for discrete $G$.\\[0.1cm]
In contrast, we do not expect to have a unitary representation of Thompson's group $F$ because a function $\beta:\cP\rightarrow\R_{>0}$ as above is not scale invariant. 
\end{remark}
\begin{remark}
\label{rem:noncptrep}
For a compact Lie group $G$, the representation of the YM$_{1+1}$ field algebra given by the dual heat-kernel state is related (via the It{\^ o} map) to the \textit{energy representation} of the Weyl algebra associated with the non-compact formulation of gauge theories, cp.~\cite{DimockCanonicalQuantizationOf, DriverYangMillsTheory, FrenkelOrbitalTheoryFor}.
\end{remark}

\subsection{Gauge transformations and observables}
\label{sec:gaugeobs}
Observables are identified as subalgebras inside the field algebras which are invariant under the action of gauge transformations, i.e.~the observables inside the semi-continuum field algebra $\fA_{\D,L}$ are given by the fixed-point algebra $\fA_{\D,L}^{\overline{G}}$ w.r.t.~the action $\alpha_{\overline{\tau}}:\overline{G}\rightarrow\Aut(\fA_{\D,L})$, see corollary \ref{cor:algsys} (an analogous statement holds for local observables \eqref{eq:localsys}). But, there is some freedom regarding the action $\alpha_{\overline{\tau}}$, because we may restrict the latter to subgroups of $\overline{G}$ depending on the choice of boundary conditions as well as regularity requirements on the gauge transformations. For the semi-comntinuum von-Neumann field algebra $\scrM_{\D,L}$ and its local subalgebras, it is necessary to require an extension of $\alpha_{\overline{\tau}}$. Such an extension will exist, if the state $\omega_{\D,L}$ is invariant under $\overline{G}$, and will be given by a unitary representation $U_{\overline{\tau}}:\overline{G}\rightarrow\cU(\cH_{\D,L})$. But, as exemplified by the dual heat-kernel state, invariance of $\omega_{\D,L}$ is not necessary to have an action of (a subgroup of) gauge transformation on $\scrM_{\D,L}$.\\[0.1cm]
Since it is known that YM$_{1+1}$ is expected to be a topological theory \cite{WittenOnQuantumGauge, DimockCanonicalQuantizationOf, DriverYangMillsTheory, CordesLecturesOn2D}, i.e.~there are no local degrees of freedom at the level of observables, let us illustrate how this is reflected by the operator-algebraic construction. According to \eqref{def:phasespacegauge} and \eqref{eq:gaugeconv}, the action of gauge transformations on basic field operators \eqref{eq:basickernels} at level $N\in\N_{0}$ reduces to:
\begin{align}
\label{eq:basicgauge}
(\tau(g^{-1})^{*}f)(h) & = f(\{g^{-1}_{\partial e_{+}}h_{e}g_{\partial e_{-}}\}_{e\in E(\gamma_{N})}), \\ \nonumber
\Ad_{U_{\tau(g)}}(\lambda_{h}) & = \lambda_{\{g_{\partial e_{\pm}}\}_{e\in E(\gamma_{N})}}\lambda_{h}\lambda_{\{g_{\partial e_{\pm}}\}_{e\in E(\gamma_{N})}}^{*},
\end{align}
for $f\in C(G^{E(\gamma_{N})})$, $h\in G^{E(\gamma_{N})}$ and $g\in G^{V(\gamma_{N})}$ (we suppress the labels of $\tau$ and $\lambda$ referring to the level).\\[0.1cm]
Since the lattice $\gamma_{N}$ defines a partition of the spatial manifold in 1+1 dimensions, invariant elements $f\in C(G^{E(\gamma_{N})})$ for the first equation are found by using the holonomy map: Definition \ref{def:discretehol} and corollary \ref{cor:discretehol} show that $f$ will be invariant under gauge transformations that vanish at $0$ and $L$ (the based loop group $\scrL_{N,L}(G)$ at level $N$) if it factorizes through the (discrete) holonomy form $0$ to $L$, see also remark \ref{rem:discretehol}:
\begin{align}
\label{eq:wilsonloop1}
f = \tilde{f}\circ\h_{L}.
\end{align}
For periodic boundary conditions $\Sigma = \T^{1}_{L}$, $f$ will be invariant under all gauge transformations at level $N$, if it factorizes to a class function on $G$:
\begin{align}
\label{eq:wilsonloop2}
\tilde{f}\circ\alpha_{g} & = \tilde{f}, & g\in G_{L}  = G.
\end{align}
Choosing $\tilde{f} = \tr_{V_{\pi}}\circ\pi$ with $\pi:G\rightarrow\textup{GL}(V_{\pi})$ the defining representation of $G$, results in the most prominent representative of these observables -- the \textit{Wilson-Wegner loop} \cite{WegnerDualityInGeneralized, WilsonConfinement}. For a compact Lie group $G$, we deduce the completeness of this type of observables from the continuity of the action $\tau:G\act\overline{\scrA}$, cp.~\cite{BaezSpinNetworksIn, AshtekarDifferentialGeometryOn} and also \cite[Section 3]{DimockCanonicalQuantizationOf}: At level $N$, the continuous functions on quotient spaces $G^{E(\gamma_{N})}/\scrL_{N,L}(G)$ and $G^{E(\gamma_{N})}/G^{V(\gamma_{N})}$ are given by the continuous functions on $G$ respectively a maximal torus $T\subset G$ invariant under the Weyl group $W = N_{G}(T)/T$. This implies the triviality of the inductive systems, i.e $\varinjlim_{N\in\N_{0}}C(G^{E(\gamma_{N})}/\scrL_{N,L}(G)) = C(G)$ and $\varinjlim_{N\in\N_{0}}C(G^{E(\gamma_{N})}/G^{V(\gamma_{N})}) = C(T/W)$. But, because of the continuity of the action $\tau$, we have $C(\overline{\scrA})^{\scrL_{N,L}(G)} = C(G)$ and $C(\overline{\scrA})^{\overline{G}} = C(T/W)$.\\[0.1cm]
Sticking to compact Lie groups at first, we analyze the implications of the second equation of \eqref{eq:basicgauge} in terms the generators of the left translation operators, i.e.~the Lie algebra $\fg$. Since $\fg$ consist of right invariant vector fields in our construction, the second equation simply expresses the fact that we are looking for bi-invariant differential operators on $G^{E(\gamma_{N})}$ that induce invariant convolution kernels (for the action of $G^{V(\gamma_{N})}$). Now, bi-invariant differential operators on $G$ correspond to the center $Z(\fg)$ of the universal enveloping algebra $U(\fg)$ \cite{VaradarajanAnIntroductionTo}. Moreover, if a direct sum of bi-invariant differential operators, $\oplus_{e\in E(\gamma_{N})}\partial_{e}$, acts on a smooth function $f\in C^{\infty}(G^{E(\gamma_{N})})$ that factorizes as in \eqref{eq:wilsonloop1}, it will correspond to the action of the single bi-invariant differential operator $\sum_{e\in E(\gamma_{N})}\partial_{e}$ acting on $\tilde{f}\in C^{\infty}(G_{L})$. This simply expresses the fact that a bi-invariant differential operator can be constantly transported to the endpoint of the holonomy. Therefore, the action of $Z(\oplus_{e\in E(\gamma_{N})}\fg_{e})$ degenerates on the $*$-subalgebra of $C^{\infty}(G^{E(\gamma_{N})})$ satisfying \eqref{eq:wilsonloop1}. For semisimple $\fg$, $Z(\fg)$ is spanned by the Casimir elements \cite{RacahGroupTheoryAnd}. If we restrict the action of the gauge group to $\scrL_{N,L}(G)$, we are free to specify non-central elements of $U(\fg)$ at $0$ and $L$.\\[0.1cm]
A special observable at each level $N$ is the Hamiltonian of YM$_{1+1}$ corresponding to the quadratic Casimir element. Thus, the existence of dynamics for the semi-continuum von-Neumann field algebra is tied to the question whether a suitable limit of the sequence quadratic Casimir elements can be defined admitting a self-adjoint extension. Theorem \ref{th:KShamiltonianstrong} shows that this is possible in the strong-coupling vacuum representation. Moreover, the restriction to observables is also well-defined because of gauge invariance of the limit operator.\\[0.1cm]
Beyond the Lie group setting, the second equation implies that observables involve only central convolution operators at each finite level $N\in\N_{0}$. In analogy with bi-invariant differential operators, the action of the latter degenerates on the $*$-subalgebra of $C^{\infty}(G^{E(\gamma_{N})})$ satisfying \eqref{eq:wilsonloop1}.\\[0.1cm]
Let also comment on the various states introduced above for the case of periodic boundary conditions and the action of $\overline{G}$. We observe that the gauge-invariant (inhomogeneous) heat-kernel states restrict to the observables. For the homogeneous heat-kernel state the restriction is given by the level-$1$ member of the coherent family:
\begin{align}
\label{eq:invstaterestriction1}
\omega_{\beta,\D,L|\fA_{\D,L}^{\overline{G}}} & = \omega^{(\gamma_{1})}_{\beta} = \omega_{\beta}\otimes\omega_{\beta}.
\end{align}
Similalrly, the inhomogeneous heat-kernel state reduces to the level-$1$ member of the coherent family:
\begin{align}
\label{eq:invstaterestriction2}
\omega_{\beta,\D,L|\fA_{\D,L}^{\overline{G}}} & =  \omega^{(\gamma_{1})}_{\beta_{0,\uL},\beta_{1,\uR}} = \omega_{\beta_{0,\uL}}\otimes\omega_{\beta_{1,\uR}}.
\end{align}
The occurrence of a two-factor state reflects the left-right structure of the cofinal sequence $\{\gamma_{N}\}_{N\in\N_{0}}$ of lattices, see also proposition \ref{prop:tensoriso}. A similar statement holds w.r.t.~the restricted gauge group $\scrL_{N,L}(G)$
\begin{remark}
\label{rem:leftrightobservables}
If we take into account the invariance of characters under the group inversion, we can also lift left-right ambiguity for the inhomogeneous heat-kernel states when passing to the observables (for the action of $\overline{G}$). More precisely, by the Peter-Weyl theorem the characters form an orthonormal basis of the $L^{2}$-class functions and, thus, any continuous class function $f$ admits an $L^{2}$-expansion invariant under inversion. It follows that $f$ and $f\circ(\ .\ )^{-1}$ belong to the same $L^{2}$-equivalence class and, therefore, only differ on a null set for the Haar measure. This implies $f = f\circ(\ .\ )^{-1}$. Since the basic observables correspond to class functions and convolution operators associated with bi-invariant differential operators, we deduce that we can naturally identify the left and right parts of the observable algebras, cp.~proposition \ref{prop:orient}. This in turn, enforces $\beta_{0,\uL} = \beta_{1,\uR}$.
\end{remark}
Concerning the dual heat-kernel state, the restriction to observables is less obvious because the state itself is not invariant under gauge group. Nevertheless, a partial results can be deduced from the discussion in the previous section: It is possible to define a representation of a restricted group of gauge transformations $H_{L}(G)$, the finite-energy paths, on the GNS representation of the commutative algebra $C(\overline{\scrA})$ using the theory of multiplicative noncommutative distributions \cite{AlbeverioNoncommutativeDistributions} for abelian and compact Lie groups $G$. Unfortunately, this representation does not admit non-trivial invariant vectors \cite{DriverTheEnergyRepresentation}, and, therefore, there is no non-trivial gauge-invariant subspace inside the GNS representation space to which we can restrict the observables.

\section{A renormalization group perspective}
\label{sec:rgp}
A conceptually appealing way to understand the constructions and results of section \ref{sec:ym11} (and to some extent those of section \ref{sec:basics}) is in terms of the Wilson-Kadanoff renormalization group \cite{KadanoffScalingLawsFor, WilsonTheRenormalizationGroupKondo, WegnerCorrectionsToScaling}. This is expected form the point of view of physics as, by construction, renormalization group theory is ideally suited to analyze continuum (or scaling) limits of lattice, i.e.~effective, models. To this end, we recall the basic setup of renormalization group theory and, afterwards, reformulate it to fit with our approach.\\[0.1cm]
Let us consider a sequence of Hamiltonian lattice models $\{\fA_{N},\cH_{N},H^{(N)}_{0}\}_{N\in\N_{0}}$. At each level $N$, $\fA_{N}$ is a $C^{*}$-algebra of basic field operators, $\cH_{N}$ is a Hilbert space on which the latter act, and $H^{(N)}_{0}$ is a Hamiltonian affiliated with $B(\cH_{N})$. The fundamental equation of renormalization group theory implicitly defines further Hamiltonians $H^{(N)}_{N'-N}$ for $N<N'$ by assuming the equivalence of the partition functions of $H^{(N)}_{N'-N}$ and $H^{(N')}$ (see, for example, Fisher's review in \cite[p. 89-135]{CaoConceptualFoundationsOf}):
\begin{align}
\label{eq:basicrg}
Z^{(N')}_{0} = \Tr_{N'}(e^{-H^{(N')}_{0}}) & = \Tr_{N}(e^{-H^{(N)}_{N'-N}}) = Z^{(N)}_{N'-N},
\end{align}
assuming that the canonical Gibbs states of $H^{(N')}_{0},H^{(N)}_{N'-N}$ are well-defined. In other words, there are transformations (or quantum operations) $\cE^{(N)}_{N'-N}:\cS^{1}_{N'}\rightarrow\cS^{1}_{N}$ on density matrices such that:
\begin{align}
\label{eq:densityrg}
\cE^{(N)}_{N'-N}(e^{-H^{(N')}_{0}}) & = e^{-H^{(N)}_{N'-N}}.
\end{align}
On the level of Hamiltonians, we have:
\begin{align}
\label{eq:hamiltonianrg}
\cR^{(N)}_{N'-N}(H^{(N')}_{0}) & = H^{(N)}_{N'-N}.
\end{align}
Since the choice of levels $N<N'$ is arbitrary, we naturally require the semi-group property:
\begin{align}
\label{eq:semigrouprg}
\cE^{(N)}_{N'-N}\circ\cE^{(N')}_{N''-N'} & = \cE^{(N)}_{N''-N}, & \textup{or} & & \cR^{(N)}_{N'-N}\circ\cR^{(N')}_{N''-N'} & = \cR^{(N)}_{N''-N}.
\end{align}
Pictorially, this is summarized by Wilson's \textit{triangle of renormalization} \cite[p. 790]{WilsonTheRenormalizationGroupKondo},
\begin{align}
\label{eq:trianglerg}
\begin{tikzpicture}
	\draw (0,0) node{$H^{(N)}_{0}$} (2,0) node{$H^{(N)}_{1}$} (4,0) node{$H^{(N)}_{2}$} (6,0) node{$\dots$};
	\draw[->] (0.75,0) to (1.25,0);
	\draw[->] (2.75,0) to (3.25,0);
	\draw[->] (4.75,0) to (5.25,0);
	\draw (2,2) node{$H^{(N+1)}_{0}$} (4,2) node{$H^{(N+1)}_{1}$} (6,2) node{$\dots$};
	\draw[->] (2.75,2) to (3.25,2);
	\draw[->] (4.75,2) to (5.25,2);
	\draw (2,1) node[left]{$\cR^{(N)}_{1}$} (4,1) node[left]{$\cR	^{(N)}_{2}$};
	\draw[->] (2,1.5) to (2,0.5);
	\draw[->] (4,1.5) to (4,0.5);
	\draw (4,4) node{$H^{(N+2)}_{0}$} (6,4) node{$\dots$};
	\draw[->] (4.75,4) to (5.25,4);
	\draw (4,3) node[left]{$\cR^{(N+1)}_{1}$};
	\draw[->] (4,3.5) to (4,2.5);
\end{tikzpicture}
,
\end{align}
and we aim at constructing the limit sequence of Hamiltonians $\{H^{(N)}_{\infty}\}_{N\in\N_{0}}$ (with a similar picture for the density matrices).\\[0.1cm]
Next, let us translate \eqref{eq:basicrg} to \eqref{eq:trianglerg} into statements involving the algebras $\{\fA_{N}\}_{N\in\N_{0}}$ and (algebraic) states on them. Using \eqref{eq:basicrg} and \eqref{eq:densityrg}, we find for $N<N'$:
\begin{align}
\label{eq:densityrgstate}
\tfrac{1}{Z^{(N)}_{N'-N}}\Tr_{N}\left(e^{-H^{(N)}_{N'-N}}a_{N}\right) & = \tfrac{1}{Z^{(N)}_{N'-N}}\Tr_{N}\left(\cE^{(N)}_{N'-N}(e^{-H^{(N')}_{0}})a_{N}\right) \\
 & = \tfrac{1}{Z^{(N')}_{0}}\Tr_{N'}\left(e^{H^{(N')}_{0}}\alpha^{N}_{N'}(a_{N})\right),
\end{align}
where $\alpha^{N}_{N'}:\fA_{N}\rightarrow\fA_{N'}$ is assumed to be the dual of $\cE^{(N)}_{N'-N}$, and $a_{N}\in\fA_{N}$. This generalizes to:
\begin{align}
\label{eq:staterg}
\omega^{(N')}_{0}\circ\alpha^{N}_{N'} & = \omega^{(N)}_{N'-N},
\end{align}
for a sequence of (initial) states $\{\omega^{(N)}_{0}\}_{N\in\N_{0}}$. A natural requirement is that $\alpha^{N}_{N'}$ is unital and completely positive (ucp) because it should map states into states and preserve probability, i.e.~the partition function \eqref{eq:basicrg}. We call the family of maps $\{\alpha^{N}_{N'}\}_{N<N'\in\N_{0}}$ the \textit{renormalization group}. In this way, our reformulation of the Wilson-Kadanoff renormalization group could be understood as being dual to the scheme of the density-matrix renormalization group (DMRG) \cite{SchollwoeckTheDensityMatrix}. The picture corresponding to \eqref{eq:trianglerg} is:
\begin{align}
\label{eq:statetrianglerg}
\begin{tikzpicture}
	\draw (0,0) node{$\omega^{(N)}_{0}$} (2,0) node{$\omega^{(N)}_{1}$} (4,0) node{$\omega^{(N)}_{2}$} (6,0) node{$\dots$};
	\draw[->] (0.75,0) to (1.25,0);
	\draw[->] (2.75,0) to (3.25,0);
	\draw[->] (4.75,0) to (5.25,0);
	\draw (2,2) node{$\omega^{(N+1)}_{0}$} (4,2) node{$\omega^{(N+1)}_{1}$} (6,2) node{$\dots$};
	\draw[->] (2.75,2) to (3.25,2);
	\draw[->] (4.75,2) to (5.25,2);
	\draw (2,1) node[left]{$\alpha^{N}_{N+1}$} (4,1) node[left]{$\alpha	^{N}_{N+2}$};
	\draw[<-] (2,1.5) to (2,0.5);
	\draw[<-] (4,1.5) to (4,0.5);
	\draw (4,4) node{$\omega^{(N+2)}_{0}$} (6,4) node{$\dots$};
	\draw[->] (4.75,4) to (5.25,4);
	\draw (4,3) node[left]{$\alpha^{N+1}_{N+2}$};
	\draw[<-] (4,3.5) to (4,2.5);
\end{tikzpicture}
\end{align}
and we are interested in the existence of sequence of limit states $\{\omega^{(N)}_{\infty}\}_{N\in\N_{0}}$. Moreover, we expect that the limit states satisfy the (projective) consistency property,
\begin{align}
\label{eq:staterglimit}
\omega^{(N')}_{\infty}\circ\alpha^{N}_{N'} & = \omega^{(N)}_{\infty},
\end{align}
because formally $\alpha^{N'}_{\infty}\circ\alpha^{N}_{N'} = \alpha^{N}_{\infty}$ for $\alpha^{N}_{\infty} = \lim_{N'\rightarrow\infty}\alpha^{N}_{N'}$. If it exists, the (projective) limit state,
\begin{align}
\label{eq:projstaterg}
\omega^{(\infty)}_{\infty} & = \varprojlim_{N\in\N_{0}}\omega^{(N)}_{\infty},
\end{align}
will be called the scaling limit of $\{\omega^{(N)}_{0}\}_{N\in\N_{0}}$. We emphasize at this point that we do not expect any uniqueness of the scaling limit unless, maybe, the initial states are on a stable manifold (with unique fixed point) of the renormalization group flow (critical couplings). This point is exemplified by the discussion of the dual heat-kernel state and its relation to the Wiener measure below (section \ref{sec:wiener}). Clearly, a necessary condition for the existence of a non-trivial sequence of limit states $\{\omega^{(N)}_{\infty}\}_{N\in\N_{0}}$ (in the sense of physical interactions) is the divergence (in terms of the lattice length) of the sequence of correlation length $\{\xi^{(N)}_{N'}\}_{N'\in\N_{0}}$ between localized operators at any level $N'\in\N_{0}$. In other words, the sequence $\{\omega^{(N)}_{0}\}_{N\in\N_{0}}$ approaches a (quantum) critical point, see \cite{SachdevQuantumPhaseTransitions} and also \cite[Chapter 4]{FernandezRandomWalksCritical} for a discussion of scaling limits and critical points in a commutative probabilistic setting.\\[0.1cm]
A simple example of quantum operations $\{\cE^{(N)}_{N'-N}\}_{N<N'\in\N_{0}}$ are partial traces $\Tr_{N'\rightarrow N}$ for $\cH_{N} = \cH_{0}^{\otimes 2^{N}}$ and $\fA_{N} = B(\cH_{N})$ such that $\alpha^{N}_{N'}(a_{N}) = a_{N}\otimes\mathds{1}_{N'\setminus N}$. There is a natural extension of the latter that we use above \eqref{eq:algop} and in \cite{BrothierConstructionsOfConformal} which is also natural in the context of the multi-entanglement renormalization ansatz \cite{MilstedQuantumYangMills}, see also equations \eqref{eq:MERAisometry} to \eqref{eq:MERAadjoint}:
\begin{align}
\label{eq:twistedpartialtrace}
\alpha^{N}_{N'}(a_{N}) & = U_{N'}(a_{N}\otimes\mathds{1}_{N'\setminus N})U_{N'}^{*},	
\end{align}
for some unitary $U_{N'}\in\cU(\cH_{N'})$ -- the disentangler. This suggest that we should consider two renormalization group transformations equivalent if
\begin{align}
\label{eq:equivalencerg}
\alpha^{N}_{N'} \sim \tilde{\alpha}^{N}_{N'} & \Leftrightarrow \alpha^{N}_{N'} = \Ad_{U_{1}\otimes ... \otimes U_{2^{N'}}} \tilde{\alpha}^{N}_{N'},
\end{align}
where $U_{j}\in\cU(\cH_{0})$, $j=1,...,2^{N'}$, solely acts on the $j$th factor of $\cH_{N'}$. This equivalence is reminiscent of the equivalence of quantum states consider in \cite{VerstraeteRenormalizationGroupTransformations}. It is clear that this notion of equivalence does not affect the locality structure defined by a renormalization group $\{\alpha^{N}_{N'}\}_{N<N'\in\N_{0}}$, cp.~remark \ref{rem:yangmillslocal}. In view of the discussion preceding \eqref{eq:MERAadjoint}, we also note that implementing a quantum operation $\cE^{(N)}_{N'-N}$ by a cut down with a proper isometry $R^{N}_{N'}:\cH_{N}\rightarrow\cH_{N'}$, $R^{N}_{N'}(R^{N}_{N'})^{*}\neq\mathds{1}_{N'}$, i.e.
\begin{align}
\Tr_{N'}(e^{-H^{(N')}_{0}}R^{N}_{N'}a_{N}(R^{N}_{N'})^{*}) & = \Tr_{N}(e^{-H^{(N)}_{N'-N}}a_{N}),
\end{align}
is not a valid option in the sense of the fundamental equation \eqref{eq:basicrg}, unless we are in an exceptional situation, for example, $e^{-H^{(N')}_{0}} = R^{N}_{N'}e^{-H^{(N)}_{0}}(R^{N}_{N'})^{*}$, cp.~\cite{JonesScaleInvariantTransfer} as well as \cite{LangHamiltonianRenormalizationI}. In other words, $\Ad_{R^{N}_{N'}}$ is not permitted as an element of a renormalization group because it is not unital. \\[0.1cm]
Let us also briefly comment on the question of dynamics in this formulation of the renormalization group: At first sight, the scheme \eqref{eq:statetrianglerg} is purely kinematical because we think of all states as defined on algebras of time-zero fields. But, if the initial family of states $\{\omega^{(N)}_{0}\}_{N\in\N_{0}}$ comes with a family of time evolution automorphisms groups $\{\alpha^{(N)}_{t}, t\in\R\}_{N\in\N_{0}}$ preserving the initial state at each level $N\in\N_{0}$, we can consider the unitary implementers $\{U^{N}_{t}, t\in\R\}$. If we assume that these implementers belong to $\fA_{N}$, for example, if $\fA_{N}=B(\cH_{N})$, we can ask for the existence of the time evolution automorphism group $\{\alpha^{(\infty)}_{t}, t\in\R\}$ in the GNS representation of the scaling limit $\omega^{(\infty)}_{\infty}$ (in the pointwise weak sense locally uniformly in $t\in\R$):
\begin{align}
\label{eq:dynamicsrg}
\lim_{N\rightarrow\infty}\pi_{\omega^{(\infty)}_{\infty}}(U^{(N)}_{t})\!\ a\!\ \pi_{\omega^{(\infty)}_{\infty}}(U^{(N)}_{t})^{*} & = \alpha^{(\infty)}_{t}(a), & a&\in\pi_{\omega^{(\infty)}_{\infty}}(\varinjlim_{N\in\N_{0}}\fA_{N})'',
\end{align}
where we consider the $U^{(N)}_{t}\in\fA_{N}\subset\varinjlim_{N\in\N_{0}}\fA_{N}$ as elements of a potential inductive limit defined by the renormalization group $\{\alpha^{N}_{N'}\}_{N<N'\in\N_{0}}$. Theorem \ref{th:KShamiltonianstrong} shows that such a limit exists for the scaling limit of strong-coupling vacua in YM$_{1+1}$. In view of remark \ref{rem:finiteprop}, we can define lattice analogues of field algebras localized in spacetime in such a case and analyze their locality properties in the scaling limit.\\[0.1cm]
Returning to the (inhomogeneous) heat-kernel and dual heat-kernel states of section \ref{sec:can11}, we interpret the corresponding coherent families of states as sequences of limit states $\{\omega^{(N)}_{\infty}\}_{N\in\N_{0}}$ according to \eqref{eq:projstaterg} and in view of the renormalization group envisioned by \eqref{eq:statetrianglerg}. An advantage of this point of view is that it tells us for which families of states $\{\omega^{(\gamma_{N})}\}_{N\in\N_{0}}$ we may expect semi-continuum von-Neumann field algebras that admit continuous extensions of the discrete symmetries of the lattice models (symmetry enhancement, see e.g.~\cite{FroehlichMasslessPhasesAnd}), e.g.~an extension of the dyadic translations $\Rot\subset T$ to continuous translations $\fS$. Namely, the states $\{\omega^{(\gamma_{N})}\}_{N\in\N_{0}}$ should define a scaling limit of (ground) states of physical systems approaching a (quantum) critical point to have a non-trivial continuum theory (in the sense of interactions).\\[0.1cm]
Reasoning along these line leads to the conclusion that the homogeneous and inhomogeneous heat-kernel states lie on a fixed-point manifold for the renormalization group flow because each choice of functions $\beta_{\uL},\beta_{\uR}:\D\rightarrow\R_{>0}$ -- corresponding to family of initial states -- directly defines a scaling limit in the sense of \eqref{eq:projstaterg}. Thompson's groups $F\subset T\subset V$ act on this fixed-point manifold via the pullback of the Jones' action on $\fA_{\D,L}$, see section \ref{sec:thompson}, and the homogeneous heat-kernel states ($\beta_{\uL},\beta_{\uR} = \textup{const.}$) form a subset of fixed-points for this action, cp.~theorem \ref{th:hklimit}. Since these states are gauge invariant and the observables of YM$_{1+1}$ do not describe local degrees of freedom (see section \ref{sec:gaugeobs}), we do not expect to have symmetry enhancement in their GNS representations. In the terminology of \cite{FernandezRandomWalksCritical}, the topological nature of YM$_{1+1}$ leads to ``boring'' scaling limits (ultra-local interactions). For $G=U(1)$, we know that for a large class of functions $\beta_{\uL},\beta_{\uR}$ the Jones' action of Thompson's group $F$ does not even extend to the semi-continuum von-Neumann field algebra (proposition \ref{prop:inhomhksingularF}). Although, such an extension is possible for the rotation subgroup $\Rot\subset T$ (corollary \ref{cor:inhomhkRot}).\\[0.1cm]
For the dual heat-kernel states the situation is rather different. By construction these states are defined on $\fA_{\D,L}$ via an extension from a commutative subalgebra, e.g.~$C(\overline{\scrA})$ in the case of a compact group, which is only fully explicit in the case of a discrete group, see the discussion of these state in section \ref{sec:timezero}. Therefore, we focus on the restriction of the dual heat-kernel states to the commutative subalgebra and illustrate the associated renormalization group flow in the following subsection.

\subsection{Scaling limits of the dual heat-kernel states}
\label{sec:wiener}
To analyze the scaling limit of the dual heat-kernel states, let us initially assume that $G$ is a Lie group of compact type, see remark \ref{rem:hkdualkakutani}. According to \eqref{eq:hkdualformula} these states are defined by families of states $\{\omega^{\gamma(t)}_{\beta_{t}}\}_{t\in\fT}$ on the inductive system $\{C(\mathring{\scrA}^{E(\gamma(t))}), \alpha^{t'}_{t}\}_{t\leq t'\in\fT}$ indexed by functions $\beta:\fT\rightarrow\R_{>0}$. Here, we denote by $\mathring{\scrA}^{E(\gamma(t))}$ the compactification of $\scrA^{E(\gamma(t))}$ resulting from separately compactifying each factor $\R$. To apply the renormalization group argument in this situation, we choose an arbitrary family of states $\{\omega^{\gamma(t)}_{\beta_{t}}\}_{t\in\fT}$ as our initial states without satisfying the coherence condition,
\begin{align}
\label{eq:wienercons}
\omega^{\gamma(t')}_{\beta_{t'}}\circ\alpha^{t}_{t'} & = \omega^{\gamma(t)}_{\beta_{t}},
\end{align}
for $t\leq t'$. On the contrary, the latter equation is replaced by a similar looking equation,
\begin{align}
\label{eq:wienerrg1}
\omega^{\gamma(t')}_{\beta_{t'}}\circ\alpha^{t}_{t'} & = \omega^{\gamma(t)}_{\beta'_{t}},
\end{align}
 as the action of the renormalization group, where the state on the left hand side belongs to our initial family of states while the state on the right hand is defined by the action of $\alpha^{t}_{t'}$ resulting in a new function $\beta'$. For complete binary trees, $t = t_{N}\leq t'=t_{N'}$, the relation between $\beta'_{t_{N}}$ and $\beta_{t_{N'}}$ is the analogue of the consistency equation states in remark \ref{rem:hkduallocality}:
 \begin{align}
 \label{eq:wienerrg2}
 \beta'_{t_{N}} = 2^{N'-N}\beta_{t_{N'}}.
 \end{align}
Clearly, the simplicity of this renormalization group equation reflects the simplicity of the model. Experience with the renormalization group formalism in other context makes it evident that it will not even be possible to restrict attention to a single class of initial states for more complicated models because the renormalization group flow will generate various interaction terms not present in this initial class (in the case of Gibbs states). As a well-known classical example of this fact, we may think of the quantum Ising chain with transverse field, see e.g.~\cite{SuzukiQuantumIsingPhases, SachdevQuantumPhaseTransitions}. Nevertheless, \eqref{eq:wienerrg2} illustrates the emergence of a scaling limit with enhanced symmetry very clearly;\\[0.1cm]
We consider any initial choice of function $\beta^{(0)}:\fT\rightarrow\R_{>0}$ with an asymptotic ($t>>t_{0}$) power-law behavior,
\begin{align}
\label{eq:powerlawasymp}
\beta^{(0)}_{t} & \sim \beta_{0}|I_{\sigma}|^{\nu}, & \sigma\in\cP_{t}, \nu\in\R, \beta_{0}\in\R_{>0},
\end{align}
in terms of the (geometric) sizes of the elements in the partition $\cP_{t}$ associated with $t\in\fT$. The renormalization group equation \eqref{eq:wienerrg2} implies the following asymptotic relation between the initial function $\beta^{(0)}$ and the resulting function $\beta^{(M)}$ after $M$ steps:
\begin{align}
\label{eq:powerlawrg}
\beta^{(M)}_{t_{N}} & \sim 2^{(1-\nu)M}\beta^{(0)}_{t_{N}}.
\end{align}
Thus, for exponents $\nu>1$ the states are driven towards the unstable fixed point $\beta=0$ while for exponents $\nu<1$ the states are driven to the stable fixed point $\beta=\infty$. In accordance with the definition of the dual heat-kernel states given in section \ref{sec:hkdual}, we find stable sequences of states for $\nu=1$ labelled by the free parameter $\beta_{0}$. A computation of two-point functions involving string-like holonomy fields, 
\begin{align}
\label{eq:2pointholonomy}
\h(\overline{g})_{(\tau,\tau']}=\h(\overline{g})^{-1}_{\tau}\h(\overline{g})_{\tau'},
\end{align}
with different endpoints $\tau<\tau'\in(0,L]$ reveals that $\beta_{0}$ plays the role of an inverse correlation length. As explained above, the holonomy $\h_{\tau}$ makes sense for arbitrary points $\tau\in(0,L]$ w.r.t. to the dual heat-kernel states because these are supported on continuous paths. Explicitly, we find the following two-point function for periodic boundary conditions w.r.t. the dual heat-kernel state with parameter $\beta_{0}$ for two unitary, irreducible representations $\pi,\pi'\in\hat{G}$ with matrix elements $\pi_{mn},\pi'_{m'n'}\in C(G)$:
\begin{align}
\label{eq:2pointfunction}
\omega_{\beta_{0},\D,L}((\pi_{mn}^{*})_{\tau}(\pi'_{m'n'})_{\tau'}) & = \delta_{\pi,\pi'}\delta_{m,n'}\delta_{n,m'}\!\!\!\sum_{\tilde{\pi},\tilde{\pi}'\in\hat{G}}\!\!\!\tfrac{d_{\tilde{\pi}}d_{\tilde{\pi}'}}{d_{\pi}^{2}}N^{\pi}_{\tilde{\pi}\tilde{\pi}'}e^{-\frac{\beta_{0}}{2}c_{\tilde{\pi}}(\tau'-\tau)}e^{-\frac{\beta_{0}}{2}c_{\tilde{\pi}'}(L-(\tau'-\tau))},
\end{align}
where we use the notation $(\pi_{mn})_{\tau}=\pi_{mn}\circ\h_{\tau}$ (and similarly for the primed objects). $N^{\pi}_{\tilde{\pi}\tilde{\pi}'}$ denotes the multiplicity of $\pi$ in $\tilde{\pi}\otimes\tilde{\pi}'$, and $c_{\tilde{\pi}}, c_{\tilde{\pi}'}$ are the eigenvalues of the quadratic Casimir element. Notably, the two-point function is manifestly invariant under continuous rotations expressing the enhancement of symmetry. Moreover, the thermodynamic limit, $L\rightarrow\infty$, can be trivially performed at the level of the two-point function because of (local) uniform convergence of \eqref{eq:2pointfunction}:
\begin{align}
\label{eq:2pointfunctionthermo}
\omega_{\beta_{0},\D,L}((\pi_{mn}^{*})_{\tau}(\pi'_{m'n'})_{\tau'}) & = \delta_{\pi,\pi'}\delta_{m,n'}\delta_{n,m'}e^{-\frac{\beta_{0}}{2}c_{\pi}.(\tau'-\tau)}
\end{align}
We conclude this section with a few remarks:
\begin{remark}[Brownian motion and 1d Ising models]
\label{rem:specialwiener}
For $G=\R$ and free boundary condtions, the formula \eqref{eq:2pointfunction} yields the covariance of the Wiener process,
\begin{align*}
\omega_{\beta_{0},\D,L}(x_{\tau}x_{\tau'}) = \beta_{0}\min(\tau,\tau').
\end{align*}
Also for $G=\R$, but with periodic boundary conditions and the Mehler kernel (see e.g.~\cite{BattleWaveletsAndRenormalization}) instead of the heat kernel, we find the covariance,
\begin{align*}
\omega_{\beta_{0},\D,L}(x_{\tau}x_{\tau'}) = \tfrac{1}{2(1-e^{-2\beta_{0}L})}(e^{-2\beta_{0}(\tau'-\tau)}+e^{-2\beta_{0}(L-(\tau'-\tau))}),
\end{align*}
of the Ornstein-Uhlenbeck process from \eqref{eq:2pointfunction}. Interestingly, the cases $G=\Z_{n}$, $n\in\N_{\geq2}$, can be treated as well, for example, by considering the embedding $\Z_{n}\subset U(1)$ via roots of unity. Moreover, as noted in section \ref{sec:hkdual} above, the dual heat-kernel states have simple extensions to the full semi-continuum field algebra $\fA_{\D,L}$ for discrete $G$.\\[0.1cm]
Amusingly, in the case $G=\Z_{2}$, the restriction of the dual heat-kernel state to the commutative subalgebra $C(\overline{\scrA}_{\Z_{2}})$ is equivalent to the 1d Ising model because of self-duality which, therefore, has a rotationally invariant two-point function:
\begin{align*}
\omega_{\beta_{0},\D,L}(\sigma_{\tau}\sigma_{\tau'}) = \tfrac{1}{(1-e^{-2\beta_{0}L})}(e^{-\beta_{0}(\tau'-\tau)}+e^{-\beta_{0}(L-(\tau'-\tau))}).
\end{align*}
To understand how this formula arises, we note that the well-know renormalization group equation of the 1d Ising model,
\begin{align*}
\beta'_{\textup{Ising}} & = \tfrac{1}{2}\log(\cosh(\beta_{\textup{Ising}})),
\end{align*}
is related to \eqref{eq:wienerrg2} by the non-linear transformation:
\begin{align*}
\beta_{\textup{Ising}} & = -\tfrac{1}{2}\log(\tanh(\beta)).
\end{align*}
We point out that the existence of the scaling limit of the 1d Ising model does not contradict the absence of a phase transition because we only require existence of an unstable critical point (here: $\beta=0$ resp. $\beta_{\textup{Ising}}=\infty$). On $\fA_{\D,L}$ the dual heat-kernel state corresponds to a weak-coupling limit of the $O(1)$-quantum rotor model (cp. section \ref{sec:o2rotor}). Interestingly, an identical duality relation as that between $\beta$ and $\beta_{\textup{Ising}}$ was found by Chatterjee \cite{ChatterjeeWilsonLoopsIn} in 4-dimensional Ising gauge theory very recently. This fits with the anticipated similarities between 2-dimensional spin systems and 4-dimensional gauge systems, cf.~\cite{KogutAnIntroductionTo}.
\end{remark}
\begin{remark}[Hunt's theorem]
\label{eq:thmHunt}
For Lie groups of compact type, the continuous nature of the scaling limit of the dual heat-kernel states can be seen as a result of Hunt's theorem and the L{\'e}vy-Khintchine formula for infinitely divisible probability measures \cite[Theorems 5.3.3 \& 5.5.1]{ApplebaumProbabilityOnCompact}. In view of this theorem, a simple derivation of \eqref{eq:2pointfunction} follows from the observation that the partition function $Z_{t}(\beta_{t})$, i.e.~the normalization of the integral \eqref{eq:hkdualformula} for periodic boundary conditions, of the dual-heat kernel state at each finite level given by $t\in\fT$ is given by the heat-kernel trace:
\begin{align*}
 Z_{t}(\beta_{t}) = \Tr_{L^{2}(G)}\left(e^{-\frac{\beta_{t}n(t)}{2}\Delta_{G}}\right) & = \rho_{\beta_{t}n(t)}(1_{G}) = \rho_{\beta_{0}L}(1_{G}) = Z_{L}(\beta_{0}),
\end{align*}
where $n(t)$ is the number of intervals in the partition $\cP_{t}$. Thus, the Laplacian $\Delta_{G}$ is seen as the Hunt generator of the convolution semi-group corresponding to the dual heat-kernel states. Now, the two-point function is obtained from an insertion of the operators $\pi_{mn},\pi'_{m'n'}$ at $\tau,\tau'$:
\begin{align*}
\omega_{\beta_{0},\D,L}((\pi_{mn}^{*})_{\tau}(\pi'_{m'n'})_{\tau'}) & = Z_{L}(\beta_{0})^{-1}\Tr_{L^{2}(G)}\left(e^{-\beta_{0}(L-\tau')\Delta_{G}}\pi'_{m'n'}e^{-\beta_{0}(\tau'-\tau)\Delta_{G}}\pi_{mn}e^{-\beta_{0}\tau\Delta_{G}}\right).
\end{align*}
\end{remark}
\begin{remark}[Compact quantum groups]
\label{rem:quantumgroupwiener}
From the preceding remark, we see that it is possible to generalize the dual heat-kernel states to a noncommutative setting in a different way than considered above: Instead of extending these states from a commutative subalgebra to the semi-continuum field algebra, e.g.~from $C(\overline{\scrA})$ to $\fA_{\D,L}$ for a compact group $G$, we may replace the (compact) group $G$ by a (compact) quantum group $\mathds{G}$ and substitute the heat-kernel trace by its analogue in terms of the quadratic Casimir element. This also makes sense from the perspective of the inductive-limit constructions because we have a coproduct $C(\mathds{G})\rightarrow C(\mathds{G})\times C(\mathds{G})$. For $\mathds{G}=SU_{q}(2)$, $q\in\R, |q|\neq 0,1$, we find \cite{KakehiLogarithmicDivergenceOf, MasudaReprersentationsOfThe}:
\begin{align*}
Z_{t}(\beta_{t}) & = \sum^{\infty}_{n=1}n^{2}e^{-\frac{\beta_{t}n(t)}{2}\left(\frac{q^{\frac{n}{2}}-q^{-\frac{n}{2}}}{q-q^{-1}}\right)^{2}},
\end{align*}
where $c_{n} = \left(\frac{q^{\frac{n}{2}}-q^{-\frac{n}{2}}}{q-q^{-1}}\right)^{2}$ is the $n$th eigenvalue of the quantum Casimir element $C_{q}$ and $d_{n} = n^{2}$ is its multiplicity. As before, the expression for $Z_{t}(\beta_{t}) = Z_{L}(\beta_{0})$ can be used to generate two-point functions of the dual heat-kernel state (and more generally $n$-point functions).
\end{remark}

\section{Conclusion}
\label{sec:con}
The results obtained on the operator-algebraic construction of scaling limits of Hamiltonian lattice gauge theories in the previous sections indicate various possibilities for generalizations and directions for further research.\\[0.1cm]
First of all, we intend to complete the analysis of section \ref{sec:ym11}, especially section \ref{sec:timezero}, for non-abelian groups (at least compact ones) to the same as extend as for abelian groups. A related generalization in view of duality is the replacement of the structure group $G$ by a (compact or discrete) quantum group $\mathds{G}$ because already the Fourier dual of a compact non-abelian group is a discrete quantum group, see section \ref{sec:fourier} and remark \ref{rem:quantumgroupwiener}. Such generalization might even be necessary to adapt the proofs of the results on the semi-continuum von-Neumann field algebras to the non-abelian case as these exploit the Fourier transform on $G$, for details see \cite{BrothierConstructionsOfConformal}.\\
Another important problem is the extension of our analysis to higher dimensions as well as models beyond pure gauge theory with richer field content, i.e. more complicated field algebras on single finite lattices, for example, including fermions as in \cite{KijowskiOnTheGauss, JarvisOnTheStructure, GrundlingQCDOnAn, GrundlingDynamicsForQCD} or scalar fields. A main reason for generalizing the model under consideration is the goal to understand and describe scaling limits of gauge theories with non-trivial local observables in an operator-algebraic formulation thereby coming entering the realm of algebraic quantum field theory more immediately. Naturally, Yang-Mills theory in 2+1 dimensions on the one hand and Higgs or Schwinger models on the other hand suggest themselves, cp. \cite{MilstedQuantumYangMills} for the former and \cite{BuyensMatrixProductStates, BuyensConfinementAndString} for the latter. In more than one spatial dimension, we expect the difference between abelian and non-abelian groups as well as discrete and non-discrete groups to become more pronounced due to the more complicated structure of the phase diagram of lattice gauge theories (confinement-deconfinement transition, ``freezing'' phase), see \cite{GuthExistenceProofOf, FroehlichMasslessPhasesAnd, GoepfertProofOfConfinement, BorgsLatticeYangMills, BorgsConfinementDeconfinementAnd}. A further complication that we expect to appear in higher dimensions is the necessity for field-strength renormalization which may render the time-zero formulation invalid in the continuum limit. A possible way to avoid this is the use of lattice approximations to spacetime localized field algebras as advocated in remark \ref{rem:finiteprop}. An essential ingredient for the extension to other models is the concrete realization of the Wilson-Kadanoff renormalization group by $*$-morphisms (or ucp maps) -- a fact that is known from the Ising model where different local algebras (or locality structures) are produced depending on whether we consider scaling limits of lattice spins or lattice fermions \cite{SchroerTheOrderDisorder, BostelmannQuantumEnergyInequality}. The realization we have chosen for this work reflects the geometric behavior of holonomies under composition of reference paths and lies as such at the heart of lattice gauge theory. The extension of the action of the renormalization group to electric fields -- necessary in the Hamiltonian formulation -- follows the MERA scheme \cite{MilstedQuantumYangMills} but can also be understood from a geometric perspective \cite{StottmeisterCoherentStatesQuantumIII}. In one spatial dimension, this choice appears to be distinguished because directed paths (without re-tracings) between two vertices in a lattice are unique, unless we allow for multiple loops in the case of periodic boundary conditions. In two or more spatial dimensions, this uniqueness is lost and there are various possibilities, see section \ref{sec:ymd1} below. In higher dimensions, our present choice can be visualized in the following way: A field operator at level $N\in\N_{0}$ associated with a holonomy along a path in a given lattice $\gamma_{N}$ is identified with the field operator   corresponding to the same path in a lattice $\gamma_{N'}$ at any level $N'>N$. In contrast, field operators associated with electric fields at level $N$ are localized ever closer to the endpoints of their respective edges by the renormalization group flow. Such behavior w.r.t.~electric fields will require additional care in higher dimensions as it will intuitively lead to pointlike fields (or rather germs) in the scaling limit and, thus, will entail distributional expectations values. To what extend it is possible to avoid distributional limits by modifications of the realization of the renormalization group will be investigated in future work. In view of the classical works by Ba{\l}aban \cite{BalabanAveragingOperationsFor} and Federbush \cite{FederbushAPhaseCell1} in the Euclidean framework, our formulation of the renormalization group is conceptually very close to that of Federbush as we think of the elements of our inductive-limit field algebras as compatible assignments of field operators along families of lattices. In contrast, Ba{\l}aban implements an averaging procedure on gauge field configurations that corresponds to a balanced division of a lattice holonomy at level $N$ among all possible holonomies in a sublattice at level $N'>N$. Clearly, both points of view are equivalent in the sense of the duality between states and field operators described in section \ref{sec:rgp}.\\
Another matter, we plan to pursue in future work, is the relation between the formulation of Wilson-Kadanoff renormalization discussed in section \ref{sec:rgp} and other operator-algebraic approaches to scaling limits, notably \cite{BuchholzScalingAlgebrasAnd1, BuchholzScalingAlgebrasAnd2} in the context of algebraic quantum field theory.\\
Concerning Jones' actions of Thompson's groups, which are unitarily implemented on the net of local semi-continuum von-Neumann field algebras of YM$_{1+1}$ in the representations induced by the (homogeneous) heat-kernel states, potential connections with the Euclidean formulation and its invariance under area preserving diffeomorphisms might be worth investigating. In relation to the latter, we are planning to investigate generalizations of these actions to suitable replacements of Thompson's groups in higher dimension \cite{BrinHigherDimensionalThompson, FunarDiffeomorphismGroupsOf}.
\vspace{0.25cm}
We close with a discussion of two generalized models involving the heat-kernel and dual-heat kernel states in an essential way.

\subsection{YM$_{1+1}$ and the $O(2)$-quantum rotor model}
\label{sec:o2rotor}
In section \ref{sec:can11}, it is shown that heat-kernel states and dual-heat kernel states are related to the strong- and weak-coupling limits of the Kogut-Susskind Hamiltonian \eqref{eq:KShamiltonian}. In particular, we argue that the dual-heat kernel states can be used to give meaning to the naively non-existent magnetic part of the Kogut-Susskind Hamiltonian in 1+1 dimensions. Applying the Trotter-Kato product formula to the convolution-kernel operators associated with the two types of states allows to introduce a model resembling the full Kogut-Susskind Hamiltonian:
\begin{align}
\label{eq:trotterkato11}
e^{-H^{(N)}_{\beta_{N},\tilde{g}_{N}}} & = \textup{strong}\lim_{n\rightarrow\infty} \left(\left(\lambda(\rho_{\beta_{N}/n})M(\rho_{\tilde{g}_{N}n/\beta_{N}})\right)^{\otimes |E(\gamma_{N})|}\right)^{n}.
\end{align}
It follows from the boundedness of the heat-kernel $\rho_{\beta}$, $\beta\in\R_{>0}$, on $G=U(1)$ (or other compact Lie groups or finite groups) and the Kato-Rellich theorem that \eqref{eq:trotterkato11} defines an essentially self-adjoint Hamiltonian $H^{(N)}_{\beta_{N},\tilde{g}_{N}}$ on $C^{\infty}(\scrA^{E(\gamma_{N})})$. For small relative coupling $\tilde{g}_{N}$, the asymptotic formula \eqref{eq:KSweakasymp} shows that the Hamiltonian (up to an energy renormalization) corresponds to that of the $O(2)$-quantum rotor model (in relative-angle coordinates between neighboring spins):
\begin{align}
\label{eq:O2hamiltonian}
H^{(N)}_{\beta_{N},\tilde{g}_{N}} & = \beta_{N}\sum_{e\in E(\gamma_{N})}\left(-\tfrac{1}{2}\Delta_{\varphi_{e}}-\tfrac{2}{\tilde{g}_{N}}\cos(\varphi_{e})\right),
\end{align}
with $g_{e} = e^{i\varphi_{e}}\in U(1)_{e}$, $e\in E(\gamma_{N})$. The analysis of this model from the operator-algebraic perspective described in this work will be interesting as it should posses a non-trivial (quantum) phase transition of Kosterlitz-Thouless type \cite{KosterlitzOrderingMetastabilityAnd, SachdevQuantumPhaseTransitions} and, thus, should share certain aspects, e.g. vortex condensation, with higher dimensional Yang-Mills theory. For a rigorous treatment in the setting of 2-dimensional Euclidean quantum field theory, we refer to \cite{FroehlichTheKosterlitzThouless,FroehlichKosterlitzThoulessTransition}. A closely related model is the principal chiral field in 1+1 dimension \cite{MilstedQuantumYangMills}, where all vertices $v\in V(\gamma)$ of a given lattice $\gamma\in\Gamma_{\D,L}$ are collapsed onto one. In both models, the gauge symmetry is reduced to a global $G$-action, i.e.~the weak-coupling part of the state is only invariant under constant elements of the gauge group $\overline{G}$ (for periodic boundary conditions). Although, it will be interesting to analyze whether there exists a unitary action of a sufficiently regular subgroup of $\overline{G}$ in the scaling limit as in the case of the dual heat-kernel state, see the end of section \ref{sec:timezero}.

\subsection{Yang-Mills theory in higher dimensions}
\label{sec:ymd1}
Finally, let us comment on the construction of canonical states for Yang-Mills theory in higher dimensions, especially 2+1 dimension (YM$_{2+1}$), and how the expected difficulties in obtaining a scaling limit are reflected in our operator-algebraic approach. Clearly, the ultimate goal is to study the family of full Kogut-Susskind Hamiltonians $\{H^{(N)}_{\pi}\}_{N\in\N_{0}}$ (see \eqref{eq:KShamiltonian}) and the renormalization group flow of associated states \eqref{eq:statetrianglerg}, e.g. that of Gibbs states. Similar to the $O(2)$-quantum rotor model, we do not expect that the renormalization group flow can be contained to a simple class of states, but instead approximation methods, such as analogues of low and high temperature expansions \cite{BorgsConfinementDeconfinementAnd, OitmaaSeriesExpansionMethods}, if we start with initial families of Gibbs states, will be essential. Nevertheless, as we have seen from the simple case of the dual heat-kernel state for $G = \Z_{2}$, not only choice of the initial families of states is important for the study of the renormalization group flow but also construction of the renormalization group $\{\alpha^{N'}_{N}\}_{N<N'\in\N_{0}}$. As argued above, the latter reflects the choice of basic field operators we intend to evaluate a potential scaling limit $\omega^{(\infty)}_{\infty}$ on in the continuum.

\subsubsection{Canonical states in 2+1 dimensions}
\label{sec:can21}
\begin{figure}[t]
	\begin{tikzpicture}
	
	
	\foreach \x in {0} 
		\foreach \y in {0,1}
		\foreach \v in {0,1} 	
		{
			\draw[->-=0.55, thick] (1+2*\x,4+2*\y) to (3+2*\x,4+2*\y);
			\draw[->-=0.55, thick] (1+2*\y,4+2*\x) to (1+2*\y,6+2*\x);
			\filldraw (1+2*\y,4+2*\v) circle (2pt);
		}
	
	\draw (3.25,5) node{$e_{1}$};
	\draw (2,6.25) node{$e_{2}$};
	\draw (0.75,5) node{$e_{3}$};	
	\draw (2,3.75) node{$e_{4}$};
	\draw (2,5) node{$p$};
	
	\draw (4,5) node{$\subset_{\uL}$};
	
	\foreach \x in {0} 
		\foreach \y in {0,1}
		\foreach \v in {0,1} 	
		{
			\draw[->-=0.6, thick] (6+2*\x,4+2*\y) to (5+2*\x,4+2*\y);
			\draw[->-=0.55, thick] (5+1*\y,4+2*\x) to (5+1*\y,6+2*\x);
			\filldraw (5+2*\y,4+2*\v) circle (2pt);
		}
	\foreach \x in {0} 
		\foreach \y in {0,1}
		\foreach \v in {0,1} 	
		{
			\draw[->-=0.6, thick] (6+2*\x,4+2*\y) to (7+2*\x,4+2*\y);
			\draw[->-=0.55, thick] (6+1*\y,4+2*\x) to (6+1*\y,6+2*\x);
			\filldraw (5+1*\y,4+2*\v) circle (2pt);
		}
	
	\draw (7.25,5) node{$e_{1}$};
	\draw (6.5,6.25) node{$e_{21}$};
	\draw (5.5,6.25) node{$e_{22}$};
	\draw (4.75,5) node{$e_{3}$};	
	\draw (6.5,3.75) node{$e_{41}$};
	\draw (5.5,3.75) node{$e_{42}$};
	\draw (5.5,5) node{$p_{2}$};
	\draw (6.5,5) node{$p_{1}$};
	\draw (4.75,6.25) node{$v_{22}$};
	\draw (4.75,3.75) node{$v_{42}$};
	\draw (6,3.75) node{$v_{5}$};

\end{tikzpicture}
\caption{Elementary (cofinal) refinement of a face $f\in F(\gamma_{N})$ with boundary $\partial p = \{e_{4},e^{-1}_{3},e^{-1}_{2},e_{1}\}\subset E(\gamma_{N})$ into intermediate faces $p_{1}$ and $p_{2}$ with boundaries $\partial p_{1} = \{e_{41},e^{-1}_{5},e^{-1}_{21},e_{1}\}$ and $\partial p_{2} = \{e^{-1}_{42},e^{-1}_{3},e_{22},e_{5}\}$. The associated holonomies are: $g_{p} = g_{e_{4}}g^{-1}_{e_{3}}g^{-1}_{e_{2}}g_{e_{1}}$, $g_{p_{1}} = g_{e_{41}}g^{-1}_{e_{5}}g^{-1}_{e_{21}}g_{e_{1}}$ and $g_{p_{2}} = g^{-1}_{e_{42}}g^{-1}_{e_{3}}g_{e_{22}}g_{e_{5}}$ .}
\label{fig:elementaryfacerefinement}
\end{figure}
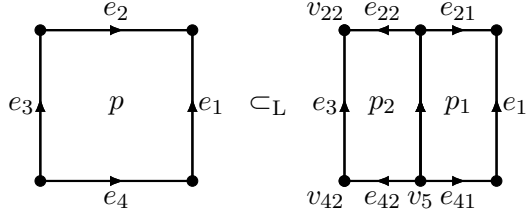

Although, we cannot provide the construction of a scaling limit starting from families of general Gibbs states of $\{H^{(N)}_{\pi}\}_{N\in\N_{0}}$ in YM$_{2+1}$, we are able to show that the Gibbs states of the strong- and the weak-coupling limit admit scaling limits for the choice of renormalization group $\{\alpha^{N}_{N'}=\alpha^{\gamma_{N}}_{\gamma_{N'}}\}_{N<N'\in\N_{0}}$ introduced in section \ref{sec:scales}, equation \eqref{eq:algsys} and following, and used throughout section \ref{sec:ym11}.\\[0.25cm]
\textit{Strong coupling limit}.
\label{sec:strong}
For compact groups $G$, the existence of scaling limits for Gibbs states of the strong-coupling limits \eqref{eq:KShamiltonianstrong} of the family $\{H^{(N)}_{\pi}\}_{N\in\N_{0}}$ follows in complete analogy with the 1+1 dimensional case, which is implicit in section \ref{sec:hk}. The key observation is that the level-$N$ Gibbs states,
\begin{align}
\label{eq:stronglimitfactor}
\tfrac{1}{Z^{(N)}(\beta_{N})}\lambda(\rho^{(N)}_{\beta_{N}}) & = \tfrac{1}{Z^{(N)}(\beta_{N})}e^{-\beta_{N}H^{(N)}_{s}} = \otimes_{e\in E(\gamma_{N})}\tfrac{1}{Z(\beta_{N}(e))}\exp\left(-\tfrac{\beta_{N}(e)g_{N}^{2}}{2a_{\gamma_{N}}}\Delta_{G_{e}}\right),
\end{align}
factorize, which reduces the consistency to the 1+1 dimensional case for our choice of renormalization group.\\[0.1cm]
More generally, according to the statement on simple tensor product states in section \ref{sec:compact}, specifically \eqref{eq:simpletensorstates} and the preceding discussion, this observation will remain valid in any dimension $d+1$ if we stick to our specific choice of renormalization group. In contrast, if we change the form of the renormalization group this conclusion will not necessarily persist (see below). \\[0.25cm]
\textit{Weak coupling limit}.
\label{sec:weak}
The existence of scaling limits for Gibbs states associated with the weak-coupling \eqref{eq:KShamiltonianweak} of the family $\{H^{(N)}_{\pi}\}_{N\in\N_{0}}$ is slightly more involved but follows from the well-known behavior of the Villain action \eqref{eq:KSweakasymp} in Euclidean YM$_{2}$ under refinements of two-dimensional partitions, see \cite{FineQuantumYangMillsRiemann, WittenOnQuantumGauge, LevyYangMillsMeasure}, also \cite{AshtekarSU(N)QuantumYang, FleischhackANewType} for a construction of the Euclidean YM$_{2}$ scaling limit in terms of the Wilson action.\\[0.1cm]
To be more explicit about this, let us consider a single face or plaquette $p\in F(\gamma_{N})$ of a lattice $\gamma_{N}$ and its elementary refinement $\{p_{1},p_{2}\}$ according to the procedure that leads to the next lattice $\gamma_{N+1}$ in cofinal sequence, cp. remark \ref{rem:cofinal} and see figure \ref{fig:elementaryfacerefinement}.

On the commutative part of the field algebra associated with $p$ and $\{p_{1},p_{2}\}$ the dual heat-kernel states take the explicit form:
\begin{align}
\label{eq:weaklimit2d}
\omega^{(\partial p)}_{\beta_{N}(p)}(M(f)) & = \int_{G^{\partial p}}\left(\prod^{4}_{j=1}dg_{e_{j}}\right)\rho_{\beta_{N}(p)}(g_{p})f(g_{e_{4}},g_{e_{3}},g_{e_{2}},g_{e_{1}}), \\ \nonumber
\omega^{(\partial p_{1}\cup\partial p_{2})}_{(\beta'_{N}(p_{1}),\beta'_{N}(p_{2}))}(M(f')) & = \int_{G^{\partial p_{1}\cup\partial p_{2}}}\left(\prod^{2}_{j=1}dg_{e_{2j-1}}\prod^{2}_{k=1}dg_{e_{(2j)i}}\right)\rho_{\beta'_{N}(p_{1})}(g_{p_{1}})\rho_{\beta'_{N}(p_{2})}(g_{p_{2}}) \\[-0.25cm] \nonumber 
 & \hspace{6cm} \times f'(g_{e_{5}},g_{e_{42}},g_{e_{41}},g_{e_{3}},g_{e_{22}},g_{e_{21}},g_{e_{1}})
\end{align}
where $f\in C_{b}(G^{\partial p})$, $f'\in C_{b}(G^{\partial p_{1}\cup\partial p_{2}})$, and $\partial p\subset E(\gamma)$ denotes the set of boundary edges of a plaquette $p\in F(\gamma)$. The renormalization group element facilitating the refinement $\partial p \subset_{\uL} \partial p_{1}\cup \partial p_{2}$ is given by:
\begin{align}
\label{eq:elementaryfacerg1}
\alpha^{\partial p}_{\partial p_{1}\cup \partial p_{2}}(M(f)) & = \lambda(\varsigma^{\partial p}_{\partial p_{1}\cup \partial p_{2}}(\delta^{\otimes\partial p}_{1_{G}}\otimes f)),
\end{align}
where
\begin{align}
\label{eq:elementaryfacerg2}
\varsigma^{\partial p}_{\partial p_{1}\cup \partial p_{2}}(\delta^{\otimes\partial p}_{1_{G}}\otimes f)(h_{e_{5}},g_{e_{5}};...;h_{e_{1}},g_{e_{1}}) & = \delta^{\otimes \partial p_{1}\cup \partial p_{2}}_{1_{G}}(h_{e_{5}},...,h_{e_{1}})f(g_{e_{41}}g^{-1}_{e_{42}},g_{e_{3}},g_{e_{21}}g^{-1}_{e_{22}},g_{e_{1}}).
\end{align}
Now, the expression for $\omega^{(\partial p_{1}\cup\partial p_{2})}_{(\beta'_{N}(p_{1}),\beta'_{N}(p_{2}))}(\alpha^{\partial p}_{\partial p_{1}\cup \partial p_{2}}(M(f)))$ can be partially evaluated because we can perform the group integral over the shared boundary edge $e_{5} = \partial p_{1}\cap \partial p_{2}$ by using the heat-kernel convolution identity:
\begin{align}
\label{eq:boundaryintegral}
\int_{G_{e_{5}}}\!\!\!dg_{e_{5}}\rho_{\beta'_{N}(p_{1})}(g_{e_{41}}g^{-1}_{e_{5}}g^{-1}_{e_{21}}g_{e_{1}})\rho_{\beta'_{N}(p_{2})}(g^{-1}_{e_{42}}g^{-1}_{e_{3}}g_{e_{22}}g_{e_{5}}) & = \rho_{\beta'_{N}(p_{1})+\beta'_{N}(p_{2})}(g^{-1}_{e_{42}}g^{-1}_{e_{3}}g_{e_{22}}g^{-1}_{e_{21}}g_{e_{1}}g_{e_{41}}) \\ \nonumber
 & = \rho_{\beta'_{N}(p_{1})+\beta'_{N}(p_{2})}(g_{e_{41}}g^{-1}_{e_{42}}g^{-1}_{e_{3}}g_{e_{22}}g^{-1}_{e_{21}}g_{e_{1}}).
\end{align}
Combining equations \eqref{eq:elementaryfacerg2} and \eqref{eq:boundaryintegral} we find the renormalization group equation:
\begin{align}
\label{eq:weaklimit2drg}
\omega^{(\partial p_{1}\cup\partial p_{2})}_{(\beta'_{N}(p_{1}),\beta'_{N}(p_{2}))}(\alpha^{\partial p}_{\partial p_{1}\cup \partial p_{2}}(M(f))) & = \omega^{(\partial p)}_{\beta'_{N}(p_{1})+\beta'_{N}(p_{2})}(M(f)).
\end{align}
This implies that we have non-trivial scaling limits whenever $\beta_{N}(p)\sim \beta_{0}A_{p}$ scales asymptotically ($N>>0$) like the area $A_{p}$ of the plaquette $p$, cp.~the discussion of equations \eqref{eq:powerlawasymp} and \eqref{eq:powerlawrg}. As for the dual heat-kernel state in 1+1 dimensions, a weak-coupling limit Gibbs state $\omega^{(N)}_{\beta_{N}}$ has a simple extension to $B(L^{2}(\scrA^{E(\gamma_{N})}))$ for discrete groups $G$. In terms of convolution kernels this extension is given by:
\begin{align}
\label{eq:weaklimit2dextension}
F^{(N)}_{\beta_{N}}(h,g) & = \delta^{\otimes E(\gamma_{N})}_{1_{G}}(h)\prod_{p\in F(\gamma_{N})}\rho_{\frac{a_{\gamma_{N}}g^{2}_{N}}{\beta_{N}}}(g_{p}), & (h,g)\in\Pi^{E(\gamma_{N})}.
\end{align}
The renormalization group equation \eqref{eq:weaklimit2drg} is not affected by this extension. A main difference to the dual heat-kernel state in 1+1 dimensions is the gauge invariance of \eqref{eq:weaklimit2d} and \eqref{eq:weaklimit2dextension}, i.e.~invariance under the action $\alpha_{\overline{\tau}}:\overline{G}\act\fA_{\D,L}$. For compact but non-discrete $G$, we cannot say anything explicit about extensions to the field algebras $B(L^{2}(\scrA^{E(\gamma_{N})}))$, $N\in\N_{0}$, at this point. But, it would clearly be interesting to understand if there were suitable Cameron-Martin spaces associated with the measures on $\overline{\scrA}$ induced by the scaling limits of the weak-coupling limits. Such Cameron-Martin space could act a replacements of the field momenta $\underline{G}$.\\[0.25cm]
\textit{Other choices of the renormalization group}.
If we choose a different realization of the renormalization group $\{\alpha^{N}_{N'}=\alpha^{\gamma_{N}}_{\gamma_{N'}}\}_{N<N'\in\N_{0}}$, for example, by changing the disentanglers $\{U_{N}\}_{N\in\N_{0}}$, see \eqref{eq:twistedpartialtrace} and \eqref{eq:MERAisometry} to \eqref{eq:MERAadjoint}, the consistency of the strong- and weak-coupling scaling limits can be easily broken. For example, the disentangler (or rather its equivalence class according to \eqref{eq:equivalencerg}) defined in \cite{MilstedQuantumYangMills} for $G = U(1), SU(2)$ leads to non-trivial coherence conditions \eqref{eq:statecoherence} for the inhomogeneous heat-kernel states and entails the invalidity of almost all homogeneous heat-kernel states. Only the strong-coupling vacuum, $\omega_{\infty,\D,L}$, is consistent for this class of disentangler because there is a choice of representative in the latter that leaves the restriction of the strong coupling vacuum to a finite level $N\in\N_{0}$ invariant. For the elementary refinement of a plaquette $p$ into plaquettes $p_{1},p_{2}$, see figure \ref{fig:elementaryfacerefinement}, this disentangler, $U^{\partial p}_{\partial p_{1}\cup \partial p_{2}}$, is defined by:
\begin{align}
\label{eq:OMdisentangler}
(U^{\partial p}_{\partial p_{1}\cup \partial p_{2}}\psi)(g_{e_{5}},...,g_{e_{1}}) & = \psi(\sqrt{g^{-1}_{p}(v_{5})}g^{-1}_{e_{42}}g^{-1}_{e_{3}}g_{e_{22}}g_{e_{5}},g_{e_{42}},g_{e_{41}}g^{-1}_{e_{42}},g_{e_{3}},g_{e_{22}},g_{e_{21}}g^{-1}_{e_{22}},g_{e_{1}}),
\end{align}
where $\psi\in C(G^{\partial p}\times G^{\{v_{5},v_{42},v_{22}\}})$ transforming in under the conjugate action of $G$ at the vertices $\{v_{22},v_{42},v_{5}\}$, and $g_{p}(v_{5}) = g^{-1}_{e_{42}}g^{-1}_{e_{3}}g_{e_{22}}g^{-1}_{e_{21}}g_{e_{1}}g_{e_{41}}$ is the holonomy around $p$ starting $v_{5}$.
\begin{align}
\label{eq:Gsqrt}
\sqrt{\ .\ } : V & \subset G\rightarrow G,
\end{align}
is a smooth, $\alpha$- and inversion-equivariant square root defined via the exponential map $\exp_{G}:U\subset\fg\rightarrow V\subset G$ on a maximal open pair of invertibility domains $(U,V)$ \cite{RieffelLieGroupConvolution}, i.e.~$\sqrt{g} = \exp_{G}(\tfrac{1}{2}\exp^{-1}_{G}(g))$ for $g\in U$, see also \cite{LandsmanMathematicalTopicsBetween, StottmeisterCoherentStatesQuantumII} where the square-root construction is discussed in the context of Weyl quantization and pseudo-differential calculus on compact Lie groups $G$. For $G = U(1), SU(2)$ the maximal invertibility domain is $G\setminus\{-1_{G}\}$ which implies that \eqref{eq:OMdisentangler} is defined almost everywhere w.r.t.~the Haar measure and, therefore, by translation invariance of the latter, defines a unitary,
\begin{align}
\label{eq:OMunitary}
U^{\partial p}_{\partial p_{1}\cup \partial p_{2}}: L^{2}(G^{\partial p}\times G^{\{v_{5},v_{42},v_{22}\}})\rightarrow L^{2}(G^{\partial p_{1}\cup\partial p_{2}}).
\end{align}
It is evident that \eqref{eq:OMdisentangler} preserves the strong-coupling vacuum:
\begin{align}
\label{eq:OMvacinv}
U^{\partial p}_{\partial p_{1} \cup \partial p_{2}}1_{\partial p_{1}\cup \{v_{5},v_{42},v_{22}\}} & = 1_{\partial p_{1}\cup \partial p_{2}},
\end{align}
where is $1_{\partial p\cup \{v_{5},v_{42},v_{22}\}}\in L^{2}(G^{\partial p}\times G^{\{v_{5},v_{42},v_{22}\}})$ and $1_{\partial p_{1}\cup \partial p_{2}}\in L^{2}(G^{\partial p_{1}\cup\partial p_{2}})$ are the respective constant unit functions. Looking at the action of the renormalization group element associated with $U^{\partial p}_{\partial p_{1}\cup \partial p_{2}}$ on a pure convolution operator, we find:
\begin{align}
\label{eq:OMconvsys}
 & (\Ad_{U^{\partial p}_{\partial p_{1}\cup \partial p_{2}}}\!\!\!(\lambda(F)\!\otimes\!\mathds{1}_{\{v_{5},v_{42},v_{22}\}})\phi)(g_{e_{5}},g_{e_{42}},g_{e_{41}},g_{e_{3}},g_{e_{22}},g_{e_{21}},g_{e_{1}}) \\ \nonumber
 & =  \int_{G^{\partial p}}\left(\prod^{2}_{j=1}dh_{e_{2j-1}}dh_{e_{(2j)1}}\right)F(h_{e_{41}},h_{e_{3}},h_{e_{21}},h_{e_{1}}) \\ \nonumber
 & \hspace{1cm}\times\phi(g^{-1}_{e_{22}}h^{-1}_{e_{3}}g_{e_{3}}g_{e_{42}}\sqrt{g^{-1}_{e_{41}}h_{e_{41}}g^{-1}_{e_{1}}h_{e_{1}}h^{-1}_{e_{21}}g_{e_{21}}g^{-1}_{e_{22}}h^{-1}_{e_{3}}g_{e_{3}}g_{e_{42}}}\sqrt{g^{-1}_{p}(v_{5})}g^{-1}_{e_{42}}g^{-1}_{e_{3}}g^{-1}_{e_{22}}g_{e_{5}},... \\ \nonumber
 & \hspace{6cm}...g_{e_{42}},h^{-1}_{e_{41}}g_{e_{41}}g^{-1}_{e_{42}},h^{-1}_{e_{3}},g_{e_{22}},h^{-1}_{e_{21}}g_{e_{21}}g^{-1}_{e_{22}},h^{-1}_{e_{1}}g_{e_{1}}),
\end{align}
where $\phi\in C^{\infty}(G^{\partial p_{1}\cup\partial p_{2}})$ and $F\in C^{-\infty}(G^{\partial p})$ (the distributional dual with suitable restrictions on its singular support). For a basic field momentum at the edge $e_{1}$, i.e.~$F(h_{e_{4}},h_{e_{3}},h_{e_{2}},h_{e_{1}}) = \delta^{\otimes 3}_{1_{G}}(h_{e_{4}},h_{e_{3}},h_{e_{2}})\delta_{h}(h_{e_{1}})$, we find:
\begin{align}
\label{eq:OMbasicfieldmomentum}
(\Ad_{U^{\partial p}_{\partial p_{1}\cup \partial p_{2}}}\!\!\!(\lambda(F)\!\otimes\!\mathds{1}_{\{v_{5},v_{42},v_{22}\}})\phi)(g_{e_{5}},...,g_{e_{1}}) & = \phi(\alpha_{g^{-1}_{e_{21}}}\!(\sqrt{h^{-1}g_{p}(\partial_{+}e_{1})}\sqrt{g^{-1}_{p}(\partial_{+}e)})g_{e_{5}},...,h^{-1}g_{e_{1}}).
\end{align}
This expression for the action of the renormalization group elements based on the proposal in \cite{MilstedQuantumYangMills} makes explicit its non-trivial implications for the coherence conditions on the heat-kernel states.\\[0.1cm]
Further complications introduced by a disentangler like \eqref{eq:OMdisentangler} are implied by the observation that it does not restrict to a map between spaces of smooth functions, based on $C^{\infty}_{c}(G)$ or $C^{\infty}(G)$, unless the exponential map is bijective.

\section*{Acknowledgements}

The authors would like to thank Roberto Longo for his constant encouragement and support. Additionally, AS would like to thank Raimar Wulkenhaar for his kind support and gratefully acknowledges generous financial support by the Alexander-von-Humboldt Foundation and the University of M{\"u}nster. Furthermore, AS would like to thank Abdelmalek Abdesselam, Daniela Cadamuro, Wojciech Dybalski, and Tobias Osborne for useful discussions. AS also acknowledges financial support and kind hospitality by the Isaac Newton Institute and the Banff International Research Station where parts of this work were developed. Both authors acknowledge financial support and kind hospitality by the Simons Center for Geometry and Physics where parts of this work were written.


\newcommand{\etalchar}[1]{$^{#1}$}

\end{document}